\documentclass[fleqn,usenatbib]{mnras}

\usepackage{newtxtext,newtxmath}

\usepackage[T1]{fontenc}

\DeclareRobustCommand{\VAN}[3]{#2}
\let\VANthebibliography\thebibliography
\def\thebibliography{\DeclareRobustCommand{\VAN}[3]{##3}\VANthebibliography}

\usepackage{graphicx}	
\usepackage{amsmath}	
\usepackage{subfigure}
\usepackage{booktabs}

\newcommand{\WH}{$W_{RP,BP-RP}-W_{K_s,J-K_s}$}
\newcommand{\MKvsWH}{$M_{K_s}$\text{ vs. }$W_{RP,BP-RP}-W_{K_s,J-K_s}$}
\newcommand{\MK}{$M_{K_s}$}
\newcommand{\Ks}{$K_s$}
\newcommand{\JKs}{$J-K_s$}
\newcommand{\M}{$M_\odot$}
\newcommand{\HI}{H\,{\small I}}
\usepackage{gensymb}
\usepackage{soul}
\usepackage{makecell}
\usepackage{subcaption}
\usepackage{ulem}
\usepackage{lscape}
\usepackage{pdflscape}  
\usepackage{booktabs}   
\usepackage{array}

\title{Machine learning reveals new Galactic C-rich and O-rich AGB sources}

\author[Kushwahaa T. et al.]{Tanya Kushwahaa,$^{1}$\thanks{E-mail: KushwahaaT@cardiff.ac.uk, 
tkushwahaa.astro@gmail.com}
Mikako Matsuura,$^{1}$
Jason A. S. Hunt,$^{2}$
Daisuke Kawata,$^{3}$
Roger Wesson,$^{1,4, 5}$  
and  \newauthor
Timothy A. Davis$^{1}$
\\
$^{1}$Cardiff Hub for Astrophysics Research and Technology (CHART),
School of Physics and Astronomy, Cardiff University, 
The Parade, Cardiff CF24 3AA, UK\\
$^{2}$
School of Mathematics \& Physics, University of Surrey, Stag Hill, Guildford, GU2 7XH, UK \\
$^3$
Mullard Space Science Laboratory, University College London, Holmbury St. Mary, Dorking, Surrey, UK \\
$^4$
Department of Physics and Astronomy, University College London, Gower Street, London WC1E 6BT, UK \\
$^{5}$ 
Department of Physics, Maynooth University, Maynooth, Co. Kildare, Ireland \\
}

\date{Accepted XXX. Received YYY; in original form ZZZ}

\pubyear{2026}

\begin{document}
\label{firstpage}
\pagerange{\pageref{firstpage}--\pageref{lastpage}}
\maketitle

\begin{abstract}

Identification of Carbon-rich (C-rich) and Oxygen-rich (O-rich) Asymptotic Giant Branch (AGB) stars is crucial to understanding the structure and dust sources of the Milky Way. We present a machine-learning (ML) classification of Galactic AGB candidates using {\it Gaia} DR3, identifying 304,090 O-rich and 4,227 C-rich sources. Our two-step approach, which involves similarity-based selection, followed by XGBoost classification that combines colour and spectroscopic features, demonstrates that a small, but high-quality training sample is sufficient for accurate identification even in massive datasets such as $ \sim$140 million sources from \textit{Gaia} DR3. The ML-identified stars occupy distinct regions in colour-magnitude and {\it Gaia}-2MASS diagrams, and comparison with stellar isochrones combined with \textit{Gaia} distances allows estimates of their masses and ages. 
Spatial distributions show that O-rich AGB stars dominate the inner Galaxy, while C-rich AGB stars are concentrated toward the anti-Galactic centre. Vertical profiles indicate that massive O-rich (5--8\M) stars remain near the Galactic mid-plane, whereas C-rich (2.3--2.5\M) and low-mass O-rich (1--1.5~\M) AGB stars are more vertically extended, with scale heights that increase with decreasing stellar mass and increasing stellar ages. C-rich AGB stars trace the Galactic warp at intermediate ages (1~Gyr), with warp amplitudes appearing larger than those traced by Cepheids. Intermediate-mass O-rich AGB stars ($\sim$300~Myr) show tentative correspondence with several spiral arm features, while the C-rich distribution is too smooth for reliable identification with the current sample. These results highlight AGB stars as powerful probes of stellar evolution and Galactic dynamical processes.

\end{abstract}




\section{Introduction}\label{introduction}

Understanding the formation and evolution of the Milky Way's current structure remains a fundamental challenge in Galactic astronomy \citep{Freeman2002, Bland-Hawthorn2016}. With the advent of large-scale stellar surveys, it is now possible to study the spatial and kinematic distribution of stars across the Galaxy in unprecedented detail \citep{Gaia2018, Skowron2019, Helmi2020, Andrae2023, Deason2024}.

The  {\it Gaia} mission \citep{Prusti2016} has revolutionised this field by providing astrometric, photometric and spectroscopic data for over a billion stars. {\it Gaia} Data Release 3 (DR3) \citep{Valenari2023, GaiaDR32023} includes precise parallax measurements for 1.8 billion stars, high-resolution spectra for 1 million stars, and low-resolution XP spectra for 220 million stars, covering wavelengths from 330 to 1100 nm. These data products enable the identification and classification of Galactic stars on a previously unattainable scale, as well as a more detailed mapping of their spatial distribution across the Galaxy.

Asymptotic Giant Branch (AGB) stars represent stellar populations ranging from a few hundred million to a few billion years in age \citep{Vassiliadis1993, Blocker1995, Karakas2007}, making them excellent tracers of past star-forming regions in the Milky Way during these epochs \citep{Marigo2011, Marigo2017}. These low- and intermediate-mass stars (1--8 \M) are in their final evolutionary stages, having evolved from the main sequence through the red giant branch (RGB) and horizontal branch phases. The AGB phase consists of an early stage (E-AGB), when shell helium burning begins, followed by a thermally pulsing stage (TP-AGB), during which episodic helium-shell flashes can dredge carbon to the surface, altering the star's carbon/oxygen (C/O) ratio. Although the AGB phase is relatively short, lasting $\leq$1\% of a star’s total lifetime \citep{Vassiliadis1993}, it contributes significantly to the chemical enrichment of the Galaxy \citep{Romano2010, Kobayashi2020} and to the dust budget of the local galaxies \citep{Dwek1980, Matsuura2009, Valiante2009}.

AGB stars are broadly divided into two major chemical types: carbon-rich (C-rich) and oxygen-rich (O-rich), based on the carbon-to-oxygen abundance ratio in their atmospheres. O-rich AGB stars typically exhibit M-type spectra, while C-rich stars correspond to the classical carbon-star sequence. 
Between these two types, S-type stars (C/O$\approx$1) represent a transitional stage during the third dredge-up \citep{Iben1983} and are characterised by ZrO and often LaO absorption bands in their optical spectra \citep{Keenan1954, VanEck1998}. They include S(M) stars, intermediate between M-type and carbon stars, and SC stars, which are closer to carbon stars and show weaker C$_2$ molecular bands \citep{Keenan1980}.
The number ratio between C-type and M-type stars serves as a tracer of the metallicity and age of the underlying stellar population \citep{Vassiliadis1993, Cioni2003, Cioni2006, Cioni2009, Boyer2013, Boyer2019}.

Nonetheless, challenges remain in understanding the spatial distribution of AGB stars, primarily due to the difficulty of identifying them within large stellar catalogues such as Gaia. Traditional selection methods, such as colour-magnitude or variability-based cuts, often suffer from contamination and incompleteness due to overlap with other evolved stellar populations, especially red giants, young dusty stars and dwarfs \citep[e.g.,][]{Lebzelter2018, Marton2019}. To address these limitations, modern machine learning (ML) techniques offer powerful tools for extracting such populations from high-dimensional survey data \citep{Ball2010, Baron2019}. Recent studies have demonstrated the potential of ML methods for classifying various stellar objects, including AGB stars, using multi-wavelength and spectroscopic data \citep{Li2018, Hernandez2021, Sanders2023, Liu2024, Zeraatgari2024}. These approaches are particularly valuable when working with complex datasets provided by {\it Gaia}, where low-resolution XP spectra and photometric measurements can be combined to classify stellar populations more efficiently and at scale.

In this paper, we identify AGB candidates from the {\it Gaia} catalogue using ML techniques. \citet{Sanders2023} previously extracted AGB stars from the {\it Gaia} DR3 long-period variable catalogue, incorporating BP/RP spectra and applying unsupervised machine learning. Their study focused on dimensionality reduction using Uniform Manifold Approximation and Projection (UMAP) and extended classification to fainter sources by combining  {\it Gaia} and 2MASS photometry.

Our work differs in both methodology and scope. We present a two-step classification pipeline: first, a similarity analysis is used to construct a candidate sample of AGB stars; second, a supervised binary classifier (in the machine learning sense of a two-class problem) is applied to distinguish between C-rich and O-rich AGB stars. This approach aims to maximise the identification of AGB candidates from {\it Gaia} DR3, offering a new dataset for studying the spatial distribution of these intermediate-age tracers and metallicity gradients across the Milky Way.

The paper is structured as follows: Sect.~\ref{gaia_xp_spectra} describes the {\it Gaia} data and the spectral characteristics of C-rich and O-rich AGB stars. The data preparation process for both known sources from Galactic catalogues and unknown sources from {\it Gaia} DR3 are detailed in Sect.~\ref{data_prep}. Sect.~\ref{two-stepmethodology} explains our two-step machine learning methodology. Results are presented in Sect.~\ref{Results}, followed by analysis and discussion in Sect.~\ref{analysis_classified_agb_and_discussions}. We conclude with a summary of our findings in Sect.~\ref{Conclusions}.

\section{Methods: database and machine learning}\label{methods}

\subsection{{\it Gaia} XP spectra}\label{gaia_xp_spectra}

\begin{figure}
\centering     
\subfigure
{\includegraphics[width=1.\linewidth]{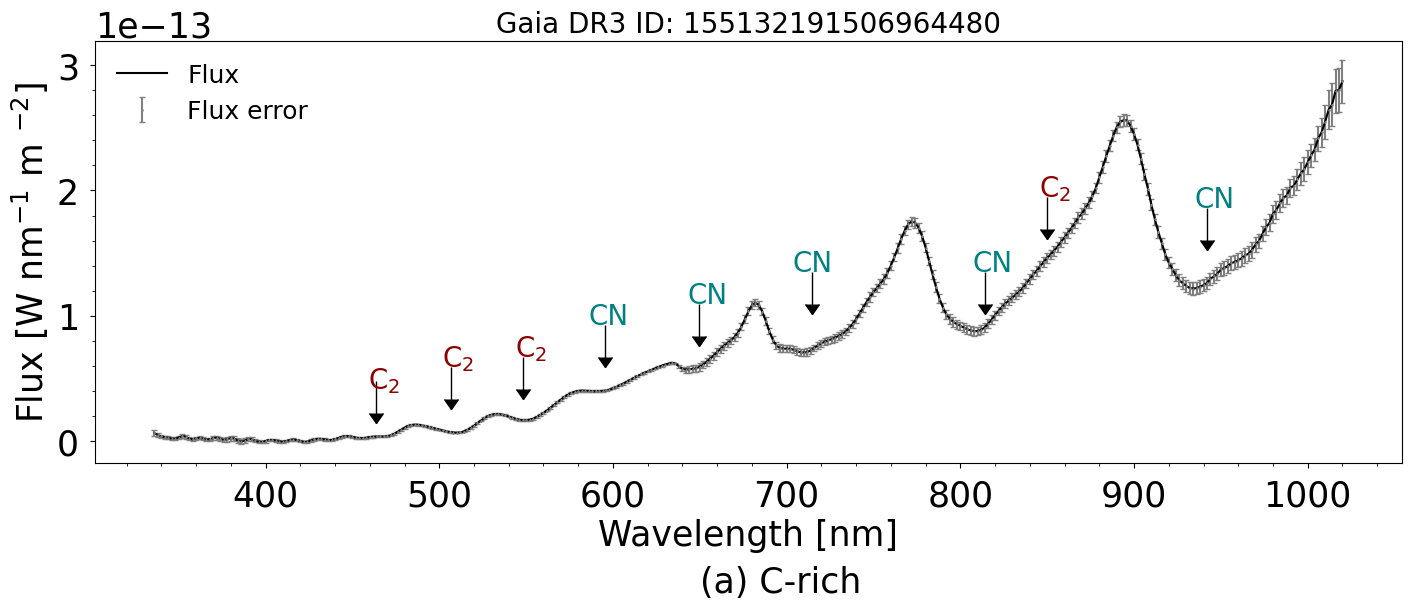}} 
\subfigure{\includegraphics[width=1.\linewidth]{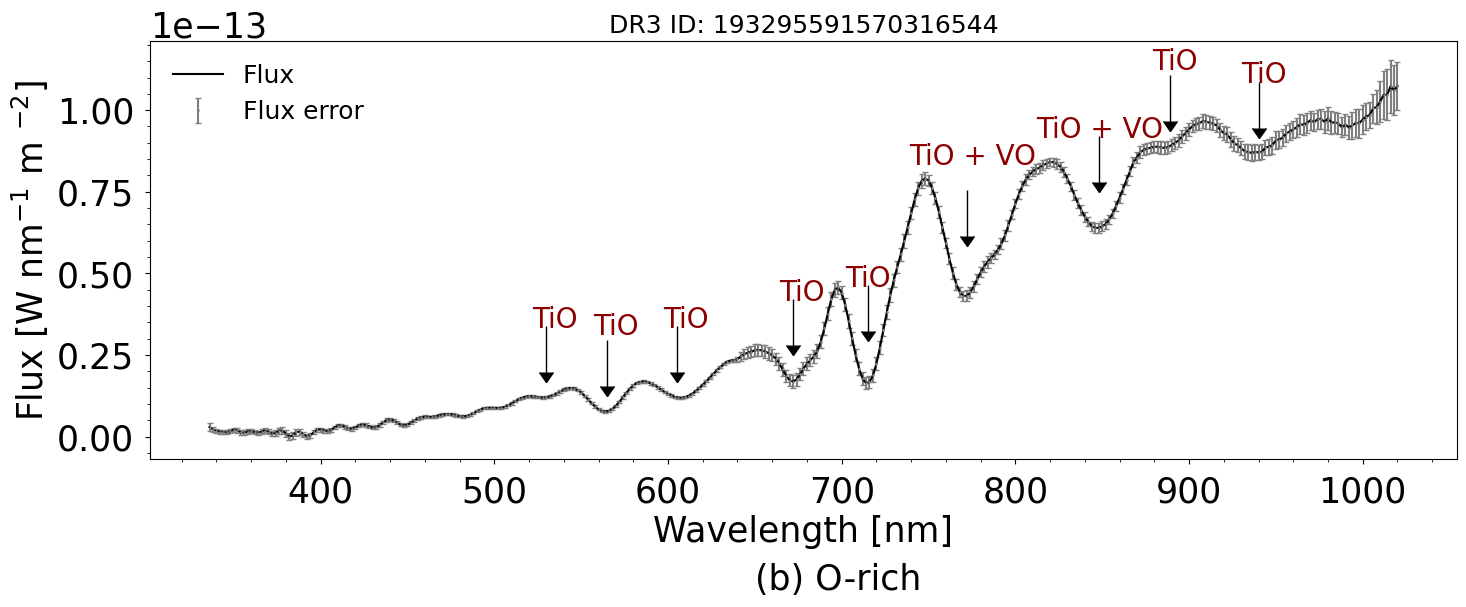}}
\subfigure{\includegraphics[width=1.\linewidth]{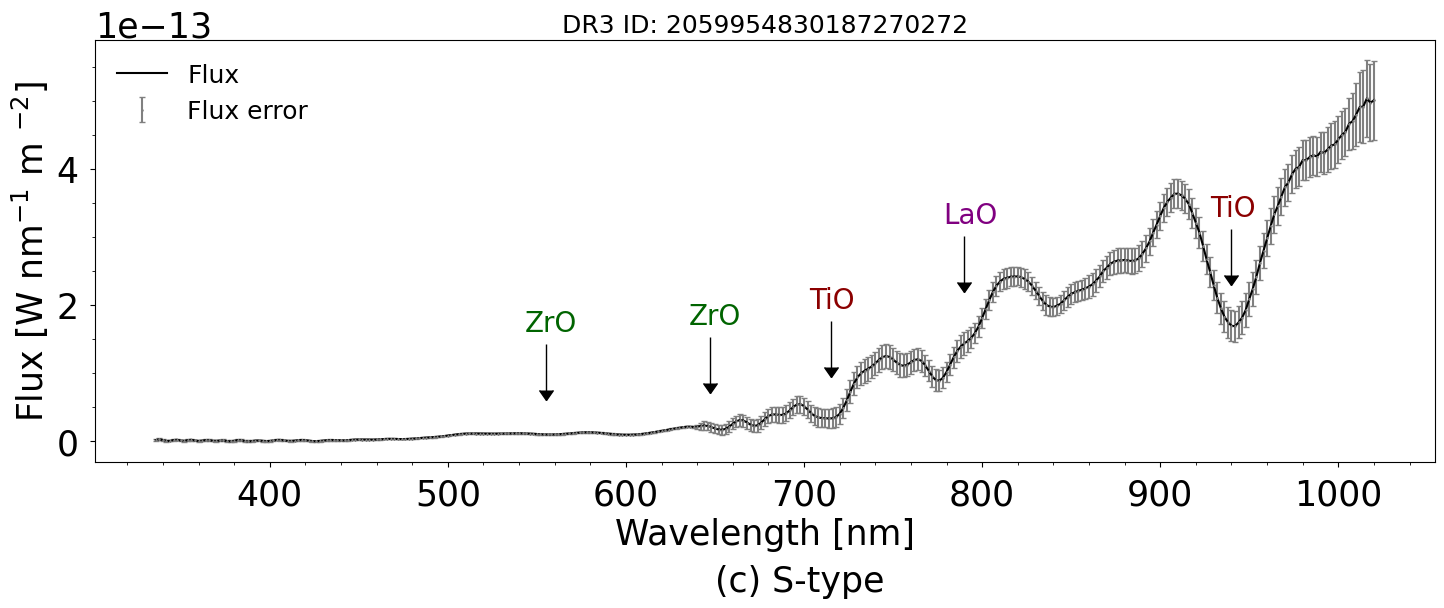}}

\caption{{\it Gaia} XP Mean Spectra of (a) Carbon star, {\it Gaia} DR3 155132191506964480, also known as TT Tau, a semi-regular variable with spectral type of C4,2--C7,4 (N3) and SiC dust \citep{Sloan1998}. Note the characteristic CN bands. (b) Oxygen-rich star, {\it Gaia} DR3 193295591570316544, also known as V$\star$V443 Aur, is a slow-irregular variable of M2 spectral type \citep{Samus2017}, showing typical TiO absorption features. (c) S-type AGB star, {\it Gaia} DR3 2059954830187270272, also known as V$\star$ AA Cyg, a semi-regular variable with spectral type S6/3 \citep{VanEck1998}, showing TiO bands together with weaker ZrO features.}
\label{C/Orich}
\end{figure}

Spectroscopically, AGB stars are classified into two main groups based on the molecular absorption features in their atmospheres. Although {\it Gaia} XP spectra have relatively low resolution (R$\sim$20-1000; \citealt{Deangeli2023}), they still detect key molecular absorption bands that are diagnostic of AGB chemistry. Examples of these spectral signatures in {\it Gaia} XP data are presented in Fig.~\ref{C/Orich}. Carbon-rich (C-type) spectra are characterised by C$_2$ and CN molecular bands, while oxygen-rich (M-type) spectra are dominated by TiO bands. This spectral difference results from the ratio of carbon to oxygen (C/O) in the stellar atmosphere. When C/O $<$ 1, oxygen dominates, and TiO features are prominent (O-rich stars). When C/O > 1, carbon dominates, and C$_2$/CN features are visible (C-rich stars). 
S-type stars are relatively rare compared with carbon stars \citep{Keenan1954}, likely because the transition from O-rich to C-rich surface chemistry is short \citep{Wright2009}.
Inspection of the \textit{Gaia} XP spectra of known S-type AGB stars, such as R~And and AA~Cyg \citep{Keenan1954, Son2025}, shows no clear S-type signature distinct from M-type spectra (Fig.~\ref{C/Orich}(c)), so S-type AGB stars are difficult to identify at the resolution of \textit{Gaia} XP spectra.

The optical spectra of O-rich AGB stars show distinctive TiO absorption bands at 0.52, 0.54, 0.72, 0.77, and 0.92 $\mu$m \citep{Phillip1969, Turnshek1985, Brett1989, Brett1990, Valenti1998}. For C-rich AGB stars, the primary molecular absorption bands span optical and NIR wavelengths (0.5-2.5$\mu$m) \citep{Loidl2001}. In the optical range, C$_2$ lines of the Swan system at 473.7 nm, 516.5 nm and 563.5 nm \citep{Swan1857} are dominant, while C$_2$ bands at 770 nm and 880 nm originate from the Phillips system \citep{Phillip1948}. Beyond 600 nm, the most prominent bandheads are from red CN \citep{Ji2016, Gonneau2016}. Similar to O-rich spectra, C-rich spectra also show a red energy spectral distribution \citep{Lancon2000}.

\subsection{Data preparation}\label{data_prep}

\subsubsection{Analysing data: {\it Gaia} DR3 XP spectra with ancillary 2MASS and WISE Data}

We retrieved {\it Gaia} DR3 sources from the {\it Gaia} source catalogue using the Gaia AIP\footnote{\url{https://gaia.aip.de/}} services, specifically selecting sources with XP spectra (\texttt{has\_xp\_continuous} = 1) which resulted in the coordinates, parallax, {\it Gaia}-band photometry and BP/RP coefficients of the {\it Gaia} DR3 XP spectra. To obtain flux-calibrated spectra in the units of W~nm$^{-1}$~m$^{-2}$, the internally-calibrated continuous mean spectra (ICS) were converted to externally calibrated spectra (ECS) using the \textit{GaiaXPy} Python package \citep{gaiaxpy}.

One of the advantages of machine learning is its ability to incorporate multiple characteristics into the analysis. Hence, in addition to {\it Gaia} XP spectra, we also use infrared colours for the machine learning analysis. Infrared colours have historically been used to identify C-rich and O-rich AGB stars \citep{Cioni2003, Matsuura2009, Woods2011}, and infrared all-sky surveys with comparable sample sizes are available. Therefore, we cross-matched {\it Gaia} DR3 sources with near-infrared (NIR) and mid-infrared (MIR) catalogues: the Two Micron All Sky Survey (2MASS) \citep{Skrutskie2006} and the Wide-field Infrared Survey Explorer (WISE) \citep{Wright2010}. This allowed us to use 2MASS magnitudes in the $J$ (1.235~$\mu$m), $H$ (1.662~$\mu$m), and $K_s$ (2.159~$\mu$m) bands, as well as \textit{WISE} magnitudes in the $W1$ (3.4~$\mu$m), $W2$ (4.6~$\mu$m), and $W3$ (12$\mu$m) bands. We excluded the $W4$ (22~$\mu$m) magnitude from \textit{WISE} due to its lower angular resolution (12$''$) whereas $W1$, $W2$ and $W3$ have resolutions of 6.1$''$, 6.4$''$, and 6.5$''$, respectively.

{\it Gaia} cross-matches with 2MASS and WISE were performed using coordinates with search radii of 5$''$\footnote{\url{https://www.ipac.caltech.edu/2mass/overview/about2mass.html}} and 6.5$''$\footnote{\url{https://notebook.community/bosscha/alma-calibrator/notebooks/2mass/13_environment-with_gaia}} \footnote{\url{https://wise2.ipac.caltech.edu/docs/release/allsky/}}, respectively. This resulted in approximately 176 million matched sources. To ensure high-quality data, we selected sources with {\it Gaia} parallax measurements more accurate than 3$\sigma$, reducing the sample to approximately 140 million sources. We will refer to this refined dataset as the \textit{G2MW} sample. This threshold retains over 80\% of the cross-matched dataset, providing a balance between completeness and the exclusion of lower-quality measurements to ensure the reliability of our analysis.

\subsubsection{Known Galactic C/O-rich AGB stars}\label{known}

To train our machine learning model to identify AGB candidates from \textit{G2MW}, we first constructed a well-characterised training sample of known AGB stars. We assembled C-rich AGB stars and O-rich AGB stars, separately. S-type stars are not included as a separate class due to their rarity \citep{Keenan1954}. Therefore, any S-type stars in the \textit{G2MW} sample may be classified as either O-rich or C-rich candidates.

For C-rich AGB stars, we started with the General Catalogue of Galactic Carbon Stars (CGCS) \citep{Alksnis2001}, an updated version of the earlier catalogues by \cite{Stephenson1973, Stephenson1989} which contains 6891 Galactic C-rich stars. The catalogue lists spectroscopically identified carbon-rich stars: characterised by C$_2$ Swan bands in the optical, CN bands in the optical and near-infrared, and 11.2 $\mu$m SiC dust emission. The catalogue indicates spectral types from the R, N, or MK classification systems if an optical spectral classification is available.

While AGB stars occupy a distinct evolutionary stage in theoretical isochrones, identifying Galactic O-rich AGB stars has historically been challenging. Traditionally, AGB stars have been identified based on their spectral type (M-type) together with long-period pulsations  \citep[Mira, semi-regular and irregular variables; ][]{1999IAUS..191..151W, Habing1996, Habing2004}. Although irregular and semi-regular variables may include some contamination from RGB stars \citep{Tabur2010}, we begin with this traditional and well-tested selection method. 
In Sect.~\ref{comparing_wesenheit_with_isochrones}, we compare the sample with theoretical isochrones, which is now possible owing to the availability of \textit{Gaia} distance measurements for Galactic stars. In Sect.~\ref{rgb_analysis}, we discuss the possible contamination by RGB stars in the machine-learning-selected sample.

For oxygen-rich AGB stars, we used the General Catalogue of Variable Stars \citep[GCVS; ][]{Samus2017}, which contains 50,000 variable sources. We filter this catalogue to select sources of long period variables and spectral type of M. AGB stars are pulsating long-period variables, and are typically divided into small-amplitude pulsations (semi-regular; SR) and Mira (M: large-amplitude periodic pulsations) \citep{Habing1996}. Additionally, they can be slow-irregular variables of late spectral types (GCVS types LB or LC; \citealt{Samus2017}). Hence, for our study, we focus on Mira, semi-regular and irregular variables, specifically selecting the M, M:, SRA, SRB, SRC, LB and LC types (see \citealt{Samus2017} for more details), narrowing the list to 4,860.

The coordinate systems of both the CGCS and GCVS catalogues were cross-matched with the \textit{G2MW} catalogue, providing positional information, {\it Gaia}-band photometry, parallax, and BP/RP coefficients, resulting in 1,623 C-rich AGB stars from CGCS and O-rich AGB stars from 4,169 GCVS sources. We retrieved flux measurements using the \texttt{calibrate} function within the \texttt{GaiaXPy} \citep{gaiaxpy} library to plot their XP spectra. We removed sources without parallax measurements - one from CGCS and two from GCVS. Additionally, 6 CGCS and 39 GCVS sources lacking magnitude data ({\it Gaia}, 2MASS or WISE) were excluded. The final sample consisted of 1,616 C-rich AGB stars from CGCS and 4,128 O-rich AGB stars from GCVS sources.

\begin{figure}
\includegraphics[trim={0.3cm 0.2cm 0.3cm 0cm},clip,width=\columnwidth]{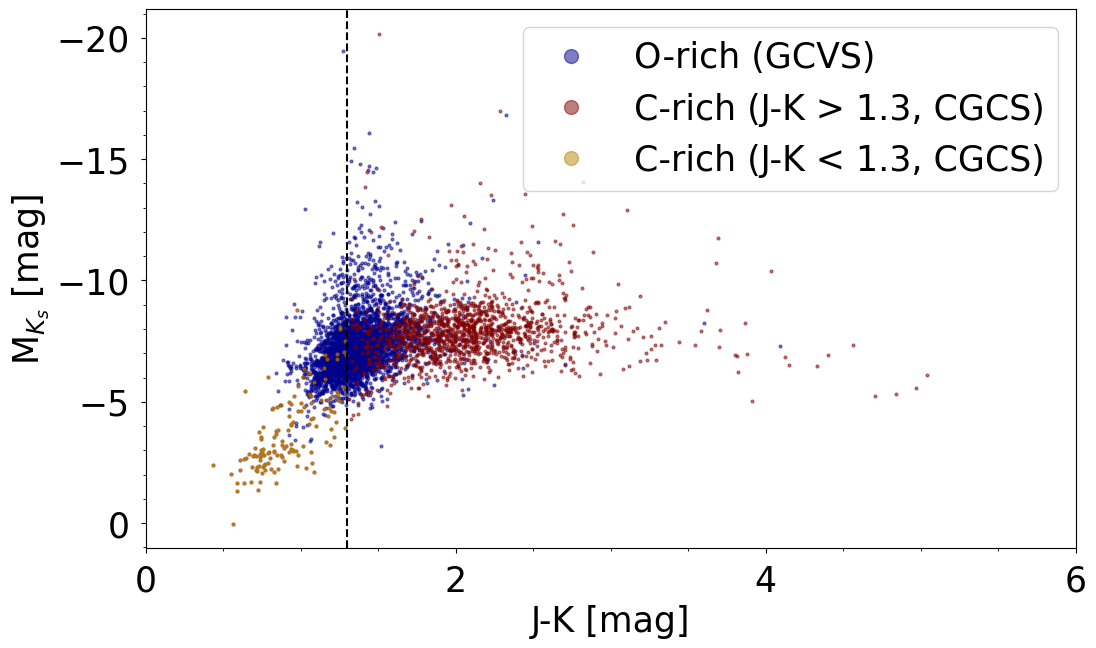}
 \caption{Absolute \Ks\  magnitude (\MK) vs. \JKs\ colour for known Galactic O-rich (blue), Galactic C-rich with \JKs$>$1.3 (red), and Galactic C-rich with \JKs$<$1.3 (yellow), extracted from the General Catalogue of Variable Stars (GCVS; \citet{Samus2017}) and General Catalogue of Galactic Carbon Stars (CGCS; \citet{Alksnis2001}). The dashed line at \JKs=1.3 separates the redder and bluer C-rich stars.}
  \label{kvsjk_known}
\end{figure}

The existing catalogues are general compilations of carbon stars and variable stars, rather than specifically targeting C-rich and O-rich AGB stars. To ensure a high-purity training sample for our machine learning analysis, we implemented additional refinement steps to eliminate potential contaminants, based on spectral classification schemes. Specifically, we computed flux ratios to identify C$_2$ and CN absorption bands in C-rich stars and TiO/VO bands in M-type stars, representing O-rich AGB stars. 

\textit{Gaia} BP/RP spectra \citep{Deangeli2023} can reveal characteristic molecular bands, such as CN in C-rich AGB stars and TiO in O-rich AGB stars (Fig.~\ref{C/Orich}). Stars with weak or absent molecular absorption bands are unlikely to be C-rich or O-rich AGB stars, and were excluded from the training sample using molecular band ratios defined as:

\begin{equation*}
    R_\lambda = \frac{F_\lambda}{F_{\lambda l}}
    \quad\mathrm{and}\quad 
    R_\lambda ^\prime = \frac{F_\lambda}{F_{\lambda r}}
\end{equation*}

where $F_\lambda$ is the mean flux averaged over three fluxes centred on the band peak wavelength $\lambda$, and $F_{\lambda l}$ and $F_{\lambda r}$ are the fluxes at the adjacent wavelengths to the left  (short wavelength) and right (long wavelength) of the peak, respectively. 
At low spectral resolution, molecular absorption bands produce a characteristic spectral shape in which a local flux peak occurs between two adjacent absorption troughs. Both $R_\lambda>$1 and $R_\lambda ^\prime>$1 confirm the presence of a genuine local flux maximum with absorption features on either side. 

For the C-rich training sample, we defined two CN molecular band ratios at 792~nm and 920~nm, computed as the ratio of the peak flux to the adjacent fluxes at 762~nm and 820~nm (for the 792~nm) and 880~nm and 952~nm (for the 920~nm). We removed 103 stars with weak CN features (ratio values below 1). For the O-rich training sample, we defined three TiO molecular band ratios at 765~nm, 835~nm, and 925~nm, computed as the ratio of the peak to the adjacent fluxes at 740~nm and 783~nm, 805~nm and 870~nm, and 849~nm and 960~nm, respectively. We removed 446 stars with weak TiO features (ratio values below 1), resulting in a final sample of 3,680 O-rich AGB stars.

The spectral index cuts alone could not identify all non-AGB contaminants in the C-rich training sample. We therefore cross-matched the CGCS sample with SIMBAD and eliminated a further 46 stars classified as CH stars, barium stars or spectroscopic binaries, which can show C-rich features at low resolution but are not AGB stars.

To further refine the selection of C-rich AGB stars and remove any non-AGB contaminants from the C-rich sample, we utilised the infrared colour-magnitude diagram (CMD). The $K_s$ vs $J-K_s$ CMD is commonly used to classify AGB stars in the Large and Small Magellanic Clouds \citep{Cioni2003}. A similar plot for our training sample, using absolute magnitude $M_{K_s}$ (calculated using distances derived from {\it Gaia} parallaxes) instead of apparent magnitude $K_s$, is shown in Fig.~\ref{kvsjk_known}. 

In this figure, O-rich stars predominantly exhibit $J-K_s$ values between 1 and 2 magnitudes. For the C-rich sample, two distinct groups are identified: one cluster with redder stars ($J-K_s > 1.3$ mag, mostly $>2$ mag) and another with bluer stars ($J-K_s < 1.3$ mag). The bluer carbon stars tend to have lower $M_{K_s}$ luminosities compared to their redder counterparts. The population with $J-K_s < 1.3$ mag exhibits characteristics similar to R-hot carbon stars, which have $M_K < -3$ mag with a median luminosity of $\langle M_K \rangle \approx -2.54$ mag \citep{Abia2020}. This corresponds to the upper part of the RGB \citep{Cioni2000}. Although the exact nature of these R-hot stars remains uncertain, they are likely non-AGB stars. Furthermore, this colour region is also populated by CH stars, which have similar spectra to normal K-giants in low-resolution spectra \citep{Zamora2009} and may not have been fully removed by the CN index cuts alone. Therefore, we excluded C-rich stars with $J-K_s < 1.3$ mag (139 sources) from our training sample, focusing on $J-K_s > 1.3$ only (1,328 C-rich AGB stars).

No ISM extinction corrections were applied to the magnitudes or colours used in the data preparation described in this section.

\subsection{Two-step machine learning analysis} \label{two-stepmethodology}
 
Machine learning (ML) provides an efficient approach for analysing multi-dimensional astronomical datasets. The choice of an ML method depends on factors such as feature quality and class separability. While supervised (SML) and unsupervised (USML) methods are widely used, both proved inadequate for our dataset:

\textit{SML Limitations:} The {\it Gaia} DR3 catalogue contains a wide variety of stellar types, many of which are not clearly defined or labelled. Since SML requires well-defined class labels for training, it becomes difficult to apply when the exact number of classes is unknown. Attempting to use a supervised approach on such heterogeneous data would likely result in misclassifications.

\textit{USML Limitations:} The unsupervised approach also proved insufficient due to considerable overlap in spectral properties between different classes, such as M-giants and M-dwarfs, making it challenging to form distinct clusters. This spectral similarity is a well-known challenge in stellar classification \citep{Bessel1988, Covey2007}. Additionally, the unknown number of classes in the {\it Gaia} dataset further complicates the clustering process, as most unsupervised algorithms require either a predefined number of clusters or clear separation between natural groupings.

To address these issues, we adopted a hybrid machine learning approach that leverages both labelled and unlabelled data. This methodology is well-suited to astronomical datasets, where comprehensive labelling is impractical but reliable reference samples exist. We first constructed a reference set of stars from \textit{G2MW} dataset that closely resembled C-rich and O-rich AGB stars in our known AGB catalogue (Sect.~\ref{known}). This was done by comparing {\it Gaia} features to labelled stars using cosine similarity (Sect.~\ref{clustering}), effectively creating a bridge between the labelled and unlabelled data.

This reference set provided a controlled dataset suitable for binary classification, as applying a trained binary model directly to the full {\it Gaia} dataset, which contains numerous stellar types, galaxies, and other objects, would not be feasible. By restricting the model to a similarity-selected subset, we focus on the specific stellar populations of interest while minimising contamination. After constructing the refined sample, we trained a binary classifier on the known AGB set and, once good performance metrics were achieved, applied the model to the similarity-selected AGB candidates. 

This two-step process ensures that the classification is accurate (by training on well-labelled data) and relevant (by applying to a carefully selected subset). Further details on the similarity analysis are provided in Sect.~\ref{clustering}, and the ML model architecture is discussed in Sect.~\ref{xgboost}.

\subsubsection{Features for machine learning}

In this study, we used two main categories of features: colours and {\it Gaia} DR3 XP spectral coefficients. These features were selected to capture the key characteristics of AGB candidates across different wavelengths for robust analysis. The colour indices were derived from multiple catalogues to provide a broad spectral coverage. From {\it Gaia} DR3, we used $BP-RP$, $G-RP$, and $BP-G$, representing the optical properties of the stars. From 2MASS, we included $J-H$, $H-K$ and $J-K$ to probe the NIR characteristics, while WISE colours ($W1-W2$, $W2-W3$, and $W1-W3$) extended coverage into the MIR region. These colour indices are essential for distinguishing between different types of AGB candidates (see Fig.~\ref{kvsjk_known}). In addition to colour indices, we used all 110 BP/RP spectral coefficients from {\it Gaia} DR3 XP spectra.
\textit{Gaia} XP spectra are represented as a linear combination of Hermite basis functions spanning 330-1100~nm \citep{Carrasco2021, Deangeli2023}, where the coefficients represent the weights of these basis functions. These coefficients consist of 55 BP (Blue Photometer) coefficients (\textit{BP1}, \textit{BP2}, …, \textit{BP55}) and 55 RP (Red Photometer) coefficients (\textit{RP1}, \textit{RP2}, …, \textit{RP55}).

In Sect.~\ref{Results}, we discuss how machine learning metrics vary with the number of spectral coefficients. Although higher-order coefficients can introduce noise into the spectra, manifesting as wavy structures \citep{Deangeli2023}, we included all 110 coefficients to maintain a consistent feature set across all data, which is crucial in machine learning to prevent information loss \citep{Carrasco2021} and avoid potential biases in the model. We also evaluate the relative importance of each feature in our final trained model (Sect.~\ref{Results}), providing insights into which colours and spectral coefficients contribute most to distinguishing C-rich and O-rich AGB candidates.

\begin{figure*}
    \centering
    \includegraphics[scale=.6]{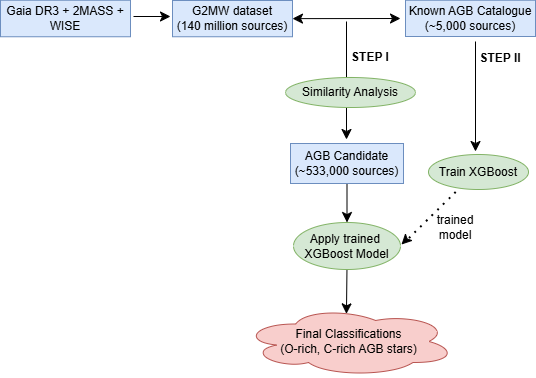}
    \caption{Schematic representation of our two-step machine learning pipeline for classifying C-rich and O-rich AGB stars from {\it Gaia} DR3. Solid arrows indicate the main classification pipeline, while the dashed arrow represents the trained {\sc XGBoost} model being applied to the AGB candidate sample. The pipeline combines data from multiple surveys ({\it Gaia} DR3 + 2MASS + WISE, forming the \textit{G2MW} dataset of $\sim$140 million sources) with known AGB catalogues (the General Catalogue of Galactic Carbon Stars (CGCS) and the General Catalogue of Variable Stars (GCVS)) to create a \textit{Known AGB Catalogue} of $\approx$5,000 C-rich and O-rich AGB stars. \textbf{Step 1:} Cosine similarity analysis identifies $\approx$533,000 AGB candidates in the \textit{G2MW} dataset by comparison with the \textit{Known AGB Catalogue}. \textbf{Step 2:} An {\sc XGBoost} classifier trained on the \textit{Known AGB Catalogue} classifies these candidates into C-rich and O-rich AGB stars, producing the labelled classification results used in the analysis.}
    \label{fig:methodology}
\end{figure*}

\subsubsection{Similarity analysis}\label{clustering}

The first step of our method was to construct a combined sample of potential C-rich and O-rich AGB candidates using cosine similarity. Cosine similarity is widely used for problems in data analysis \citep{Kim2019}, natural language processing \citep{Ahmad2024}, text analysis \citep{Gomaa2013, Yogish2020}, and glitch or plagiarism detection \citep{Ferreira2022}. It is particularly effective for classification tasks where the datasets contain similar characteristics and features.

If there are two vectors $\Vec{A}$ = ($A_1$, $A_2$, $A_3$, …) and $\Vec{B}$ = ($B_1$, $B_2$, $B_3$, …), then the cosine similarity ($S_c$) between the two vectors will be

\begin{equation}
S_c = \cos \theta = \frac{\Vec{A} \cdot \Vec{B}}{\|\Vec{A}\| \|\Vec{B}\|}
\end{equation}

The cosine similarity can also be used as a distance metric, which is defined as:

\begin{equation}
    d(\Vec{A}, \Vec{B}) = 1 - S_c
\end{equation}

Cosine similarity measures the cosine angle between two vectors in a multidimensional space, providing a measure of their alignment. Values of $S_c$ close to 1 indicate that the vectors are nearly parallel, signifying high similarity. This metric is particularly useful in high-dimensional datasets, where absolute distance measures (e.g., Euclidean) can be less informative due to data dispersion. A smaller angle corresponds to greater similarity, whereas a larger angle reduces the similarity between vectors. For our analysis involving 110 spectral coefficients, cosine similarity focuses on the pattern of features rather than their absolute magnitudes.

We computed the cosine similarity between the sample of known Galactic C-rich and O-rich AGB stars (constructed in Sect.~\ref{known}) and the \textit{G2MW} sources. We used the 9 colour indices and 110 BP/RP spectral coefficients (as discussed above), with the \texttt{cdist} function from the \texttt{scipy} library. This process selected \textit{G2MW} sources with properties similar to the known Galactic C-rich and O-rich stars, providing a targeted sample for subsequent binary classification and further investigation into the properties of AGB candidates in the MW (Sect.~\ref{similarity_analysis_results}).

\subsubsection{Binary classification of machine learning}\label{xgboost}

In this section, we describe the architecture, parameters, and optimisations of the machine learning model used to classify C-rich and O-rich AGB candidates after the cosine similarity step. The known Galactic AGB sample (\textit{KnAGB}; Sect.~\ref{known}) consists of 5,008 sources and is imbalanced, with an O-rich to C-rich ratio of 2.8:1. While not as extreme as some cases in the literature \citep{Akras2019}, this imbalance can bias model training toward the majority class. Standard approaches to address imbalance, such as undersampling, oversampling, or synthetic data generation \citep{Weiss2003, Haibo2009, Krawczyk2016}, were unsuitable for our dataset as they could reduce coverage in parameter space or introduce inconsistencies in XP spectral coefficients. Therefore, we retained the full \textit{KnAGB} dataset and applied a higher penalty for misclassifications to improve C-rich identification.

Colour-magnitude diagrams (CMDs) are commonly used to classify AGB stars as C-rich or O-rich. However, CMDs are two-dimensional and do not capture the multidimensional nature of the classification. To explore whether machine learning models using spectral coefficients or photometric colours alone can distinguish between the two classes more effectively, we created two subsamples from the \textit{KnAGB}: one containing only 9 colours and another containing only spectral features (55 for \textit{BP} and 55 for \textit{RP} coefficients). The full \textit{KnAGB} dataset combines both colour and spectral features (119 features), providing the widest parameter space for classification. We also assigned binary labels to the two classes: 1 for C-rich and 0 for O-rich.

We split the dataset into a 70-30 train-test set using the \texttt{train$_-$test$_-$split()} function from \textit{scikit-learn} \citep{scikit-learn2011}, with 70\% of the data used for training and 30\% for testing. We shuffled the \textit{KnAGB} sources before splitting to prevent positional bias. To address class imbalance, we employed stratified sampling using the \texttt{stratify} parameter, ensuring that the training and test sets preserved the original class distribution. This approach prevents the model from overfitting to the majority class and improves generalisation to unseen data.

For classification, we used XGBoost, a high-performance gradient boosting technique widely used in astrophysical studies \citep{Bethapudi2018, Cunha2022, Andrae2023, Lucey2023, Yao2024}. To account for the inherent class imbalance in the training sample (2,576 O-rich vs. 929 C-rich instances), we set the \texttt{scale\_pos\_weight} parameter to 2.77, computed as the ratio of O-rich to C-rich AGB stars. This balances the contribution of each class during training.

We implemented hyperparameter tuning using \texttt{RandomizedSearchCV} from \textit{sklearn}'s \textit{model\_selection} module. This approach systematically explored the hyperparameter space for the combined, colour-only, and spectroscopic-only datasets to identify optimal configuration parameters. To prevent overfitting, we incorporated two complementary regularisation techniques: \\
1. \textbf{L1 Regularisation} (\texttt{reg\_alpha}): Introduces sparsity by selectively reducing less significant feature weights to zero, effectively performing feature selection.\\
2. \textbf{L2 Regularisation} (\texttt{reg\_lambda}): Constrains model complexity by penalising large coefficient magnitudes and promotes a balanced distribution of feature importance.

\begin{table*}
\centering
\caption{XGBoost hyperparameter values used for training the binary classifier to distinguish C-rich and O-rich AGB candidates. This table lists each hyperparameter, its description, the search range used during optimisation, and the optimal value found.}
\label{hyperparameter_tuning}
\begin{tabular}{lllr}
\hline
Hyperparameter & Description & Search Range & Optimal Value \\
\hline
\texttt{n\_estimators} & Number of boosting trees in the ensemble & 50--200 & 160 \\
\texttt{learning\_rate} & Step size for weight updates & 0.01--0.03 & 0.0374 \\
\texttt{gamma} & Minimum loss reduction for tree splitting & 0--0.5 & 0.307 \\
\texttt{subsample} & Fraction of training data per tree & 0.6--0.8 & 0.682 \\
\texttt{colsample\_bytree} & Fraction of features per tree & 0.6--0.8 & 0.830 \\
\texttt{max\_depth} & Maximum depth of individual trees & 3--8 & 4 \\
\texttt{min\_child\_weight} & Minimum sum of instance weights per leaf & 1--20 & 1 \\
\texttt{reg\_alpha} & L1 regularisation strength & 0.2--1.0 & 0.926 \\
\texttt{reg\_lambda} & L2 regularisation strength & 8--12 & 15.116 \\
\hline
\end{tabular}
\end{table*}

Our hyperparameter optimisation process extended beyond the initial search boundaries, consistent with adaptive search methods \citep{Bergstra2012}. The optimal \texttt{learning\_rate} (0.0374) exceeded our initial upper bound of 0.03, indicating that the classifier benefited from more aggressive weight updates than initially anticipated. Similarly, the optimal \texttt{reg\_lambda} value of 15.116 surpassed our initial maximum of 12, suggesting that stronger L2 regularisation was required to prevent overfitting and achieve optimal generalisation. These expanded parameter values significantly improved model performance, as validated on the test set and confirmed by application to the similarity-selected sample, where results aligned with expected classifications. Table~\ref{hyperparameter_tuning} presents the ranges and optimal values of XGBoost hyperparameters determined through tuning.

Each model was trained using five-fold stratified cross-validation (\texttt{StratifiedKFold}), a technique that ensures reliable model performance across different data subsets. The training set was divided into five subsets (folds) while preserving the original class distribution, and the model was iteratively trained on four folds and validated on the remaining fold. Cross-validation scores are presented in Sect.~\ref{binary_classification_results}.

We evaluated our models primarily using the Receiver Operating Characteristic - Area Under the Curve (ROC-AUC), which is particularly suitable for imbalanced datasets like ours. The ROC curve plots the True Positive Rate (TPR) against the False Positive Rate (FPR) at different classification thresholds, with ROC-AUC values ranging from 0 to 1 (1 indicating perfect discrimination). In our binary classification, True Positives (TP) are correctly identified C-rich sources, False Positives (FP) are O-rich stars misclassified as C-rich, True Negatives (TN) are correctly identified O-rich sources, and False Negatives (FN) are C-rich stars misclassified as O-rich. We chose ROC-AUC as the primary evaluation metric because it remains informative even with class imbalance, unlike accuracy, which can be misleading in unevenly distributed datasets.

\begin{equation}
    \text{TPR} = \frac{TP}{TP+FN}
\end{equation}

\begin{equation}
    \text{FPR} = \frac{FP}{FP + TN}
\end{equation}

For completeness, we also calculated accuracy, defined as the proportion of correctly classified sources relative to the total. However, model optimisation was based on ROC-AUC using the \texttt{roc\_auc\_scorer} in \texttt{scikit-learn}. The best-performing models were evaluated on the test set, with results and confusion matrices presented in Sect.~\ref{binary_classification_results}.

Fig.~\ref{fig:methodology} illustrates our two-step approach for classifying Galactic C-rich and O-rich AGB stars. \textbf{Step I:} A similarity analysis is performed between the \textit{G2MW} catalogue and the \textit{Known AGB Catalogue}, producing a set of AGB candidates. \textbf{Step II:} An {\sc XGBoost} classifier is trained on the \textit{Known AGB Catalogue} (with known labels) and applied to the candidate set to classify stars as C-rich or O-rich.

\section{Results}\label{Results}

\subsection{Performance of two-step ML}\label{results_performance_of_two_step_ML}

We applied a two-step machine learning approach to extract O-rich and C-rich AGB candidates. The analyses and their results are described separately below.

\begin{figure*}
\centering     
\subfigure{\includegraphics[width=.5\linewidth]{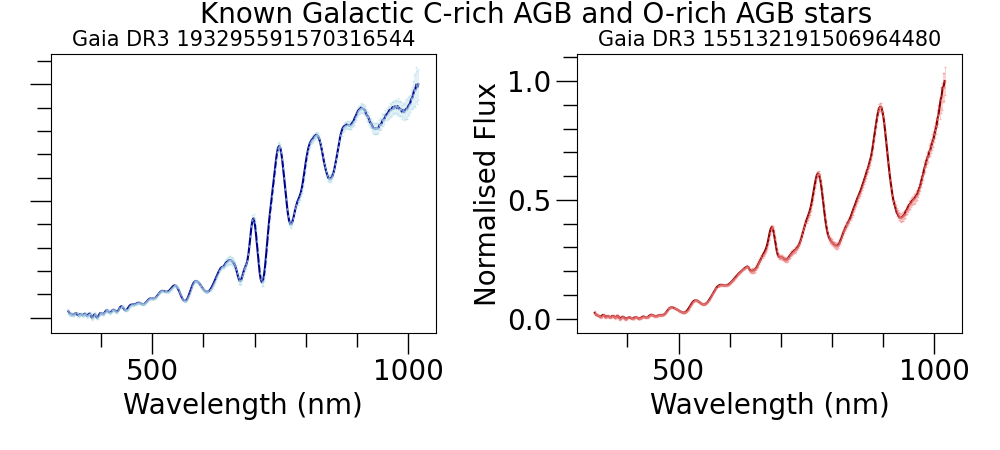}}
\subfigure{\includegraphics[width=0.99\linewidth]{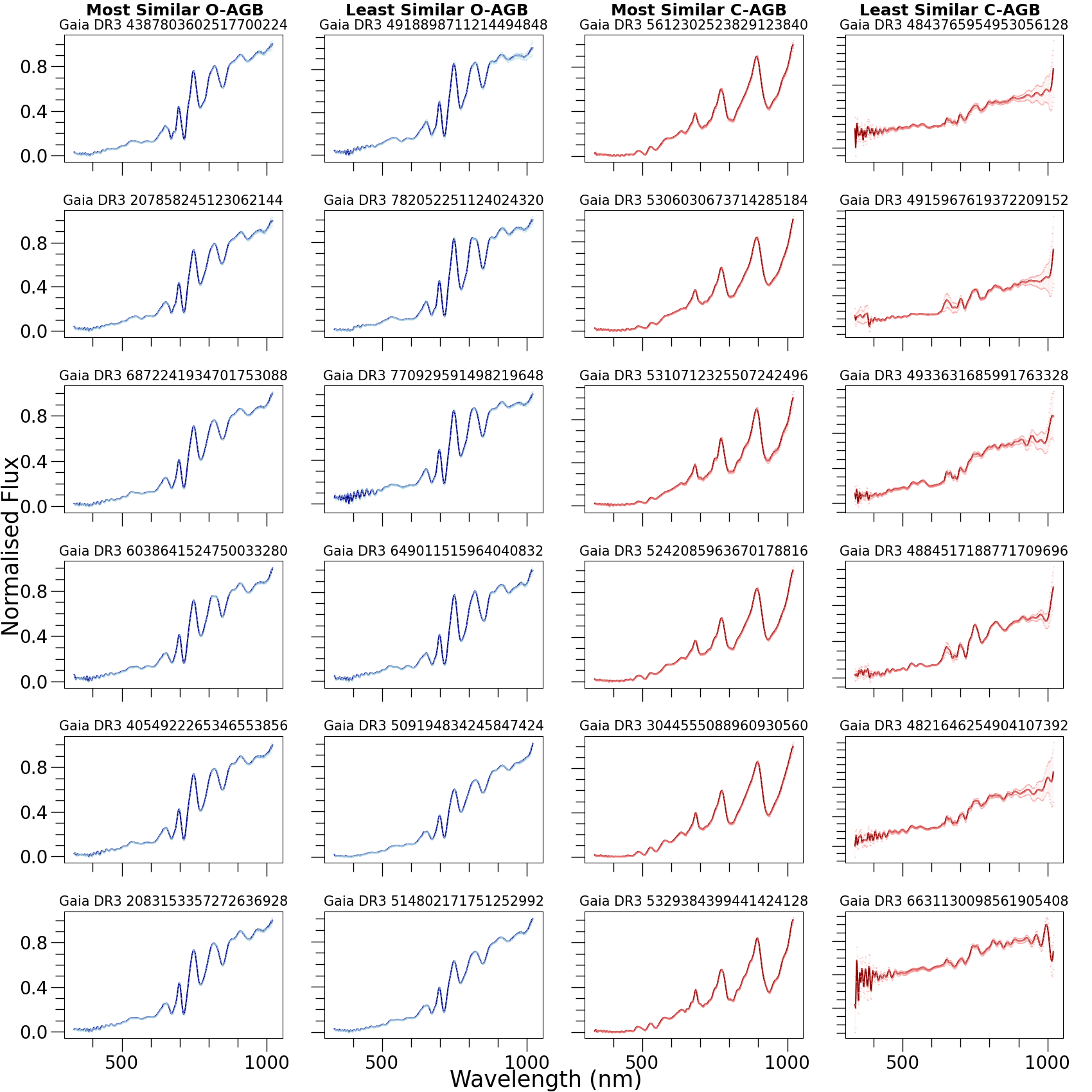}}
\caption{Examples of the results from the similarity analysis. The top panels show a Galactic O-rich (left) and C-rich (right) AGB stars from the KnAGB sample, which were compared to the processed \textit{G2MW} sample. In the first and second columns, the most and least similar sources found to the O-rich AGB star (top-left) within the 99.999 percentile similarity threshold are shown. The third and fourth columns show the most and least similar sources to the C-rich AGB star (top right).}
\label{similar_sources}
\end{figure*}

\subsubsection{Determining similarity threshold}\label{similarity_analysis_results}

The first step of our two-step methodology involved similarity analysis, which effectively identifies new sources in a large, heterogeneous survey that resemble known samples. In this study, it was applied to construct a sample of AGB candidates, which were subsequently classified as C-rich or O-rich. 

A critical aspect of the first step was setting a similarity threshold, which defines the minimum similarity level at which a {\it Gaia} source is identified as an AGB candidate. To establish an initial threshold, we begin with an order-of-magnitude estimate of the number of AGB stars. We consider two aspects: first, the number of AGB stars identified in previous all-sky surveys, and second, the typical lifetime of AGB stars.

Previous surveys have identified 5,000–25,000 AGB stars \citep{Ishihara2011, Lewis2020, Suh2021, Sanders2023}. These studies focused primarily on late-stage AGB stars: \citet{Ishihara2011}, \citet{Lewis2020}, and \citet{Suh2021} selected stars with mid-infrared dust excess, while \citet{Sanders2023} focused on high-amplitude long-period variables. 
Our selection, based on the GCVS and CGCS catalogues, focuses on Mira (M), semi-regular (SR) and irregular (LB, LC) variables (Sect.~\ref{known}), which are predominantly AGB stars.
The lifetime of AGB stars is estimated to range from $10^5$ to $10^7$  years \citep{Vassiliadis1993}, compared to a total stellar lifetime of 10$^{8}$--10$^{10}$~years, depending on initial mass \citep{Karakas2014}. This suggests that roughly 0.001\%-0.1\% of stars in a stellar population may be in the AGB phase at any given time. Therefore, we set the threshold at the 99.999th percentile (selecting the top 0.001\% of sources most similar to each known AGB star), corresponding to the lower end of this estimate.

We conducted sanity checks by testing thresholds one order of magnitude higher and lower than the 99.999th percentile. We note that the primary purpose of the similarity analysis is to reduce the data set from 140 million sources to a computationally manageable size for binary classification, rather than to define the final sample directly. A stricter threshold would limit matches to fewer than 1,000 matches per known AGB star, potentially excluding important AGB stars. This could result in an incomplete sample, overlooking the diversity of AGB evolutionary stages. 

Conversely, lowering the threshold from 99.999th to 99.99th percentile risks including non-AGB stars in the candidate sample, such as C-rich stars with C$_2$ and CN features that are not AGB stars (e.g., CH-type stars), as well as M-dwarfs, which share spectral similarities with M-giants but are intrinsically fainter and numerous. We found that the 99.999th percentile threshold provides the best balance between accuracy and robustness.

The similarity analysis selects the top 0.001\% of sources most similar to a \textit{KnAGB} star from the entire \textit{G2MW} catalogue. As a result, a source in \textit{G2MW} can be selected by multiple \textit{KnAGB} stars,  and the total number of unique candidates is smaller than the number of selected candidates.

To demonstrate the performance of our similarity method, we present two examples: an O-rich AGB star ({\it Gaia} DR3 193295591570316544), which has an M2 spectral type with irregular variability, according to the \textit{HIPPARCOS} catalogue \citep{Perryman1997} and the GCVS \citep{Samus2017}, and a C-rich AGB star, {\it Gaia} DR3 155132191506964480 \citep{Alksnis1962}. Using a 99.999th percentile similarity threshold, we identified 1,415 and 1,416 sources as corresponding to these two AGB stars from \textit{G2MW}, respectively. Fig.~\ref{similar_sources} shows the spectra of the most and least similar sources for each AGB star. The first and second columns show the six most and six least similar sources for the O-rich AGB star among the 1,415 identified similar sources. The third and fourth columns show the same for the C-rich AGB star. Within each column, spectra are arranged in descending similarity scores from top to bottom.

The spectra in the first column of Fig.~\ref{similar_sources} demonstrate that the similarity analysis effectively identifies O-rich AGB stars with high similarity scores, exhibiting characteristic O-rich signatures, such as strong TiO and VO absorption bands. Likewise, the spectra in the third column show distinct C-rich spectral characteristics, with prominent C$_2$ and CN bands, especially toward the red end of the {\it Gaia} XP spectra. 

Even at lower similarity scores (second column), O-rich spectra retain visible TiO and VO absorption bands, although they appear noisier. 
However, this is not the case for C-rich AGB stars. As similarity scores decrease, sources similar to C-rich AGB stars often correspond to O-rich AGB stars, as shown in the fourth column of Fig.~\ref{similar_sources}. Once the catalogue runs out of sources similar to C-rich AGB stars, the similarity analysis begins selecting O-rich AGB stars, which are the next most similar to C-rich stars. This occurs due to the imbalance in the numbers of C-rich and O-rich AGB stars, with previously reported ratios ranging from 1:1.5 \citep{Ishihara2011, Suh2021} to 1:14 \citep{Lewis2020}. One way to mitigate this issue could be by introducing different thresholds, but this might also introduce sampling bias. Even though sources similar to C-rich AGB stars may actually be O-rich AGB stars, they are still AGB stars. Therefore, we apply the same threshold for both O-rich and C-rich AGB stars, with their classification relying on the next step: binary classification.

The similarity analysis identified an AGB candidate sample of approximately 533,000 stars from the 140 million \textit{G2MW} dataset. 

\subsubsection{Binary Classification for C-rich and O-rich AGB Stars}\label{binary_classification_results}

\begin{figure*}
   \includegraphics[trim={2.cm 2.cm 1.2cm 1.9cm},clip, scale=.15]{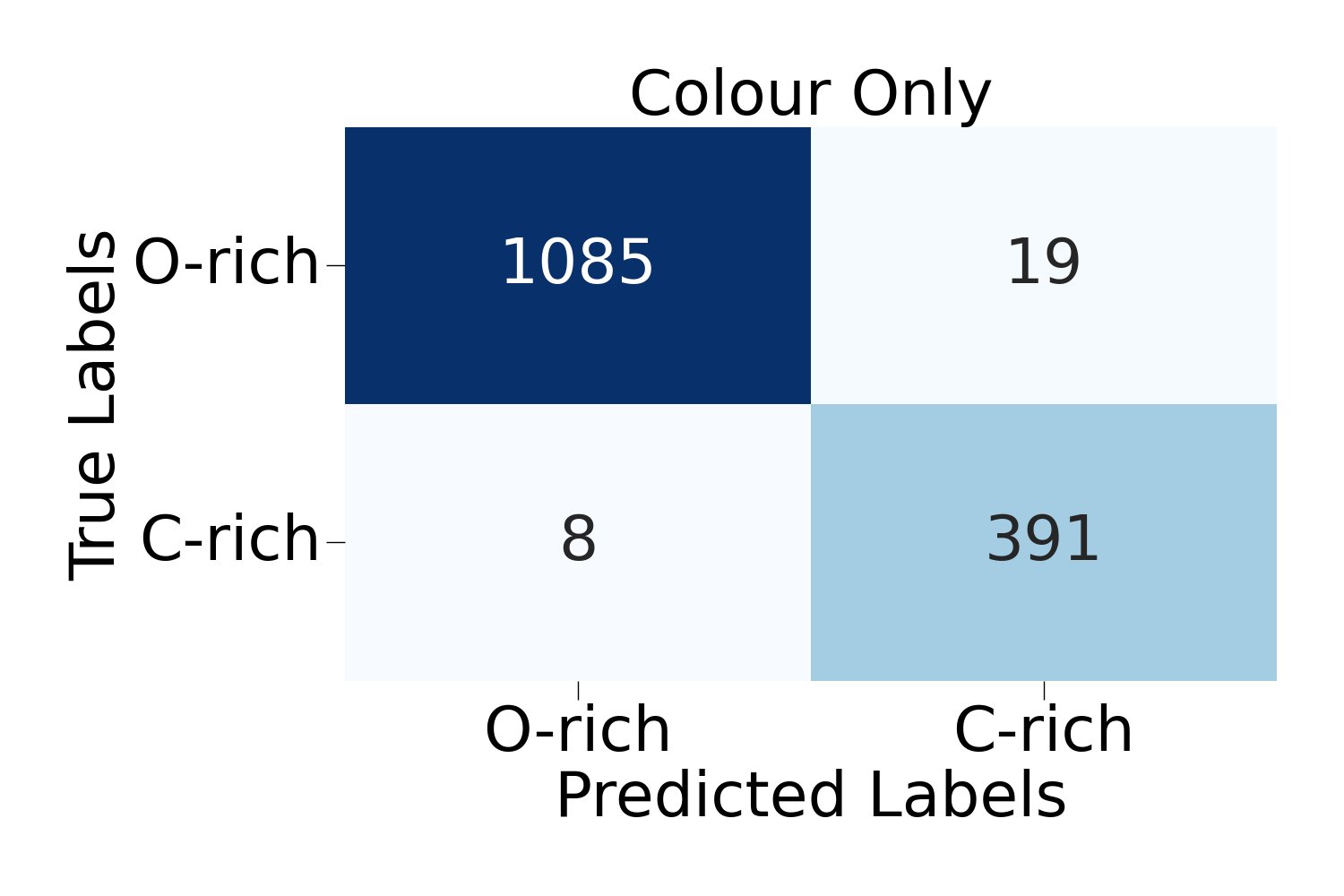}
  \includegraphics[trim={2.cm 2.cm 1.2cm 1.9cm},clip, scale=.15]{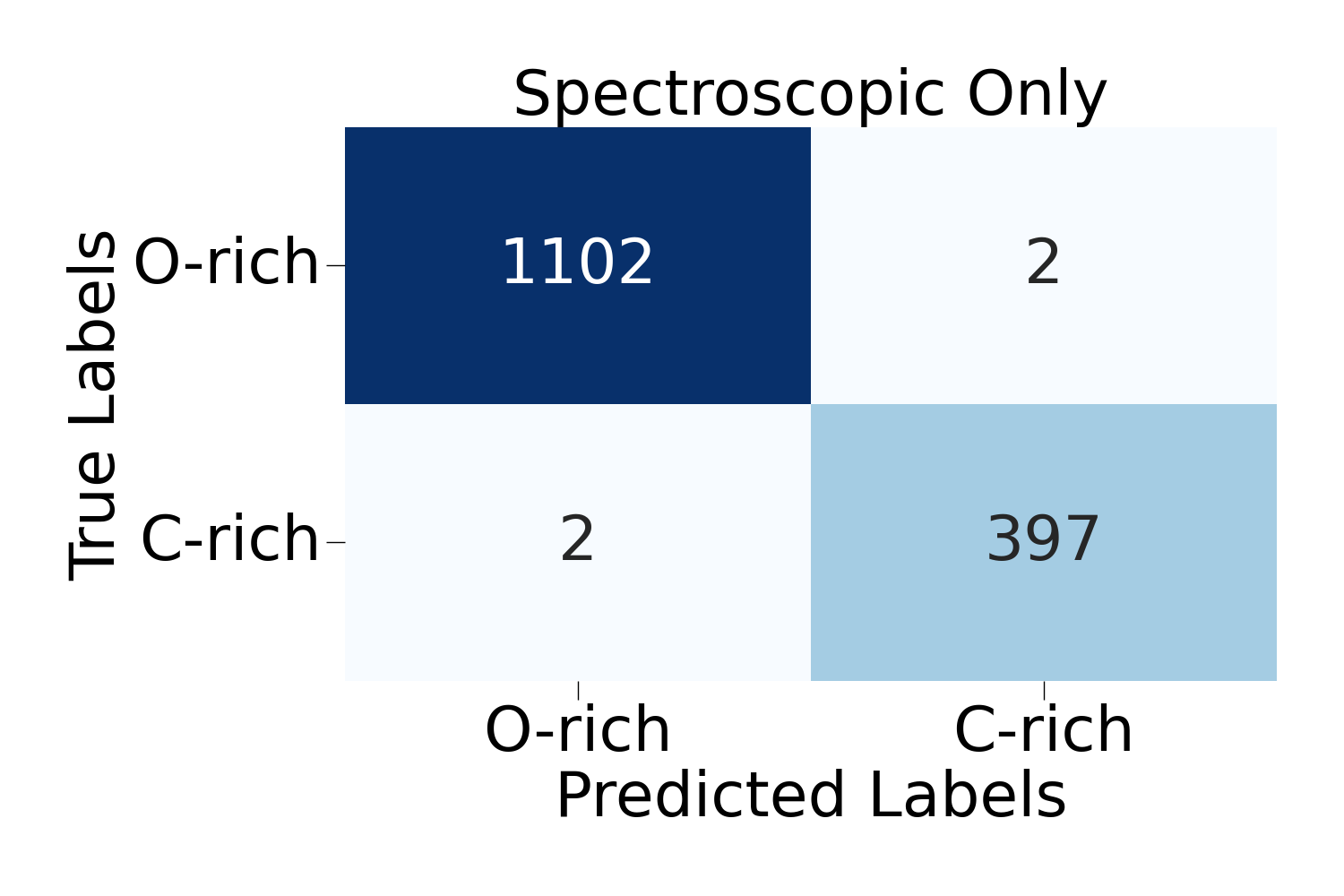}
 \includegraphics[trim={2.cm 2.cm 1.2cm 1.9cm},clip, scale=.15]{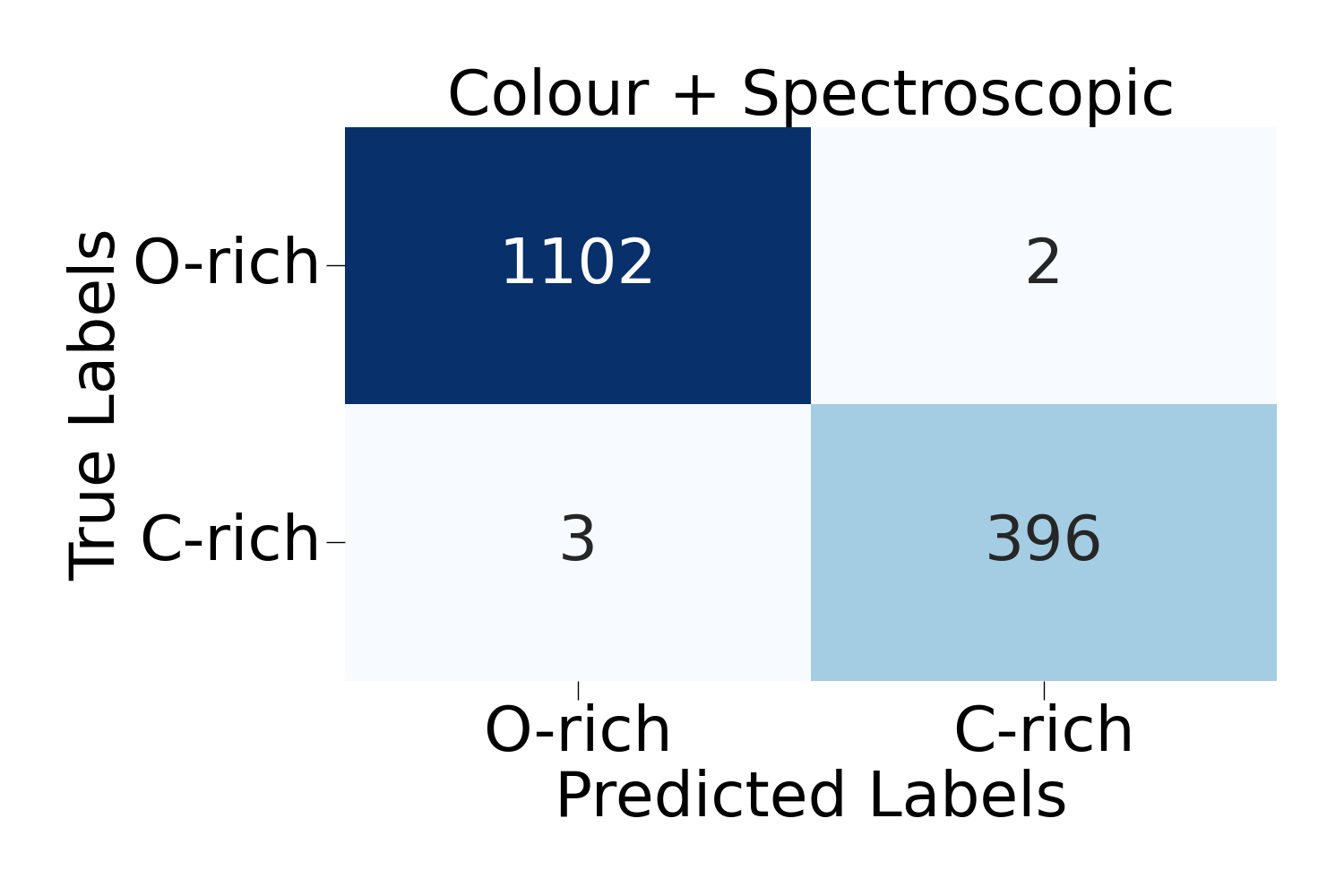}
 \caption{Confusion matrices of the test set for the best XGBoost models trained with (from left to right): colour (photometric) features, spectroscopic features, and a combination of both. The rows represent the true labels, and the columns represent the predicted labels.}
 \label{confusion_matrix}
\end{figure*}

As the second step of our methodology, we applied an optimised {\sc XGBoost} model (Sect.~\ref{xgboost}) to classify the similarity-selected sample as C-rich or O-rich AGB candidates. The model was trained and tested on the \textit{KnAGB} dataset using three feature sets: colour features, spectroscopic features, and a combination of both. 

Model performance was primarily evaluated using ROC-AUC, following the methodology outlined in Sect.~\ref {xgboost}. A 70/30 train-test split of the \textit{KnAGB} dataset ($\sim$5,000 known AGB stars) was used to ensure robust model validation.

To determine the optimal number of spectroscopic features for {\sc XGBoost}, we tested different numbers of input BP/RP coefficients: 1, 5, 25, and 55.
We found that using a single coefficient led to poor spectroscopic performance (ROC-AUC: 0.7649, Accuracy: 78.24\%), increasing the number of coefficients to 5 significantly improved the model's ability to distinguish between C-rich and O-rich AGB stars (ROC-AUC: 0.9992, Accuracy: 99.33\%). Performance plateaued at 25 and 55 coefficients (ROC-AUC: 0.9998, Accuracy: $\geq$99.60\%). While reducing the number of features could improve computational efficiency, we used all 110 BP/RP coefficients to maximise predictive power and maintain consistent performance on the unseen dataset.

To complement our ROC-AUC performance analysis, we visualised the model's classification outcomes using confusion matrices for the 30\% test set of the \textit{KnAGB} dataset. The confusion matrices in Fig.~\ref{confusion_matrix} provide a detailed breakdown of the model's predictions across three feature scenarios: colour-only ({\it Gaia}, 2MASS, and WISE colour indices), spectral coefficients-only (BP/RP coefficients), and combined colour and spectral coefficients. The {\sc XGBoost} model consistently achieved high classification accuracy (>98\%) across all feature sets, with ROC-AUC exceeding 99\% when trained with spectroscopic coefficients. Integrating both colour and spectroscopic features further minimised misclassifications, yielding performance comparable to using spectroscopic features alone.

Our results reveal that combining colour and spectroscopic features provides the most robust classification. Very red C-rich stars often lack distinct spectroscopic features, challenging machine learning models, while colour indices offer critical additional discrimination. The misclassification analysis supports this approach: using colour-only features resulted in 27 misclassified stars (2.1\%), primarily due to overlapping optical-NIR colours, whereas spectroscopic features reduced misclassifications to just 4 stars (0.2\%). Integrating both feature sets resulted in only 5 misclassifications (0.3\%), effectively compensating for the limitations of individual feature types.

The predictions from the XGBoost model are probabilistic, which allows further refinement of classifications by adjusting the probability threshold \textit{p} provided by XGBoost. For each source, XGBoost outputs two probabilities: \textit{p}(O-rich) and \textit{p}(C-rich)=1-\textit{p}(O-rich). Table~\ref{compare_probability_threshold} presents the number of classified C-rich and O-rich AGB candidates for a range of probability thresholds. At the default threshold of $p=0.5$ (i.e. sources are classified as C-rich if $p$(C-rich)$\geq$0.5, and as O-rich otherwise), all the sources in the similarity-selected dataset are assigned to one of the two classes. Increasing the probability threshold decreases the number of classified sources, reflecting a higher confidence in the classifications. For example:
\begin{itemize}
    \item At $p$>0.8 for both C-rich and O-rich probabilities, i.e., $p$(C-rich)>0.80 and $p$(O-rich)>0.80, 440,371 sources are classified as O-rich and 15,618 as C-rich, while approximately 60,000 sources remain unclassified with probabilities below 0.8.
    \item At $p$>0.9, the number of classified sources decreases to 338,530 O-rich and 6,111 C-rich AGB candidates.
    \item At the highest threshold of $p$>0.99, only 2,669 sources are classified as C-rich and 164,644 as O-rich, showing a highly selective classification.
\end{itemize}

\begin{table}
    \centering
    \caption{Number of C-rich and O-rich AGB candidates classified at different XGBoost probability thresholds ($p$). Columns list the chosen thresholds ($p$=0.5, 0.8, 0.9, 0.95, 0.99) and the resulting number of O-rich and C-rich AGB candidates.}
    \label{compare_probability_threshold}
    \begin{tabular}{c c c}
    \hline
Probability threshold & Number of O-rich &  Number of C-rich\\
\hline
    $p$>0.5 & 471,085 & 45,384 \\
    $p$>0.8 & 440,371 & 15,618 \\
    $p$>0.9 & 338,530 & 6,111 \\ 
    $p$>0.95 & 307,191 & 4,227 \\
    $p$>0.99 & 164,644 & 2,669 \\
    \hline
    \end{tabular}
\end{table}

\begin{table}
    \centering
    \caption{Number of C-rich and O-rich AGB candidates classified with probability threshold $p>$0.95 using colour, spectroscopic, and combined feature sets.}
    \label{phot_spect_comb}
    \begin{tabular}{l c r}
    \hline
Features & Number of O-rich &  Number of C-rich\\
\hline
    Colour & 398106 & 5,407 \\
    Spectroscopy & 247,404 & 4,521 \\
    Combined & 307,191 & 4,227 \\ 
    \hline
    \end{tabular}
\end{table}

\begin{figure*}
 \includegraphics[trim={2.cm 2.cm 2.cm 1.cm},clip, width=.33\linewidth]{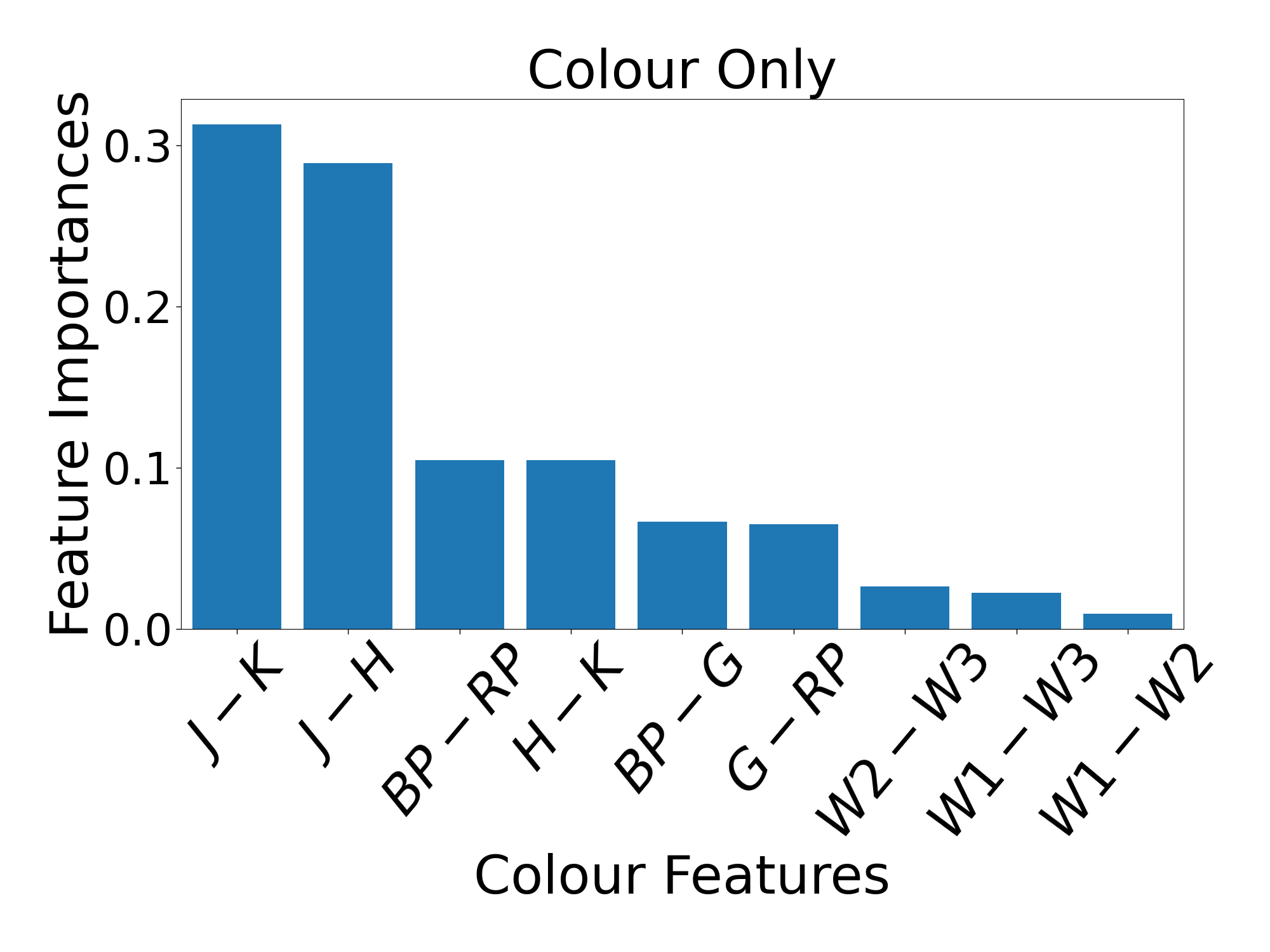}
 \includegraphics[trim={2.cm 2.cm 2.cm 1.cm},clip, width=.33\linewidth]{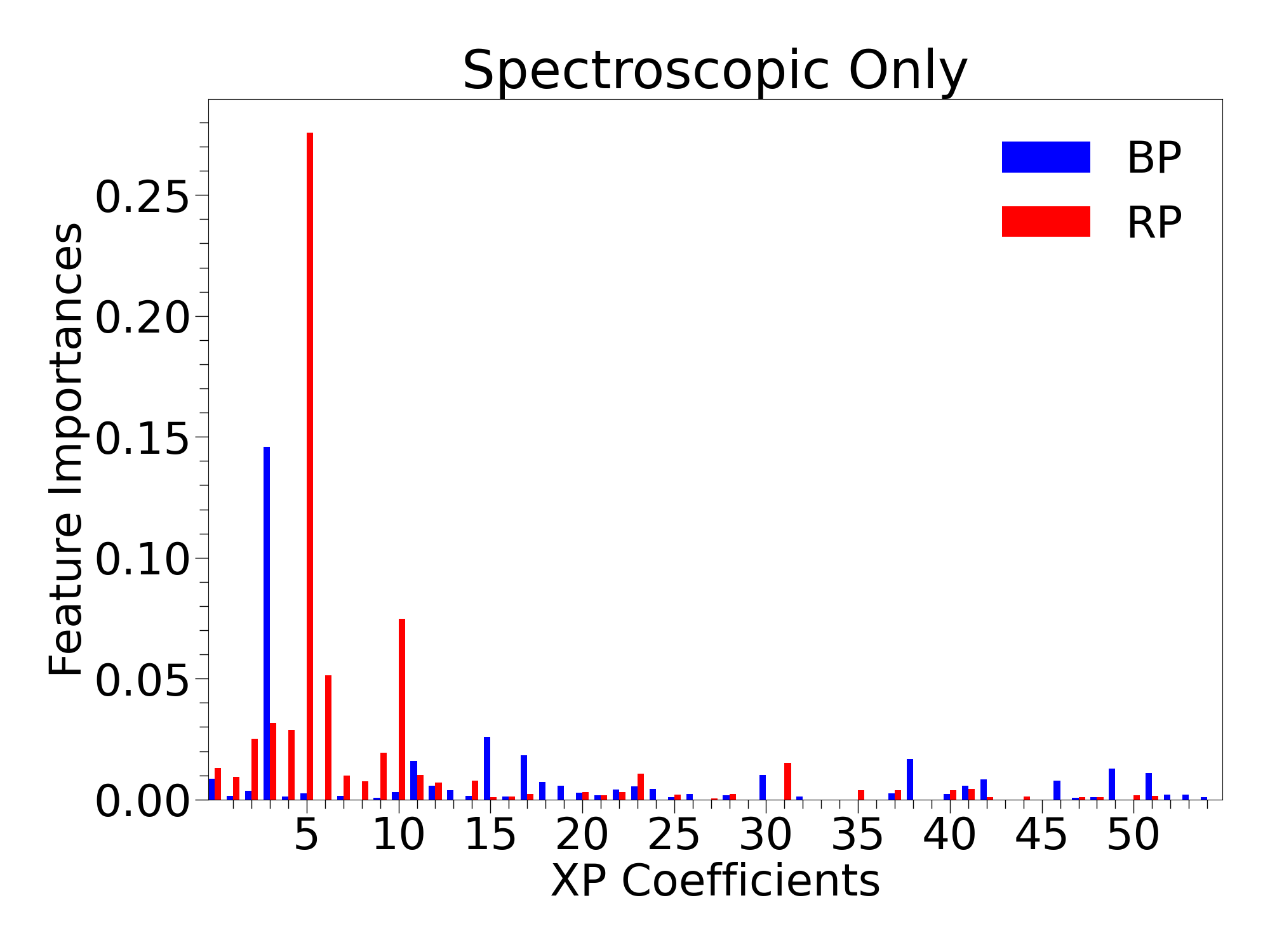}
\includegraphics[trim={2.cm 2.cm 2.cm 1.cm},clip, width=.33\linewidth]{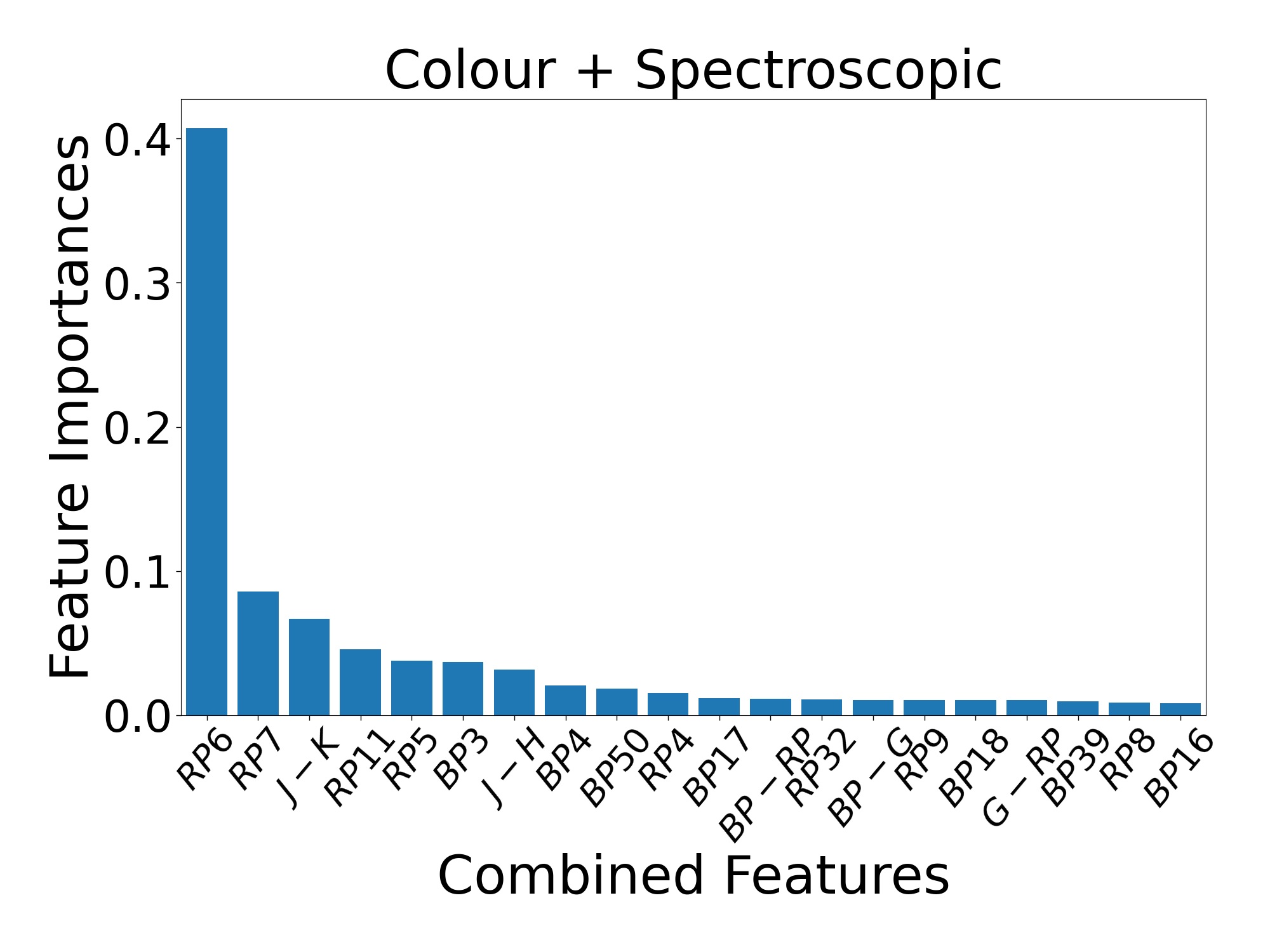}
 \caption{Feature-importance histograms for the colour features (top), spectroscopic BP/RP coefficients (middle), and the combined feature set (bottom), showing how each feature set contributes to the XGBoost classification.}
 \label{feature_importance}
\end{figure*}

\begin{figure}
    \centering
    \includegraphics[trim={1.9cm 2cm 1.5cm 0cm},clip,width=1.\linewidth]{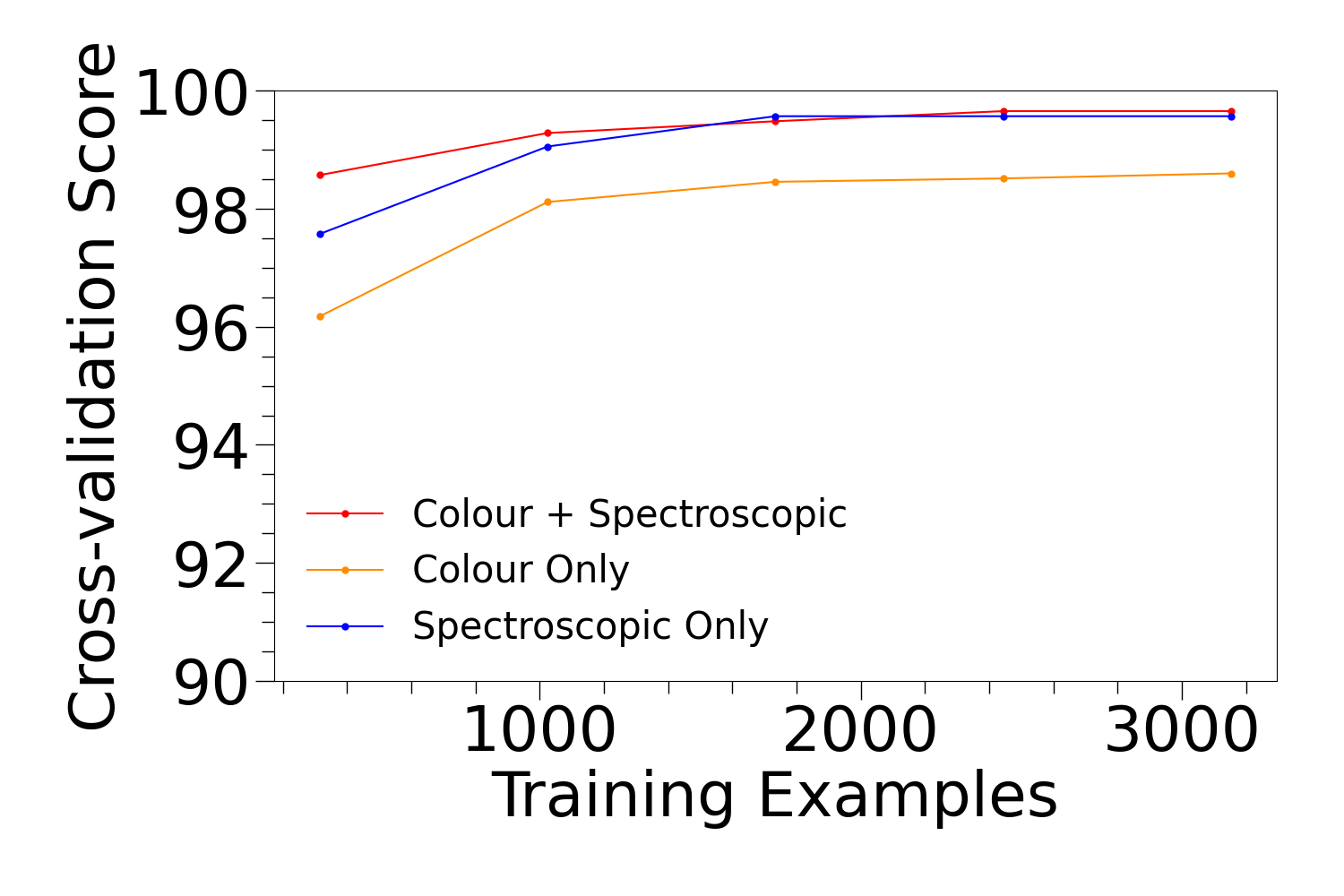}
    \caption{Learning curves showing five-fold cross-validation scores for different training-set sizes using colour, spectroscopic, and combined features. The limited number of points reflects the discrete training subset sizes defined by the \textit{train\_sizes} parameter.}
    \label{crossvalidation}
\end{figure}

A higher probability threshold increases the confidence in classifying a source as C-rich, O-rich, or unclassified. We selected $p$>0.95 as the optimal threshold to balance accuracy and completeness, resulting in 307,191 O-rich and 4,227 C-rich AGB candidates. 

We further compared the predictive capabilities of different feature sets at this fixed probability threshold. Table \ref{phot_spect_comb} presents the number of classified C-rich and O-rich AGB candidates using different input features: colour-only, spectroscopy-only, and a combination of both. When using only colour features, we classified 398,106 O-rich and 5,407 C-rich sources. The larger O-rich count (Table \ref{phot_spect_comb}) compared to the combined feature set may be due to contamination from RGB stars, which occupy overlapping regions in colour-magnitude space, as discussed in Sect.~\ref{rgb_analysis}. In contrast, using spectroscopic features alone, 247,404 O-rich and 4,521 C-rich sources were identified. Spectroscopy is more selective, effectively identifying sources with clearer spectral bands but potentially overlooking those with subtler absorption features, a common limitation for many {\it Gaia} DR3 XP sources. The combined feature set offers a more balanced approach, predicting 307,191 O-rich and 4,227 C-rich sources. This balance arises from the complementary nature of the two data types: colour features provide greater discrimination in the optical to infrared colour space, while spectroscopic features offer more precise classification based on molecular absorption patterns. By integrating both, we reduce potential contamination and improve overall classification, particularly in cases where spectroscopy alone might miss important spectral characteristics.

In ML, the importance of features quantifies the relative contribution of each input feature to the model's classification decisions. Fig.~\ref{feature_importance} presents the most important features for XGBoost classification. The top panel shows the case for colour features only, the middle panel for spectroscopic features only, and the bottom panel for the combined approach. For the colour features, $J-K$ is the most important, contributing $\sim$30\% of the total feature importance, followed by $J-H$ at 28\%, $BP-RP$ at 10\%, and $H-K$ at 10\%. The $BP-G$ and $G-RP$ colours contribute 7\%, while WISE colours play a minimal role, each contributing less than 5\%. In the spectroscopic feature set, the first 10 RP coefficients dominate the classification. The sixth RP coefficient alone contributes over 25\% of the importance, with BP3 providing about 15\%. Other BP coefficients play a relatively minor role in the classification. In the combined feature set, the sixth RP coefficient ($RP6$) remains dominant, contributing more than 40\%, followed by $RP7$ at approximately 8\%. The dominant $RP6$ coefficient captures the broad spectral shapes across the full RP wavelength range (640--1050~nm), where O-rich and C-rich AGB stars' spectra differ most: CN absorption bands dominate in C-rich stars while TiO/VO absorption bands dominate in O-rich stars, making $RP6$ the single most diagnostic feature of the two chemical types. 
Colour index $J-K$, along with $RP11$, and $RP5$, each contributes only 5-6\%. This demonstrates that spectroscopic-only classification, especially with RP coefficients, is the most effective method for distinguishing O-rich and C-rich AGB stars, while the $J-K$  colour provides complementary discriminative power. 

Using the training samples described in Methods (Sect.\ref{xgboost}), we evaluated cross-validation scores (Fig.~\ref{crossvalidation}). In general, increasing the number of training samples leads to higher validation scores. For colour features only, the validation score starts at $\sim$96\% with 700 training samples and increases to $\sim$98\% with 1,000 samples, which is the maximum accuracy achievable using only colour features. In contrast, spectroscopic features can reach 99\% accuracy with over 1,000 training samples. The highest validation score is obtained when using both colour and spectroscopic features, reaching 99.8\%. These results confirm that combining colour and spectroscopic features provides the highest classification performance.

After classification, the CMD of absolute $G$ magnitude vs. $BP-RP$ revealed a small contamination of M-dwarfs among the O-rich AGB candidate population. Here, absolute magnitudes are computed using distances derived from {\it Gaia} DR3 parallaxes, consistent with the parallax quality cut applied during data preparation (Sect.~\ref{data_prep}). The O-rich sources with an absolute \textit{G} magnitude greater than 5 are likely to be M-dwarfs and were therefore removed from further analysis. This corresponds to 1\% of the O-rich sources. Further details on this contamination removal process are provided in Appendix~\ref{removal_of_mdwarfs}. This reduced our classified O-rich sources from 307,191 to 304,111. Removing this contamination ensures that our final sample of O-rich AGB candidates is more reliable for subsequent analysis.

\section{Analysis of the classified C-rich and O-rich AGB stars}\label{analysis_classified_agb_and_discussions}

Using similarity analysis and \textsc{XGBoost}, we identified 304,111 and 4,227 stars as candidates for O-rich and C-rich AGB stars, respectively, from the {\it Gaia} catalogue. We apply ISM extinction corrections as detailed in Appendix~\ref{ism_extinction_correction} to all magnitudes used in the following analysis. To validate our O-rich and C-rich classifications, we visually inspected spectra of a representative subset of classified stars and examined their positions on the colour-magnitude diagram. We further validate the sample using the {\it Gaia}-2MASS CMD. Finally, we compare these extinction-corrected diagrams with theoretical isochrones to estimate the ages and masses of the stars. Throughout Sect.~\ref{analysis_classified_agb_and_discussions}, absolute magnitudes are computed using geometric distances from \citet{Bailer-Jones2021}. They are inferred from \textit{Gaia} parallaxes and their associated uncertainties using a Bayesian approach, which provides more reliable estimates than direct parallax inversion, particularly for farther stars with large fractional distance uncertainties. Of the 304,111 O-rich AGB candidates, 304,090 have reliable Bailer-Jones distance estimates and are used in subsequent analyses; the other 21 are excluded. All 4,227 C-rich AGB stars have Bailer-Jones distances available.

\subsection{Absolute \texorpdfstring{$K_s$}{Ks} vs. \texorpdfstring{\textit{J-$K_s$}}{J-Ks} CMD of Classified C/O-rich AGB Stars} 

\begin{figure*}
\centering     

\subfigure{\label{kvsjk_co}\includegraphics[scale=.33]{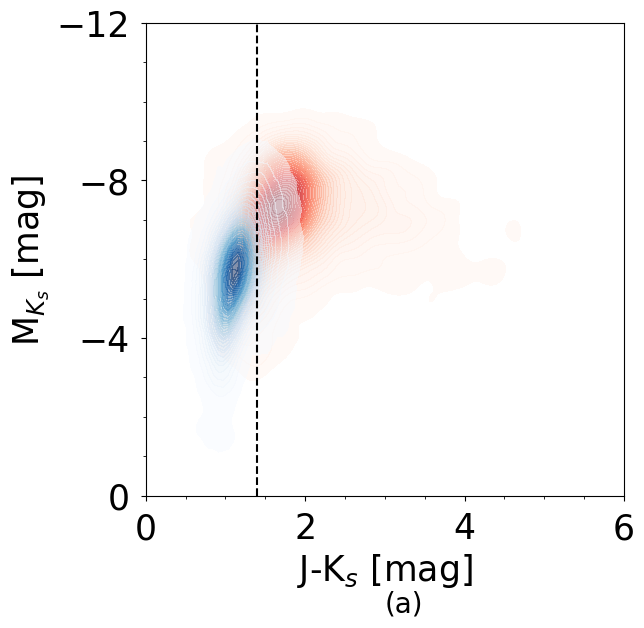}}
\subfigure{\label{crich_spectra}\includegraphics[scale=.33]{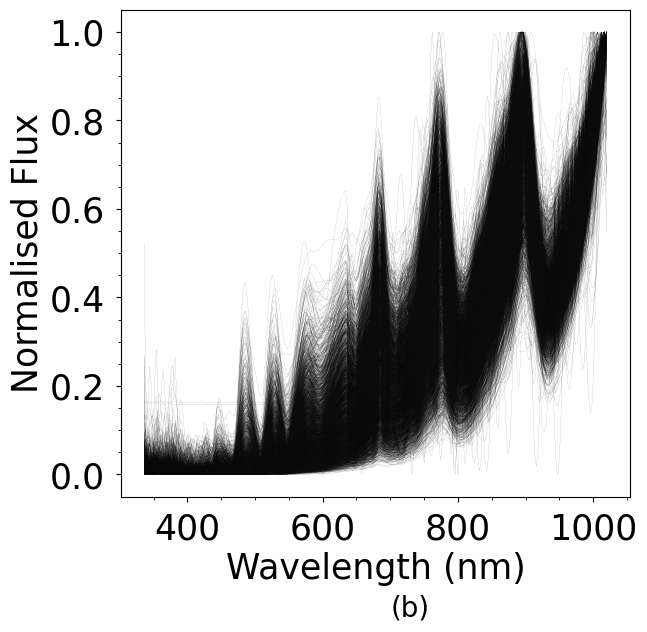}}
\subfigure{\label{orich_spectra}\includegraphics[scale=.33]{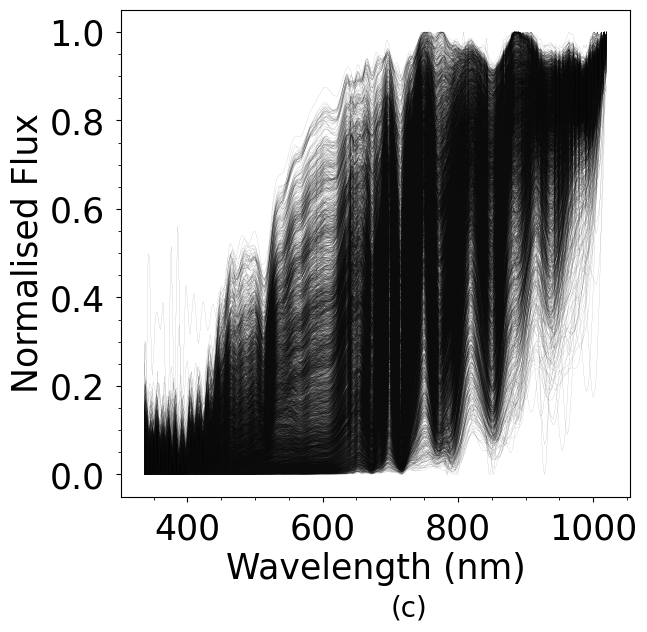}}
\caption{Panel (a) shows a kernel density estimation (KDE) plot of absolute \Ks\ magnitude versus \JKs\ for the classified C-rich (red) and O-rich (blue) AGB stars. The KDE represents the smoothed density distribution, where higher colour intensity indicates a greater concentration of stars. Panel (b) shows the spectra of all classified C-rich stars, while Panel (c) shows 5,000 randomly selected spectra of O-rich AGB stars from the O-rich cluster.}
\label{plot_discussion_cmd_w_spectra}
\end{figure*}

We validate our machine learning classification by constructing the absolute \textit{K$s$} magnitude (\MK) versus {\JKs} colour-magnitude diagram, a diagnostic widely used to distinguish O-rich and C-rich AGB stars in the Large Magellanic Cloud \citep{Nikolaev2000, Cioni2003, Cioni2014}.  Nikolaev reported that E-AGB stars typically have \JKs$>0.6$, while \citet{Cioni2003} found that O-rich AGB stars have \JKs$<1.4$ and C-rich AGB stars have \JKs$>1.4$, in the DENIS filter system. The DENIS and 2MASS filter systems are convertible \citep{Cabrera2003}.

The left panel of Fig.~\ref{plot_discussion_cmd_w_spectra} shows the \MK\ versus \JKs\ CMD as a kernel density estimate (KDE) distribution. The KDE distributions provide smoothed representations of the number densities of classified C-rich and O-rich AGB stars. The majority of our AGB candidates are indeed found at \JKs\ $>0.6$. O-rich AGB stars tend to lie at \JKs\ $<1.4$, and C-rich AGB stars at \JKs\ $>1.4$, although the dividing line is not strictly linear.  The CMD results validate the classification of AGB candidates obtained from our two-step machine learning approach.

To further support the classification, the middle and right panels of Fig.~\ref{plot_discussion_cmd_w_spectra} show {\it Gaia} XP spectra for the identified AGB stars: the middle panel displays spectra of stars classified as C-rich AGB stars, while the right panel shows a randomly selected subset of 5,000 spectra from the 304,090 classified O-rich AGB stars. C-rich AGB stars exhibit multiple absorption bands due to C$_2$ and CN molecules, whereas O-rich AGB stars show strong TiO and VO absorption bands (Fig.~\ref{C/Orich}).

\begin{figure}
    \centering
    \includegraphics[ width=1.\linewidth]{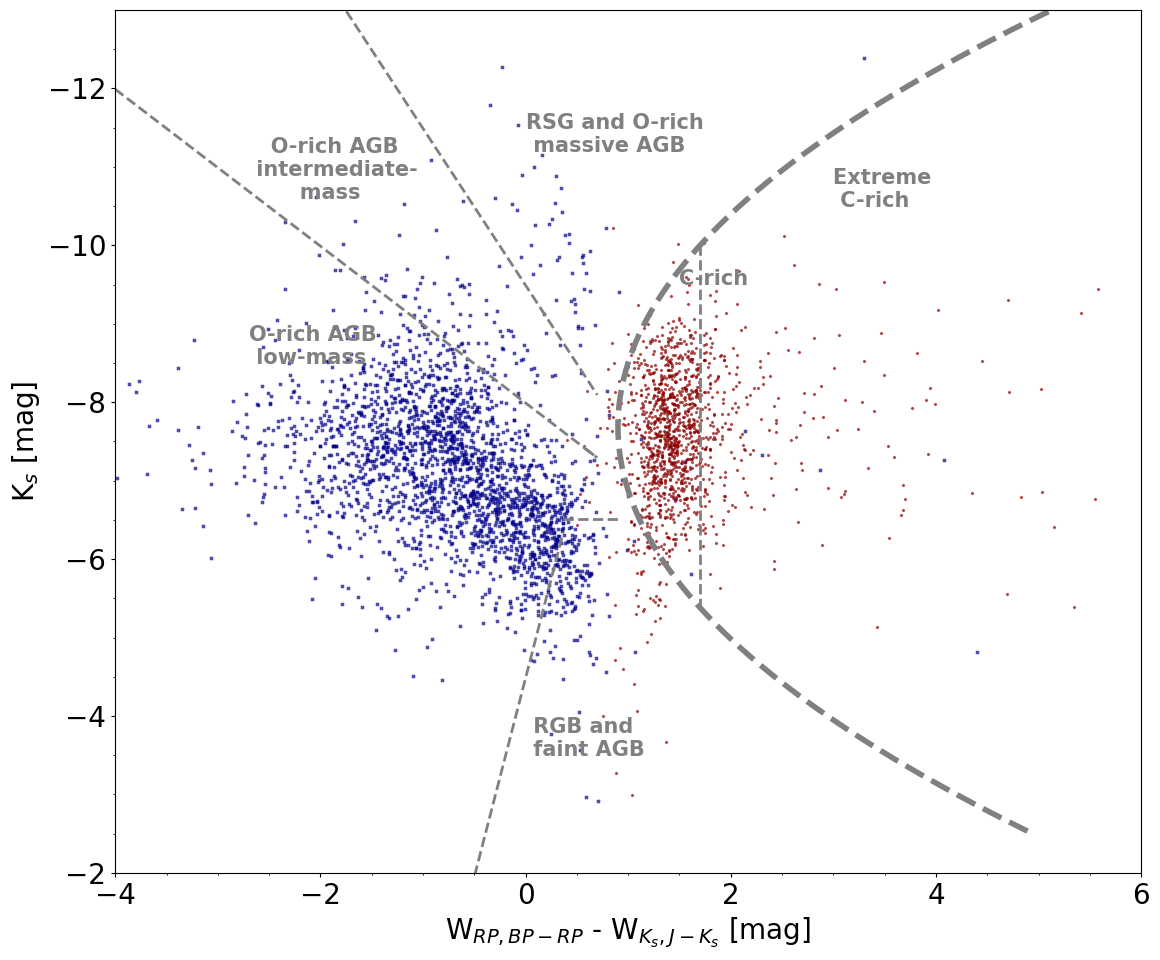}
    \caption{\MKvsWH\ diagram of the known Galactic AGB training sample (\textit{KnAGB}), showing C-rich (red) and O-rich (blue) AGB stars. The grey dashed parabolic boundary separating C- and O-rich AGB stars, and the extreme C-rich limit, are adapted from \citet{Abia2020} for the MW. Dashed lines denote the stellar sub-groups following \citet{Lebzelter2018}, rescaled to absolute magnitudes using the LMC distance (49.59~kpc; \citealt{Pietrzynski2019}). The training sample predominantly occupies the expected AGB branches of the diagram, with the majority of O-rich stars in the low-mass and intermediate-mass O-rich branches and C-rich stars in the C-rich and extreme C-rich regions.}
    \label{fig:training_agb_wesenheit}
\end{figure}

\begin{figure*}
\centering
\includegraphics[trim={.2cm, 4.2cm, .25cm, 2cm}, clip,width=1.\linewidth]{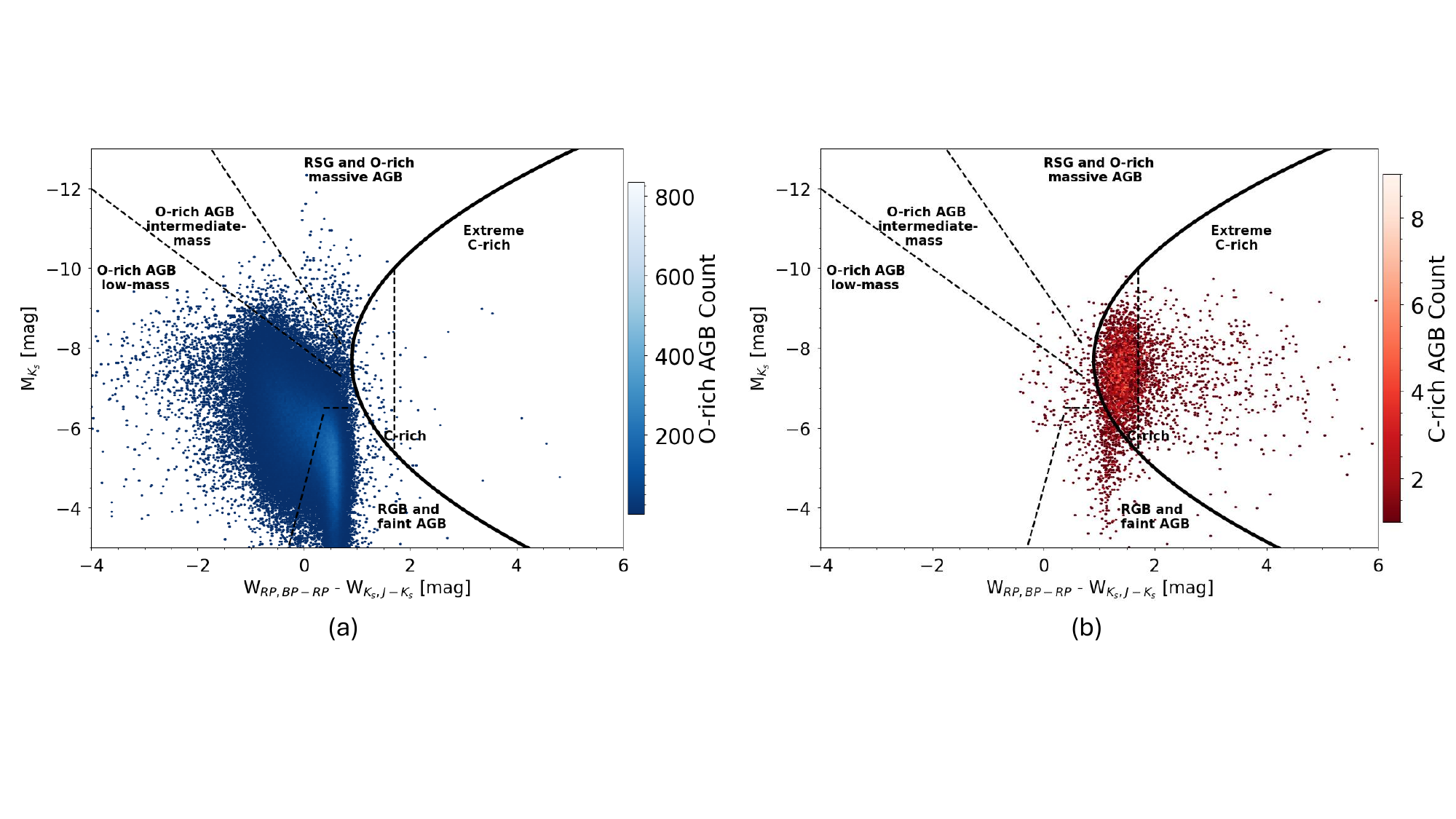}
\caption{{\it Gaia}-2MASS (\MK vs. \WH) diagram \citep{Lebzelter2018} of the classified AGB candidates obtained from our two-step classification approach. Panel (a) shows O-rich AGB stars (blue) and panel (b) shows C-rich AGB stars (red), with colour bars indicating the number counts. Different subgroups: low-mass, intermediate-mass, and massive O-rich AGB stars, as well as C-rich and extreme C-rich AGB stars, occupy distinct regions of the diagram. The solid curve marks the empirical boundary separating the two populations, with O-rich AGB stars on the left and C-rich stars on the right. Dashed lines denote the stellar sub-groups indicated in the plot.}
\label{wessenheit_c/o}
\end{figure*}

\begin{figure*}
\centering     
\subfigure{\includegraphics[width=.49\linewidth]{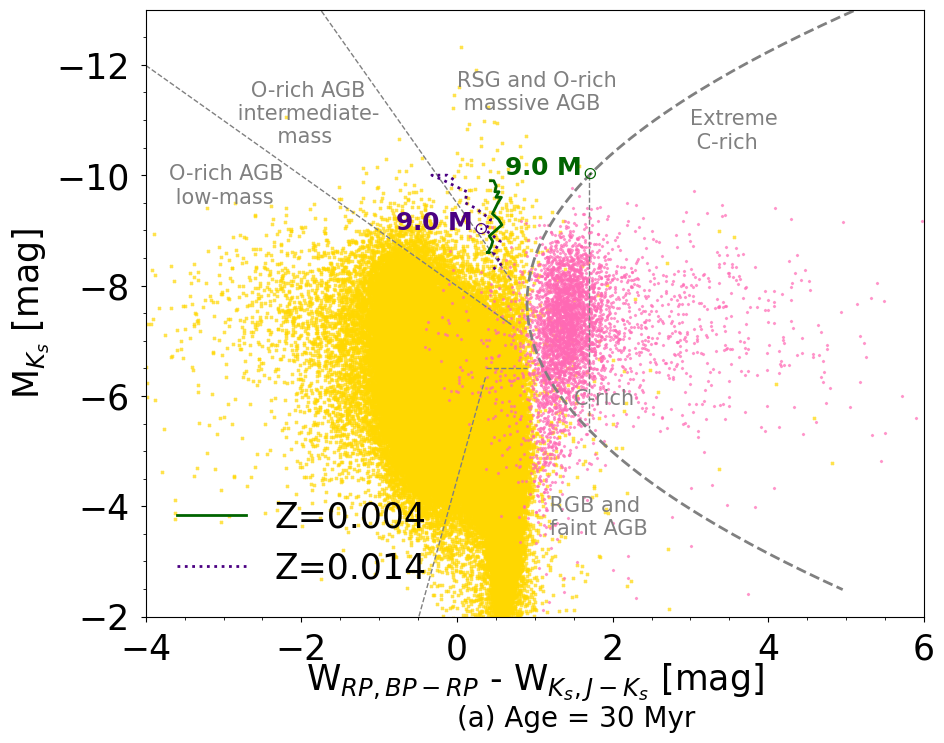}} 
\subfigure{\includegraphics[width=.49\linewidth]{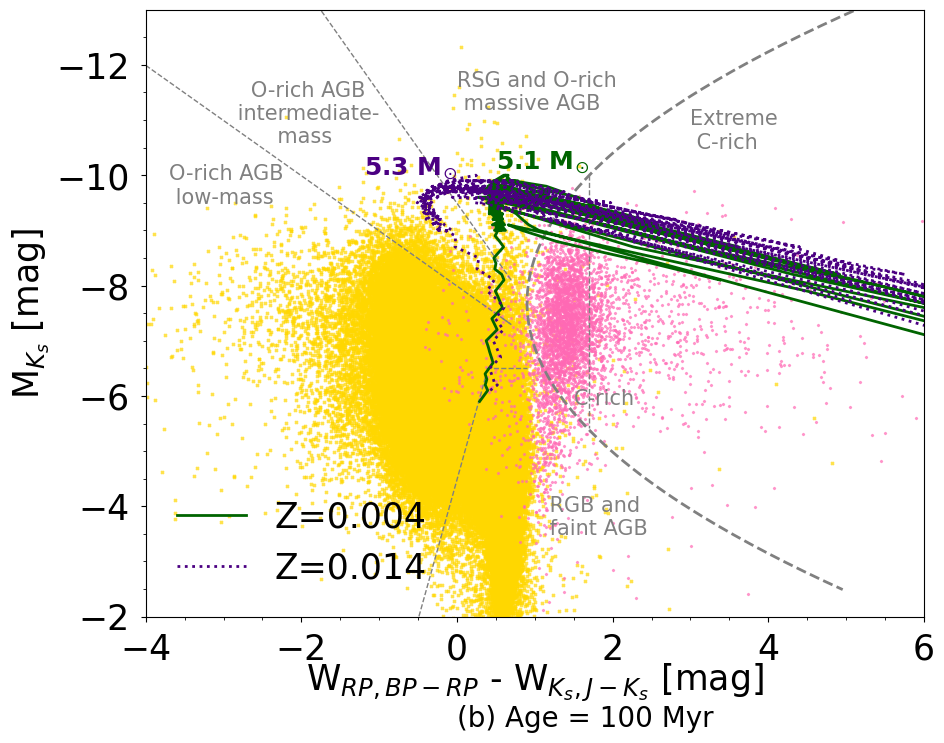}}
\subfigure{\includegraphics[width=.49\linewidth]{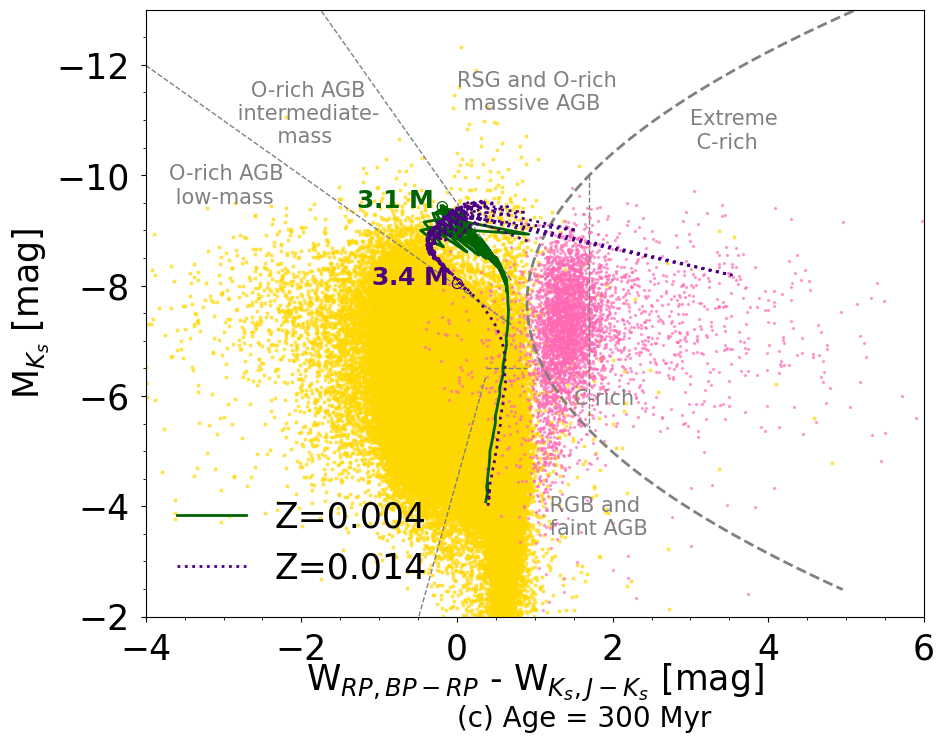}}
\subfigure{\includegraphics[width=.49\linewidth]{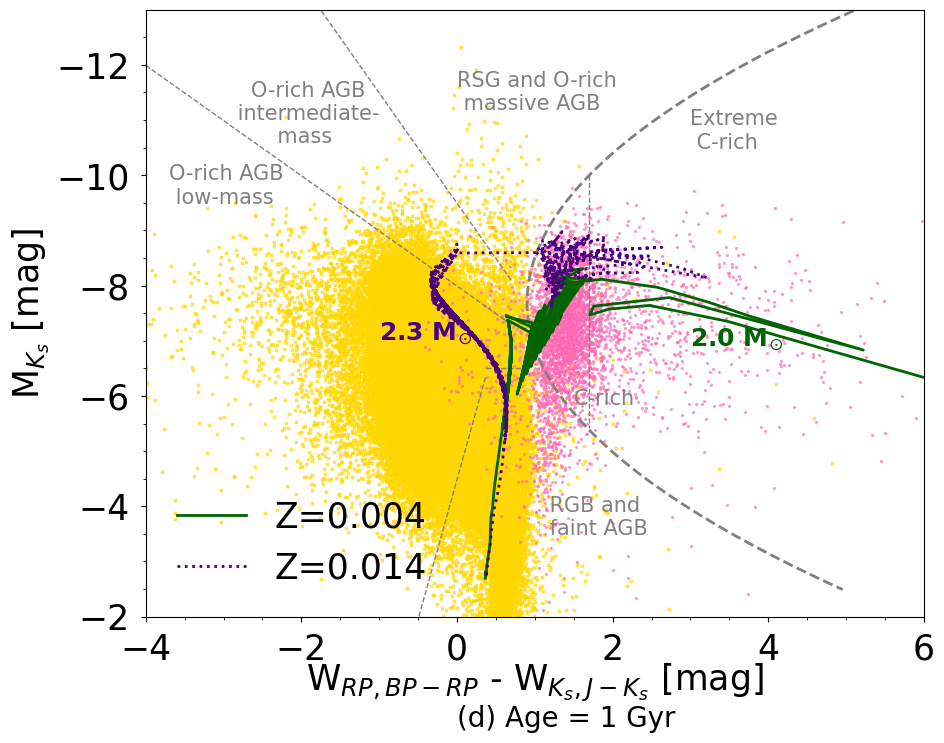}} 
\subfigure{\includegraphics[width=.49\linewidth]{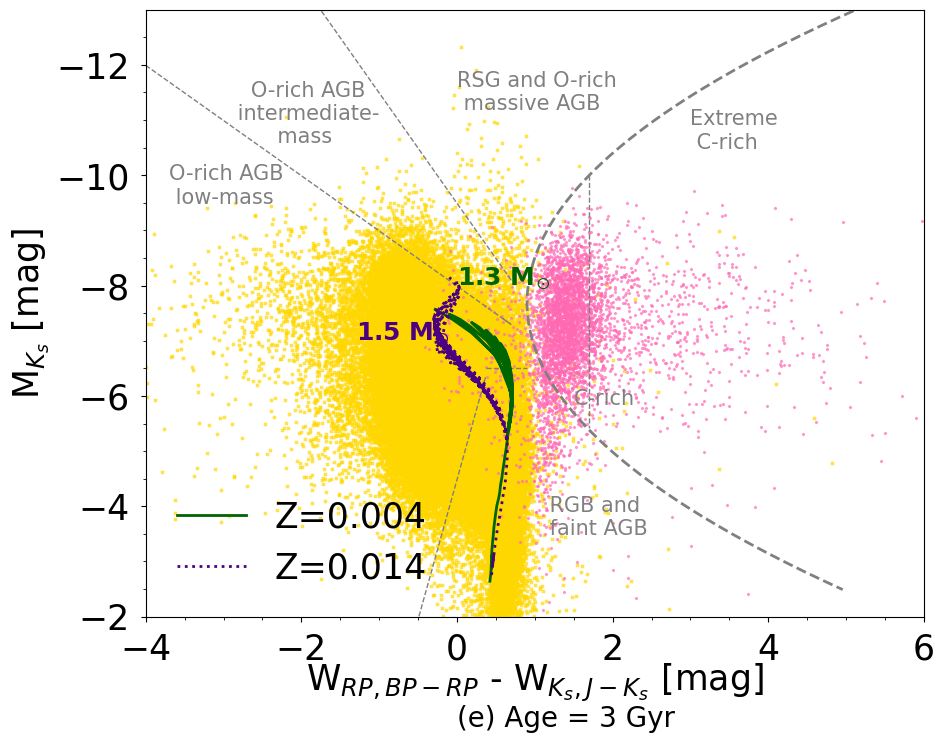}}
\subfigure{\includegraphics[width=.49\linewidth]{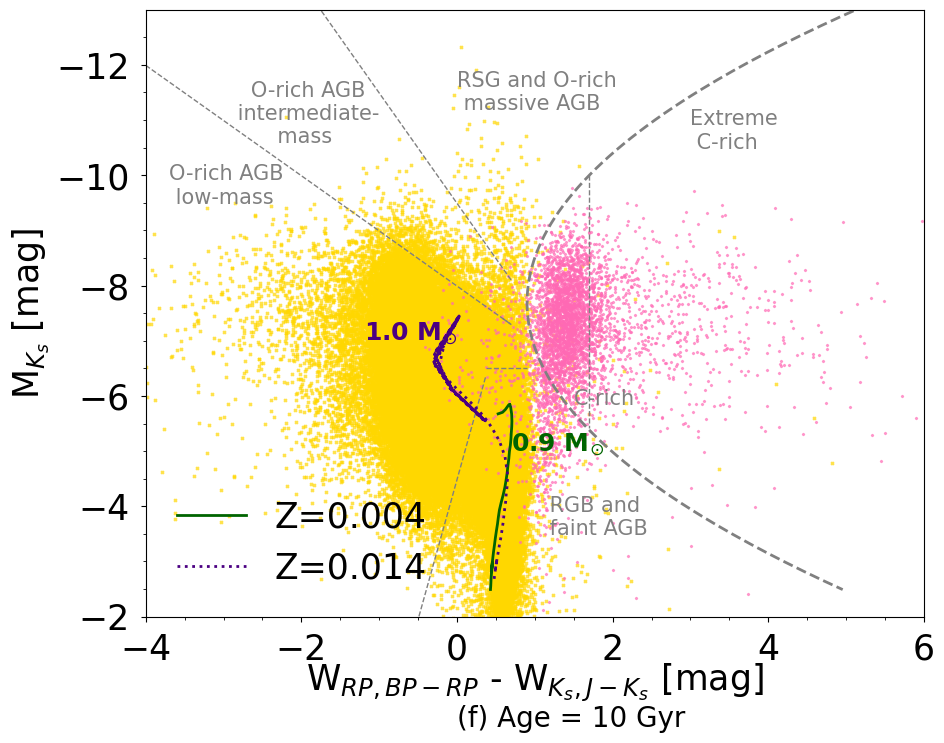}} 
\caption{\MKvsWH\ diagrams showing classified C-rich (pink) and O-rich (yellow) AGB candidates, overlaid with AGB evolutionary models. Grey dashed lines mark AGB sub-groups corresponding to ages of 30~Myr, 100~Myr, 300~Myr, 1~Gyr, 3~Gyr, and 10~Gyr, adapted from \citet{Lebzelter2018} and rescaled for absolute magnitudes. The grey dashed parabolic boundary separating C- and O-rich AGB stars, and the extreme C-rich limit, follow \citet{Abia2020}. Stellar isochrones for Z = 0.014 (purple solid) and Z = 0.004 (green solid) represent MW and LMC metallicities, respectively, and cover only the E-AGB and TP-AGB phases, to allow comparison of observed distributions with AGB evolutionary models \citep{Marigo2017}. Initial masses of the isochrones are indicated on the plot in matching colours. Overall, the diagram shows that ML-classified C-rich and O-rich AGB candidates occupy regions consistent with stellar evolutionary model predictions for their masses and ages.}
\label{wessenheit_c/o_isochrone}
\end{figure*}

\subsection{{\it Gaia}-2MASS Diagram with Wesenheit Function}\label{wesenheit_intro}

\citet{Lebzelter2018} developed a classification scheme for AGB stars in the LMC based on a diagram that combines {\it Gaia} and 2MASS photometry. The diagram relates the absolute \Ks\ (\MK) magnitude to a Wesenheit index ($W_{RP-K_s}$) constructed from {\it Gaia} ($BP-RP$) and 2MASS (\JKs) colours. Although this diagram is not a classical CMD, it provides an effective separation between C-rich and O-rich AGB stars, which occupy distinct regions. The locations of these regions were identified empirically from the observed LMC population and subsequently compared with stellar evolution models to interpret the stellar types and evolutionary stages associated with each region \citep[e.g.][]{Girardi2005, Marigo2017}. The Wesenheit function accounts for interstellar dust extinction (assuming $R_V = 3.1$). For 2MASS colours, it is defined as $W_{K_s,J-K_s} = K_s - 0.686 \times (J - K_s)$ \citep{Soszynski2005}, and for {\it Gaia} $BP-RP$, it is defined as $W_{RP} = G_{RP} - 1.3 \times (G_{BP} - G_{RP})$ \citep{Lebzelter2018}.

The Wesenheit function, which combines colour and magnitude, is largely insensitive to extinction because the colour term cancels most of the effect. However, the absolute $K_s$ magnitude is still affected by interstellar extinction, which we correct using the maps of \citealt{Vergely2022} (see Appendix~\ref{ism_extinction_correction}).

To construct the {\it Gaia}-2MASS diagram for classified AGB stars, we compute the Wesenheit function on the x-axis directly from apparent magnitudes. On the y-axis, we plot absolute \Ks magnitudes, computed using geometric distances from \citet{Bailer-Jones2021}. This enables a direct comparison with Lebzelter’s LMC results by adopting an LMC distance of 49.59 kpc \citep{Pietrzynski2019}.

\citet{Lebzelter2018} showed that the Wesenheit diagram places the bluest objects near the centre, with redder stars toward both sides, unlike classical CMDs where colour varies monotonically. The O-rich region contains branches corresponding to low-mass, intermediate-mass, red supergiants (RSGs) and massive O-rich AGB stars, and RGB/faint AGB stars, while the C-rich region comprises typical and extreme C-rich categories \citep{Lebzelter2018}. In Fig.~\ref{fig:training_agb_wesenheit} and Fig.~\ref{wessenheit_c/o}, we mark different AGB branches with dashed lines, while a solid curve separates C-rich and O-rich populations. We adapted these boundaries from Table 1 of \citet{Lebzelter2018} and rescaled them to absolute magnitudes using the LMC distance modulus, except for the solid curve and the extreme C-rich boundary, which were modified by \citet{Abia2020} for the MW.

We show the \textit{KnAGB} training sample on the \textit{Gaia}-2MASS diagram in Fig.~\ref{fig:training_agb_wesenheit}. The sample largely occupies the expected AGB branches, with O-rich AGB stars located in the low-mass and intermediate-mass O-rich branches and C-rich AGB stars in the C-rich and extreme C-rich regions. This agreement with the classification scheme of \citet{Lebzelter2018} is consistent with a training sample dominated by AGB stars, although the semi-regular variables below the tip of the RGB may include some RGB stars (Sect.~\ref{rgb_analysis}).

After validating the training sample, we examine our ML-classified C-rich and O-rich AGB candidates in Fig.~\ref{wessenheit_c/o}. They form two distinct clusters in this diagram, separated by the solid curve. A large number of O-rich AGB candidates (Fig.~\ref{wessenheit_c/o}(a)) populate the low-mass branch, which appears more continuous and less clearly separated from the intermediate-mass branch than in the LMC \citep{Lebzelter2018}, with fewer stars in the intermediate- and massive O-rich branches.

There is no clear boundary separating low- from intermediate-mass stars. The low-mass O-rich AGB stars extend further downward into the RGB and faint AGB region, with a high density between $-7<M_{K_s}<-4$. 

C-rich AGB candidates (Fig.~\ref{wessenheit_c/o}(b)) are concentrated to the right side of the solid separation curve, with a subgroup located at \WH $\gtrsim1.7$ corresponding to stars with particularly red NIR colours (\JKs$\gtrsim$2; \citealt{Blum2006}). These have often been referred to as ``extreme'' C-rich AGB stars in the literature \citep{Blum2006}. However, as discussed by \citet{Sloan2016} and \citet{Groenewegen2022}, this label is somewhat misleading, as these stars are not necessarily unusual in their evolutionary stage, but rather are ordinary carbon stars with mass-loss rates of $\sim$10$^{-6}$~\M yr$^{-1}$. By contrast, some C-rich AGB show mass-loss rates exceeding $10^{-4}$~\M yr$^{-1}$ \citep[e.g.,][]{Gruendl2008, Sloan2016, Groenewegen2018}, and these may be more appropriately described as truly ``extreme'' C-rich AGB stars. We also observe a notable presence of ML-classified C-rich stars in the RGB and faint AGB branches. 

Out of 4,227 C-rich AGB candidates identified by ML, we obtained 2,552 (60\%) stars in the C-rich region of \MKvsWH\ diagram, 958 (23\%) in the extreme C-rich region, 550 (13\%) in the RGB and faint AGB branch and the remaining 167 (4\%) on the left side of the solid curve. In contrast, out of 304,090 classified O-rich candidates, 141,839 (47\%) are in the low-mass O-rich AGB region, 160,577 (53\%) in the RGB and faint AGB region, 1,306 (0.4\%) in the intermediate O-rich AGB region, 275 (0.1\%) in the RSG and massive O-rich region, and the rest about (0.03\%) are found on the right of the solid curve.

While the majority of ML-identified C-rich and O-rich AGB candidates fall into the corresponding regions in the \MKvsWH\ diagram, some minority stars do not. We examine these minority stars by manual inspection to determine whether the issue is machine learning’s classification or CMD classification. We inspected the spectra of 167 ML-classified C-rich stars located within the O-rich branches in the \MKvsWH\ diagram. Of these, 53 (1.3\% of all ML C-rich stars) exhibit strong C$_2$ and CN molecular bands, confirming their C-rich nature despite their location. Many of these stars lie close to the boundary curve, highlighting the difficulty of achieving perfect separation using a 2D colour-based classification. While most O-rich and C-rich AGB stars form distinguishable clusters, there remains a region of overlap where the two populations intermingle. This overlap appears more prominent in our Milky Way sample than in the LMC \citep{Lebzelter2018}, where the corresponding populations are more distinctly separated. This suggests that although the colour-based classification scheme performs well overall, it may fail to identify some minority stars, such as C-rich stars located within O-rich regions in the \MKvsWH\ diagram. In such cases, {\it Gaia} BP/RP spectra provide more accurate classifications.

The remaining 114 ML C-rich stars that fall in the O-rich region in the \MKvsWH\ diagram showed either featureless profiles or faint molecular bands that proved difficult to make visual classification. These stars have inconclusive classifications based on their spectra.

Conversely, only 0.03\% (93 stars) of ML-classified O-rich AGB stars fall within the C-rich region of the \MKvsWH\ diagram. Spectral inspection confirms these stars are O-rich, as indicated by strong TiO and VO bands. Their redder near-to-mid-infrared colours, caused by thermal emission from dust in their circumstellar envelopes, shift them into the C-rich regime (as discussed in detail in Sect.~\ref{comparing_wesenheit_with_isochrones}). These rare, exceptionally dust-enshrouded O-rich AGB stars show how ML can predict unusual objects that deviate from typical photometric expectations. 

Our feature importance analysis (Fig.~\ref{feature_importance}) showed that the \JKs\ colour index was the most important feature in the colour-only classification, contributing approximately 30\%, followed by {\it Gaia} $BP-RP$ colour (about 10\%). Interestingly, these features correspond closely to the ($BP-RP$)- and (\JKs)-based Wesenheit indices used by \citet{Lebzelter2018}, demonstrating that the machine learning approach independently recovers the key photometric dimensions relevant for AGB star classification. In the combined feature set, the {\it Gaia} BP/RP spectral coefficients proved particularly valuable, with the sixth RP coefficient (RP6) alone contributing over 40\% (Fig.~\ref{feature_importance}). This is especially critical for stars near the C/O boundary, where colour information alone may be insufficient for accurate classification. 

\subsection{Comparing {\it Gaia}-2MASS Diagram with Stellar Isochrones}\label{comparing_wesenheit_with_isochrones}

\begin{table}
    \centering
    \begin{tabular}{l  r  c  r}
    \hline
    \thead{Classified \\ AGB Region} &  \thead{Number\\ of Stars} & \thead{Mass Range\\ (\M)} & Age\\
    \hline
    \textbf{O-rich} & & & \\
    \hline
    RGB and faint AGB     & 160,577 & $\sim$1 & 10~Gyr\\
    Low-mass AGB          & 141,839 & 1--1.5  & 3--10~Gyr \\
    Intermediate-mass AGB &   1,306 & 3--5    & 100--300~Myr \\
    RSG and massive AGB   &     275 & $\sim$9 & 30~Myr \\
    \hline
    \textbf{C-rich} &  & & \\
    \hline
    C-rich (including\\ extreme C-rich) & 3,510 & 2--2.5 & 1~Gyr \\
    \hline
    \end{tabular}
    \caption{Summary of classified AGB candidates from the \MKvsWH\ diagram (Fig.~\ref{wessenheit_c/o_isochrone}), grouped by approximate mass ranges and evolutionary phases. Masses and ages are taken from PARSEC--COLIBRI E-AGB and TP-AGB isochrone tracks that intersect the visually identified branches in the diagram, based on their overlap with the observed distributions.}
    \label{tab:agb_stats}
\end{table}

Stellar evolution models can produce isochrones on colour--magnitude diagrams. In this case, we computed the \WH\ function values from {\it Gaia} $BP-RP$ and 2MASS $J-K$ colours provided in the isochrone tables and plotted them against the \MK. These isochrones allow us to estimate the age, mass, and metallicity of stars. We used the brightness of E-AGB and TP-AGB stars, as modelled by PARSEC--COLIBRI \citep{Bressan2012, Marigo2013, Tang2014, Chen2015, Marigo2017, Pastorelli2019, Pastorelli2020}, to calculate magnitudes in the \MK\ vs. \WH\ plane. These models also account for thermal emission from circumstellar dust, following the prescriptions of \citet{Groenewegen2006}. A metallicity of $Z = 0.014$ is used to represent the MW conditions, while $Z = 0.004$ corresponds to the metallicity of the LMC. The models additionally provide carbon and oxygen surface abundances, which indicate whether a star is O-rich or C-rich. The plotted isochrones overlaid on the distributions of classified C-rich and O-rich AGB stars are shown in Fig.~\ref{wessenheit_c/o_isochrone}, with subplots of isochrones corresponding to 30~Myr, 100~Myr, 300~Myr, 1~Gyr, 3~Gyr, and 10~Gyr. A summary of their approximate masses, ages and evolutionary phases is provided in Table~\ref{tab:agb_stats}.

In Fig.~\ref{wessenheit_c/o_isochrone}(a), we show the isochrone for stars with an age of 30~Myr, corresponding to an initial mass of $\sim$9~\M. In the PARSEC-COLIBRI models, stars at this mass are labelled as entering the TP-AGB phase, although their subsequent evolution remains uncertain, as this mass lies close to the upper mass limit of the AGB regime \citep[e.g.,][]{Poelarends2008, Doherty2017}. Indeed, in Fig.~\ref{wessenheit_c/o_isochrone}(a), $\sim$9~\M stars trace the high-mass region occupied by red supergiants (RSGs) and massive AGB stars. Notably, the LMC metallicity isochrone appears nearly vertical around \WH\ $\sim 0$, while the MW metallicity isochrone is tilted approximately 45 degrees to the left. The observed distribution of MW stars aligns more closely with the LMC-metallicity isochrone, suggesting that theoretical models may require refinement in their colour and magnitude predictions.

In Fig.~\ref{wessenheit_c/o_isochrone}(b), we show the isochrone for stars with an age of $\sim$100~Myr, corresponding to an initial mass of approximately 5.3~\M. These stars evolve through the E-AGB and TP-AGB phases. The isochrone for LMC metallicity extends into the region associated with RSGs and massive O-rich AGB stars, while the MW metallicity isochrone follows a slightly different trajectory, curving into the region already predicted by stellar evolution models for intermediate-mass O-rich AGB stars.

In Fig.~\ref{wessenheit_c/o_isochrone}(c), the isochrone corresponding to an age of $\sim$300~Myr and initial masses of 3.4~\M, representing O-rich AGB stars, also extends toward the C-rich region in the \MKvsWH\ diagram. Despite this, the C/O ratios along the isochrone remain below unity, confirming that these sources are still O-rich in chemical composition. 

\subsubsection{Finding of O-rich AGB stars with circumstellar envelope}

In Sect.~\ref{wesenheit_intro}, we identified roughly 93 O-rich sources that appear in the region of the \MKvsWH\ diagram typically associated with C-rich and extreme C-rich AGB stars. Of these, 70 lie in the C-rich region and 23 in the extreme C-rich regime. Visual inspection of their {\it Gaia} DR3 XP spectra confirms molecular bands such as TiO and VO in many cases, supporting their O-rich classification. However, these spectra also show a redder continuum than typical O-rich AGB stars, suggesting additional reddening.

To investigate these red O-rich sources, we compared their distribution to PARSEC-COLIBRI isochrones (Fig.~\ref{wessenheit_c/o_isochrone}). Isochrones with ages of 100~Myr (Fig.~\ref{wessenheit_c/o_isochrone}(b)) and 300~Myr (Fig.~\ref{wessenheit_c/o_isochrone}(c)) extend into the C-rich region of the diagram, with \WH values exceeding 2. The carbon-to-oxygen (C/O) ratios from the isochrone abundance tables remain below unity along these evolutionary phases. This comparison indicates that while this region of the \MKvsWH\ diagram is typically associated with C-rich stars, a minority of O-rich AGB stars identified by our ML classification also occupy this region.

Previously, \citet{Lebzelter2019} found that six known OH/IR stars in the LMC sample are located in the extreme C-rich AGB regime, while \citet{Sanders2023} showed that some O-rich stars can reach \WH\ colours as red as 2. Our machine-learning classification identifies such O-rich stars that fall within the C-rich colour regime in the Galactic sample, and comparisons with isochrones indicate that these objects are O-rich AGB stars surrounded by circumstellar dust envelopes.

\subsection{Comparison of Galactic and LMC stars on the \MKvsWH\ diagram}

There are notable differences between the AGB populations of the Milky Way and the LMC in the \MKvsWH\ diagram. The 1~Gyr isochrones (Fig.~\ref{wessenheit_c/o_isochrone}(d)) reveal that, at LMC metallicity, AGB stars reach a C/O ratio greater than 1, causing them to enter the C-rich regime in the \MKvsWH\ diagram. In contrast, the MW isochrone remains mostly O-rich, although some stars appear to extend into the C-rich regime of the \MKvsWH\ diagram, due to increased mass loss or circumstellar dust extinction, despite retaining C/O$<$1.

These differences in the isochrones help explain the observed differences in the \MKvsWH\ diagrams for the LMC and MW. In the LMC, intermediate-mass and low-mass O-rich AGB stars form two clearly separated branches, with a noticeable gap in stellar densities between them, marked by a dashed grey line (see \citet{Lebzelter2018}). This gap arises because, at LMC metallicity, stars with initial masses around 1.8--2.5~\M\ \citep{Karakas2007} transition into the C-rich phase, resulting in few stars remaining in the corresponding O-rich regime.

In contrast, the MW at the same age has a narrower mass range ($\sim$2.3--2.8~\M; \citealt{Karakas2007}) over which AGB stars evolve into C-rich stars. As a result, the distribution of stellar densities in the \MKvsWH\ diagram is more continuous, and the separation between low- and intermediate-mass O-rich AGB stars is less distinct.

Although uncertainties in ISM extinction corrections may contribute to uncertainties in the \WH\ and \MK\ measurements, the overall distribution \textemdash\ especially the presence of a gap in the number density of stars between RSGs/massive O-rich AGB stars and intermediate-mass O-rich AGB stars \textemdash\ suggests that ISM extinction alone does not account for the continuous distribution of low- to intermediate-mass O-rich AGB stars. Instead, this reflects the effect of metallicity on AGB stellar evolution.

At older ages, the $\sim$3~Gyr isochrone (Fig.~\ref{wessenheit_c/o_isochrone}(e)) shows that stars with masses of 1.5~\M\ reach the E-AGB and TP-AGB phases. These low-mass stars have not undergone significant mass loss or third dredge-up events, consistent with their location along the low-mass O-rich branch of the \MKvsWH\ diagram as predicted by the MW isochrone. Similarly, for stars with ages of 10~Gyr (Fig.~\ref{wessenheit_c/o_isochrone}(f)), very low-mass stars ($<$1~\M) populate the same low-mass O-rich region. Overall, the isochrones indicate that a substantial population of stars with initial masses below 1~\M\ occupies the RGB and faint AGB, as well as the low-mass O-rich regions of the diagram.

Our ML classification results are broadly consistent with \citet{Lebzelter2018}, with O-rich and C-rich AGB stars occupying their expected regions of the \MKvsWH\ diagram. A key finding is the identification of redder O-rich AGB stars, which are also predicted by intermediate-age isochrones and are successfully captured by our classification.

\subsubsection{Modified boundaries of {\it Gaia}-2MASS for Milky Way}

Based on our ML classifications, we suggest a slight modification to the \MKvsWH\ diagram. The existing classification scheme for AGB stars in this diagram is based primarily on studies of the LMC and SMC \citep{Lebzelter2018}. However, as discussed in Sect.~\ref{comparing_wesenheit_with_isochrones}, a subset of the ML-classified MW AGB stars does not fall exactly within the original regions of the \MKvsWH\ diagram. To investigate these discrepancies, we visually inspected the corresponding {\it Gaia} DR3 spectra of these sources. Based on this inspection, we introduce a slight modification to the primary C-O separation boundary of \citet{Lebzelter2018} to better reflect the MW AGB population.

We revise the boundary between O-rich and C-rich AGB stars, which was originally defined by a parabolic curve. While \citet{Abia2020} suggested a slight modification to better match MW data, their revised curve still remains too narrow: 667 C-rich AGB stars (as identified through ML classification and confirmed using visual inspection of {\it Gaia} DR3 XP spectra) still fall in the O-rich region of the diagram. To address this, we propose a modified boundary with a wider parabola: from $0.9 + 0.15 \times (K_s - 10.8)^2$ \citep{Abia2020} to $0.9 + 0.05 \times (K_s - 10.8)^2$. The updated classification is illustrated in the Appendix in Fig.~\ref{modified_wesenheit}.

A sample of the final catalogue of 308,317 ML-classified AGB candidates (4,227 C-rich and 304,090 O-rich), including coordinates, parallaxes, photometric magnitudes, Bailer-Jones distances, classification probabilities, and the occupied region in the Wesenheit diagram, is presented in Appendix~\ref{appendix:catalogue} (Table~\ref{tab:catalogue}), with the full machine-readable version available online at \url{https://doi.org/10.5281/zenodo.20336870}.

\subsection{Comparison with literature}\label{comparison_lit}

We compared our ML-classified C-rich sample against three recent carbon star catalogues: \citet{Abia2022} and \citet{Sanders2023}, both based on \textit{Gaia} data, and \citet{He2024} from Large Sky Area Multi-Object Fiber Spectroscopic Telescope (LAMOST) spectroscopy, to evaluate how our C-rich AGB sample relates to previously identified carbon star catalogues.

We crossmatched our ML-classified C-rich sample of 4,227 stars against the two catalogues presented in \citet{Abia2022}: their Table 1 sample of 827 previously known carbon stars, assembled from infrared carbon stars \citep{Chen2012}, LAMOST DR2 survey \citep{Ji2016} and J-type Galactic carbon stars \citep{Chen2007}, supplemented by the sample from \citet{Abia2020}, and their newly identified C-rich stars catalogue (NC) of 518 stars identified using the \textit{Gaia}-2MASS diagram. We find 174 stars in common with Table 1 and 324 stars in common with the NC catalogue, giving a total of 497 unique matches across both catalogues. The remaining 3,730 stars in our sample are not found in either Abia catalogue, extending to redder colours and slightly fainter magnitudes, likely representing C-rich AGB stars with dust excess.

\begin{figure}
    \centering
    \includegraphics[width=1.\linewidth]{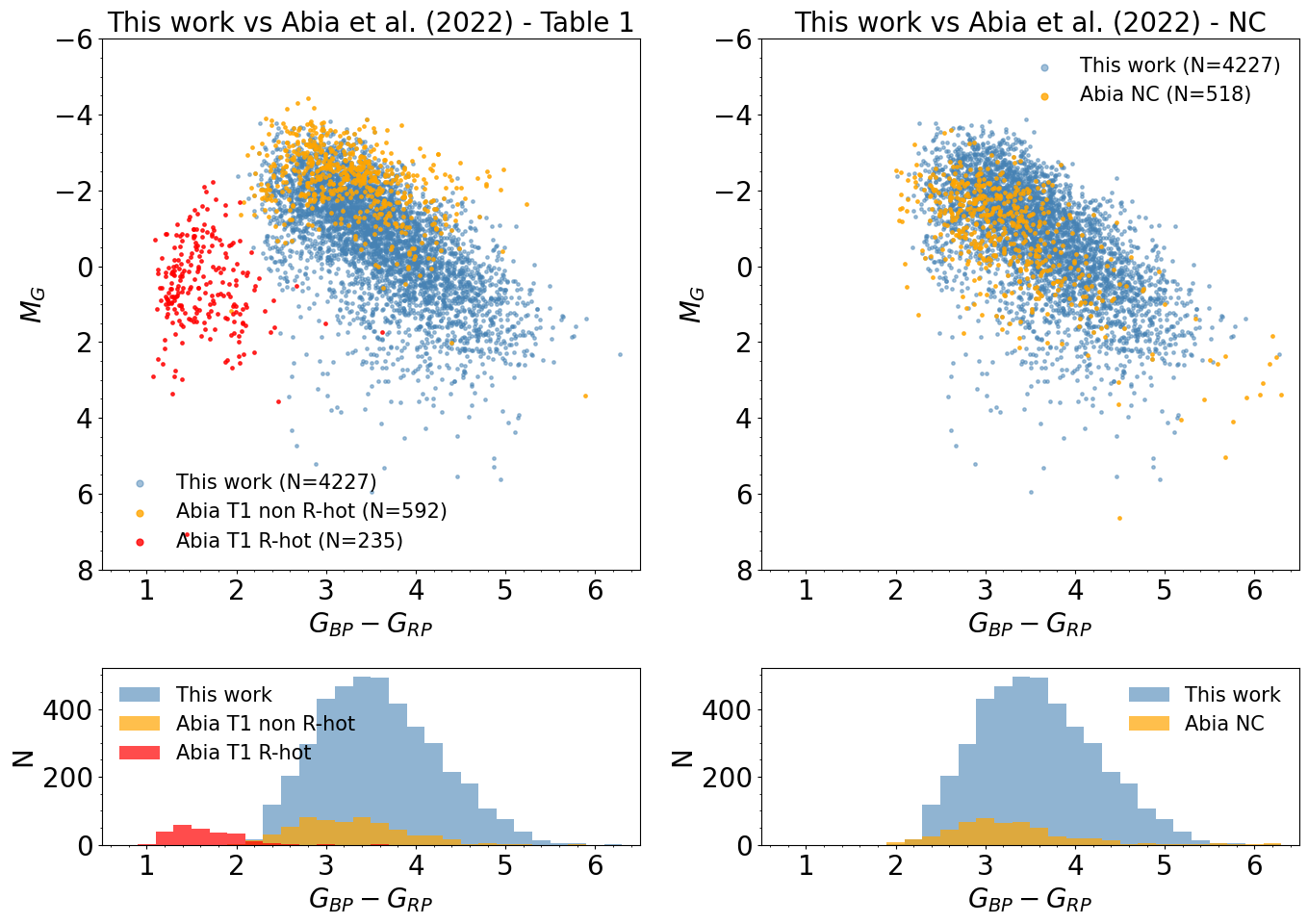}
    \caption{$M_{G}$ vs. $G_{BP}-G_{RP}$ colour-magnitude diagrams (top) and $G_{BP}-G_{RP}$ colour distributions (bottom) comparing our ML-classified C-rich sample (blue, N=4,227) with the \citet{Abia2022} catalogues: Table 1 Carbon stars (left, N=827) and newly identified NC stars (right, N=518). The overlap between our sample and Table 1 is 174, and 324 stars with the NC catalogue. 
    In the left panel, Table 1 stars are split by spectral type: R-hot stars (red, N=235) occupy distinctly bluer colours ($G_{BP}-G_{RP}\lesssim$2), while the remaining non R-hot stars (orange, N=592) overlap with our sample. R-hot type stars are not AGB stars and were explicitly excluded from our training sample as contaminants.}
    \label{fig:comparison_w_abia}
\end{figure}

The \textit{Gaia} CMD is a powerful tool for distinguishing stellar populations and evolutionary stages, as different classes of stars occupy characteristic regions in the absolute $G$ magnitude ($M_G$) vs $G_{BP}-G_{RP}$ plane \citep{Gaia2018}.
Fig.~\ref{fig:comparison_w_abia} compares the \textit{Gaia} CMD ($M_G$ vs $G_{BP}-G_{RP}$) and $G_{BP}-G_{RP}$ colour distributions of both samples. The Table 1 stars with $G_{BP}-G_{RP}>2.3$ roughly occupy the same region as our sample. However, Table 1 stars at bluer colours ($G_{BP}-G_{RP}\lesssim2$) correspond to R-hot type stars \citep{Abia2022}. R-hot stars occupy the same region of the HR diagram as red clump (RC) stars and are He-core burning stars rather than AGB stars \citep{Knapp2001, Zamora2009}, and were explicitly excluded from our training sample to avoid their contamination with AGB stars. Their absence from our ML classifications, therefore, demonstrates the intended behaviour of the classifier rather than incompleteness. In contrast, the strong overlap of our C-rich AGB stars with the NC catalogue of \citet{Abia2022} ($\sim$63\% of their NC stars recovered by us) shows that both samples occupy the same $2.5<G_{BP}-G_{RP}<6$ region in the CMD, characteristic of C-rich AGB stars. The NC catalogue contains a mix of intrinsic and extrinsic carbon stars \citep{Abia2022}. Of the 194 NC stars not recovered by our classifier, 119 (61\%) are absent from our \textit{G2MW} parent catalogue due to missing WISE photometry, and 22 (11\%) were removed during the similarity filtering stage. The remaining 53 stars entered our XGBoost pipeline, with 32 classified as O-rich and 21 falling marginally below our confidence threshold of $p(C)>$0.95.

We compared our results with \citet{Sanders2023}, who used an unsupervised learning approach to classify C-rich long-period variables (LPVs) from \textit{Gaia} DR3 BP/RP spectra. Although the methodologies differ, both studies aim to identify C-rich AGB stars from \textit{Gaia} DR3 BP/RP spectral data, making this a useful consistency check. We find 3,703 C-rich stars in common with the \citet{Sanders2023} catalogue of 23,737 C-rich LPVs, representing 89\% of the $\sim$4,157 \citet{Sanders2023} stars present in our \textit{G2MW} catalogue. The remaining 19,580 stars are absent from our \textit{G2MW} catalogue due to our parallax quality cut ($\varpi/\sigma_\varpi>3$) and 2MASS and WISE cross-match requirements. The high recovery rate within our selection criteria shows strong consistency between the two independent ML classification approaches.

We further compared our results with the carbon catalogue of \citet{He2024}, constructed by applying a Convolutional Neural Network (CNN) to LAMOST DR9 data, which contains 4,383 carbon stars of various subtypes. Like the Abia catalogues, their classification also includes C-H, C-R, and Ba type stars in addition to C-rich AGB stars. We find 98 stars in common, of which 96 are found to be C-N type, and 2 are C-R type. The remaining 4,285 \citet{He2024} stars absent from our sample are predominantly Ba (N=1,971), CH (N=1,481), and C-R (N=468) stars, subtypes explicitly excluded from our training sample as non-AGB contaminants. Of the 119 C-N stars in \citet{He2024} not recovered by our classifier, 108 are absent from our \textit{G2MW} catalogue entirely due to their missing 2MASS and/or WISE magnitudes. The remaining 11 C-N stars present in our \textit{G2MW} catalogue are correctly identified as C-rich but fall below our XGBoost confidence threshold of $p(C)>$0.95.

The comparisons with \citet{Abia2022}, \citet{Sanders2023}, and \citet{He2024} confirm that our ML-classified C-rich sample successfully identifies C-rich AGB candidates within our selection criteria. Stars absent from our sample are predominantly non-AGB carbon sub-types (CH, R-hot, Ba), or are excluded due to our infrared cross-match and parallax quality requirements, rather than misclassifications. 

\subsection{RGB contamination in the colour--magnitude and Wesenheit diagrams}\label{rgb_analysis}

\begin{figure*}
    \centering
    \includegraphics[width=.8\linewidth]{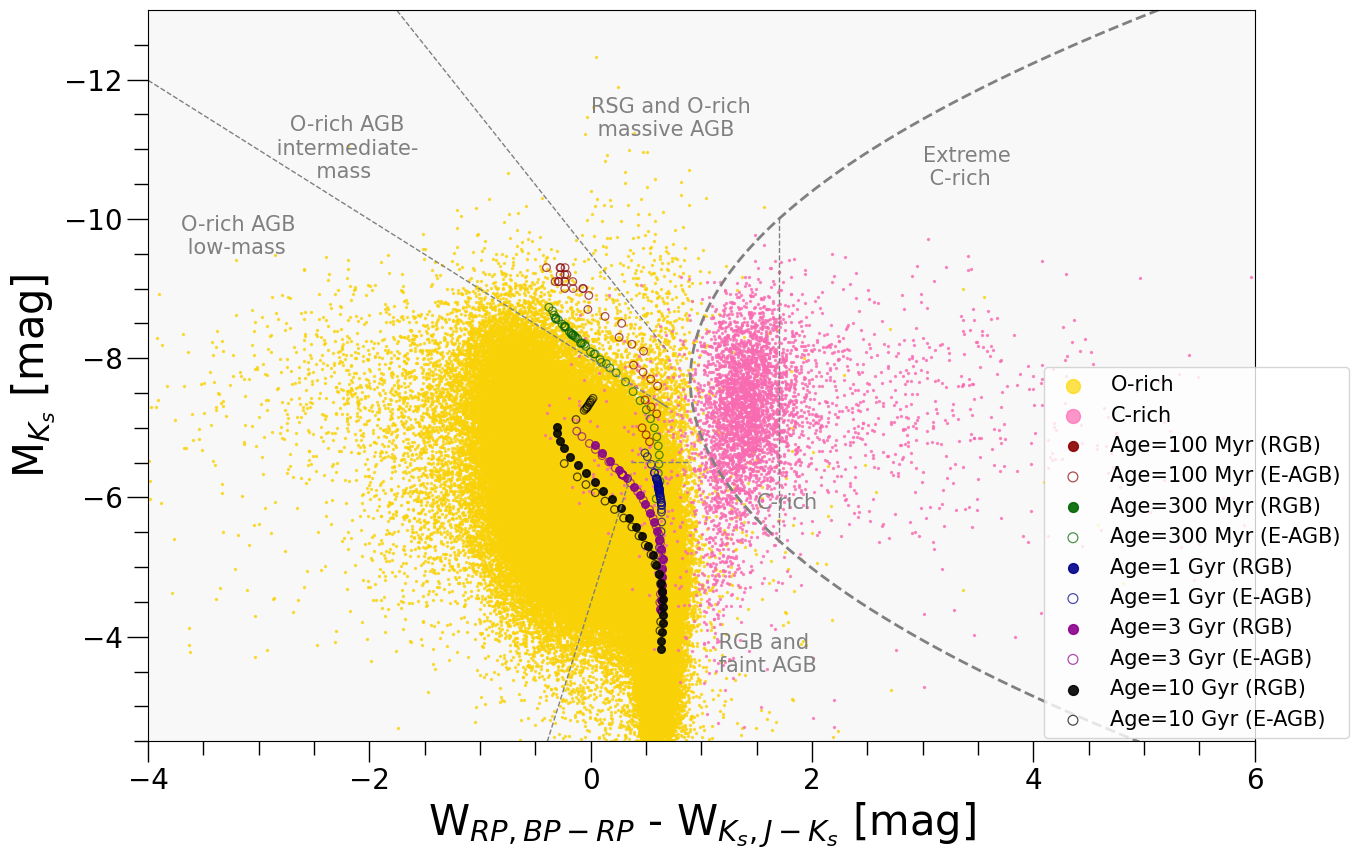}
    \caption{M-type giants ($T_\mathrm{eff}\leq$4000~K; \citealt{Dyck1996}) in the RGB (filled circles) and E-AGB (open circles) evolutionary phases for PARSEC-COLIBRI isochrones of ages 100~Myr, 300~Myr, 1~Gyr, 3~Gyr, and 10~Gyr, overlaid on the Wesenheit diagram of ML-classified O-rich (yellow) and C-rich (pink) AGB candidates to evaluate the RGB contamination in our ML-classified AGB candidates. Older RGB isochrones (3 and 10~Gyr) mainly occupy the ``RGB and faint AGB region'', with some overlap into the ``low-mass O-rich AGB'' region, indicating potential contamination of RGB in low-mass O-rich AGB branch.}
    \label{fig:rgb_analysis}
\end{figure*}

Separating Galactic field AGB and RGB stars with spectral type of M is inherently difficult because they overlap in colour–magnitude space. 
RGB stars can have spectral type M and can also be long-period variables \citep{Tabur2010}. As a result, they can contaminate catalogues intended to select AGB stars.

To evaluate the extent of RGB contamination in our ML-classified sample, we used PARSEC-COLIBRI isochrones in the \textit{Gaia}–2MASS diagram. PARSEC-COLIBRI isochrones provide stellar effective temperatures and evolutionary phases, and we selected RGB and AGB isochrones with effective temperatures below 4000~K, corresponding to M giants \citep{Dyck1996}. Fig.~\ref{fig:rgb_analysis} shows the evolutionary tracks during the RGB and E-AGB phases. RGB stars at older ages (3 and 10~Gyr) predominantly occupy the ``RGB and faint AGB'' region of the \textit{Gaia}–2MASS diagram, with some evolving into the ``low-mass O-rich AGB'' region. The AGB branch runs nearly parallel to the RGB branch but extends to higher luminosities. This indicates that the ``RGB and faint AGB'' region contains a substantial fraction of RGB stars, while the ``low-mass O-rich AGB'' region may also include some RGB stars.

In our ML-classified O-rich sample, a substantial fraction of the sources are located in the ``RGB and faint AGB'' region (160,577 stars, corresponding to 53\% of the classified O-rich sample). This is expected because RGB, as well as E-AGB evolution, is relatively slower than the tip of the AGB, causing stars to spend a larger fraction of their lifetimes in this region. 


\begin{figure*}
\centering
\subfigure[]{\includegraphics[width=0.6\linewidth]{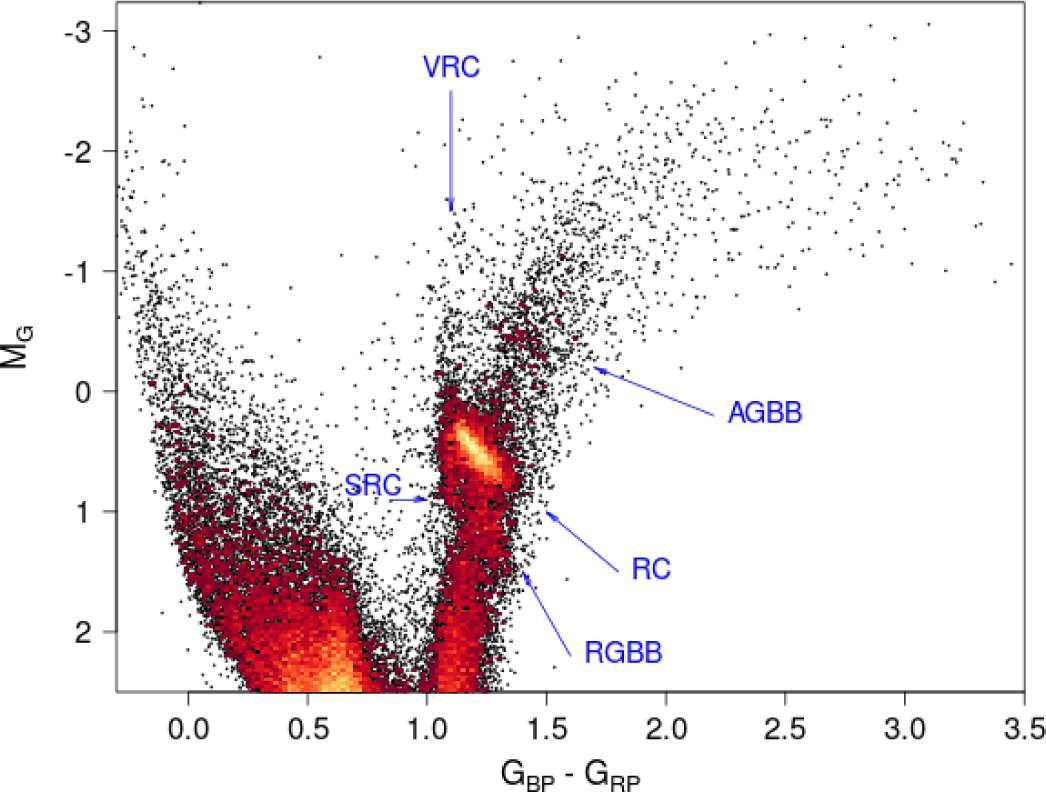}}\\
\subfigure[]{\includegraphics[width=0.49\linewidth]{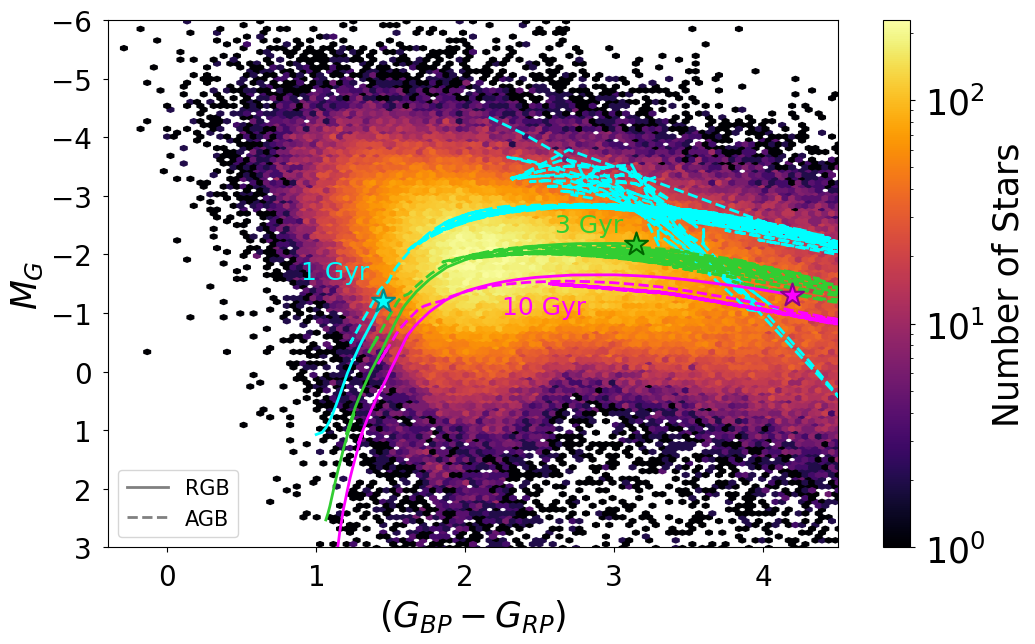}}
\subfigure[]{\includegraphics[width=0.49\linewidth]{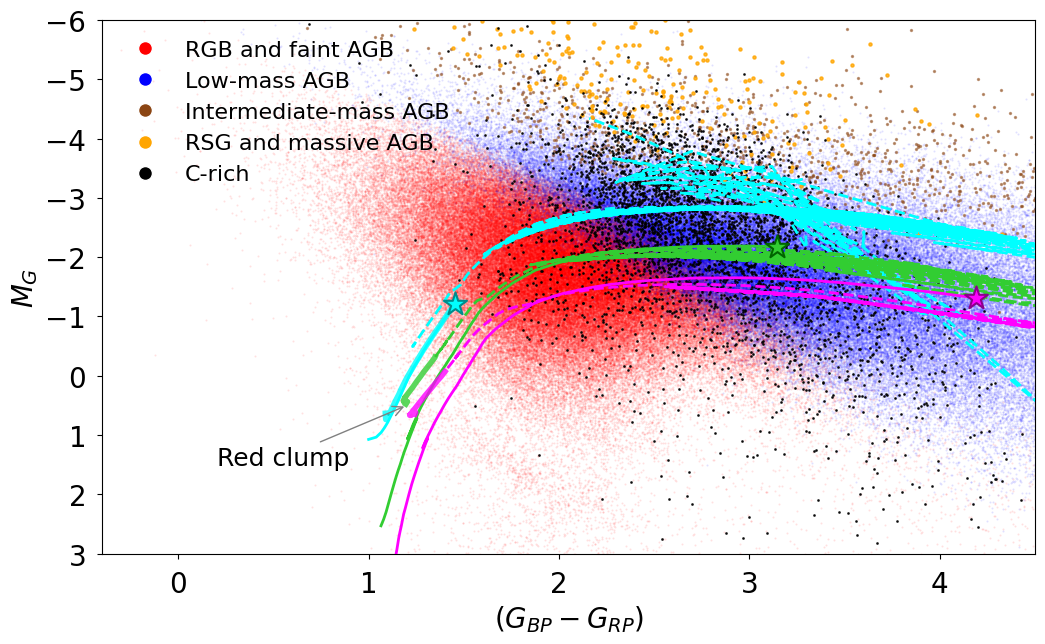}}
\caption{$M_G$ versus $(G_{\rm BP}-G_{\rm RP})$ colour--magnitude diagrams. (a) The \textit{Gaia} HR diagram from \citet{Gaia2018}, with the Red Giant Branch Bump (RGBB), Red Clump (RC), AGB Bump (AGBB), Sub-Red Clump (SRC), and Very Red Clump (VRC) labelled. (b) Number density of the ML-classified O-rich AGB candidates in extinction-corrected absolute $G$ magnitude ($M_G$) and colour $(G_{\rm BP}-G_{\rm RP})$, overlaid with PARSEC--COLIBRI isochrones ($Z=0.014$) of 1~Gyr (cyan), 3~Gyr (green), and 10~Gyr (magenta). Solid lines show the RGB, while dashed lines show the E-AGB and TP-AGB phases. Star symbols mark the most massive RGB star at each age (about to ignite helium), at $M_G \approx -1.2$, $-2.2$, and $-1.3$~mag for the 1, 3, and 10~Gyr isochrones, respectively. The stellar density peaks at approximately $M_G \simeq -1.9$~mag and declines towards higher luminosities, with a particularly steep decline above $M_G \simeq -2.8$~mag. (c) All ML-classified AGB candidates, colour-coded according to the regions they occupy in the \textit{Gaia}-2MASS diagram (Fig.~\ref{wessenheit_c/o}), with the same isochrones and star symbols as in panel~(b). Thick line segments indicate the red clump. Some C-rich stars lie fainter than $M_G=-2$~mag, likely because strong CN and C$_2$ bands suppress the $G$-band magnitudes.}
\label{fig:rgb_gaia_hr}
\end{figure*}

We further compared the extinction-corrected $M_G$ vs. $G_{BP}-G_{RP}$ distribution of our ML-classified AGB candidates with the \textit{Gaia} HR diagram of \citet{Gaia2018} (Fig.~\ref{fig:rgb_gaia_hr}(a)) and with PARSEC-COLIBRI isochrones of 1, 3 and 10~Gyr (Fig.~\ref{fig:rgb_gaia_hr}(b) and (c)). Each isochrone represents stars of different initial masses at a fixed age.
On each isochrone, we mark the most massive RGB star at that age, i.e. the star about to ignite helium in its core, with initial masses of about 2.1, 1.5 and 1.0~M$_\odot$ for the 1, 3 and 10~Gyr isochrones, respectively.

For the 1 and 3~Gyr isochrones, the most massive RGB stars are also the brightest in $M_G$, at $M_G\approx-1.2$ and $-2.2$~mag, respectively. This is not the case for the 10~Gyr isochrone: its brightest RGB stars lie at $M_G\approx-1.65$~mag, whereas the most massive RGB stars at this age are fainter, at $M_G\approx-1.3$~mag. These stars have almost identical initial masses ($\approx$1.03~M$_\odot$), but the most massive RGB star is cooler ($T_\mathrm{eff}\approx3100$ compared with 3400~K) and more luminous ($\log L/L_\odot\approx3.4$ compared with 3.1). Its lower $G$-band flux is likely due to two effects. At lower effective temperatures, the peak of the spectral energy distribution shifts to longer wavelengths, so that a smaller fraction of the flux falls within the $G$ band. In addition, TiO absorption becomes the dominant factor affecting the optical flux below $T_\mathrm{eff}\approx3700$~K \citep{Kucinskas2005}. As both stars have $T_\mathrm{eff}$ below this limit, the stronger TiO absorption in the cooler star is probably the main cause of its lower $G$-band flux.

For the ML-classified AGB candidates, the stellar density peaks at $M_G\approx-1.9$~mag (Fig.~\ref{fig:rgb_gaia_hr}(b) and (c)). The density then decreases towards higher luminosities, with a particularly steep decline above $M_G\approx-2.8$~mag. The density peak may reflect the relatively long time stars spend at these luminosities on the RGB, while the decline above $M_G\approx-2.8$~mag is likely related to the rapid evolution of stars close to the tip of the RGB \citep{Valenti2004}. Uncertainties in distance and extinction corrections also scatter stars' brightness.

The steepest decline occurs at a brighter magnitude than the most massive RGB stars in the individual isochrones. This difference may reflect the range of ages and metallicities of the RGB stars in our sample, as well as observational uncertainties. The decline around $M_G\approx-2.8$~mag therefore provides an approximate indication of the bright end of the RGB population, rather than a precise measurement of the RGB tip.

Fig.~\ref{fig:rgb_gaia_hr}(c) shows the ML-classified AGB candidates colour-coded by the region they occupy in the \textit{Gaia}-2MASS diagram (Fig.~\ref{wessenheit_c/o}). 
At magnitudes fainter than the approximate bright end of the RGB population ($M_G\approx-2.8$~mag), the isochrones show that RGB and AGB stars cannot be distinguished. Given the longer evolutionary timescale of the RGB \citep{Valenti2004}, the ``RGB and faint AGB'' region of the \textit{Gaia}-2MASS diagram (red in Fig.~\ref{fig:rgb_gaia_hr}(c)) is likely to contain more RGB stars than AGB stars.

The isochrones also show that the RGB of populations aged 3--10~Gyr extends into the colours and magnitudes of the candidates in the low-mass O-rich AGB region of the \textit{Gaia}-2MASS diagram (blue in Fig.~\ref{fig:rgb_gaia_hr}(c); see also Fig.~\ref{fig:rgb_analysis}), suggesting that this region also contains some RGB stars. 
As RGB and AGB stars overlap in both regions, the RGB contamination is difficult to quantify with the current data; we therefore refer to the full ML-classified sample as AGB candidates. Furthermore, the Galactic structure analysis presented in Sect.~\ref{galactic_distribution} is limited to the AGB regions of the \textit{Gaia}-2MASS diagram, excluding the ``RGB and faint AGB'' region.

\subsection{Galactic distributions of C-rich and O-rich AGB stars}\label{galactic_distribution}

\begin{figure*}
\centering     
\includegraphics[width=.95\linewidth]{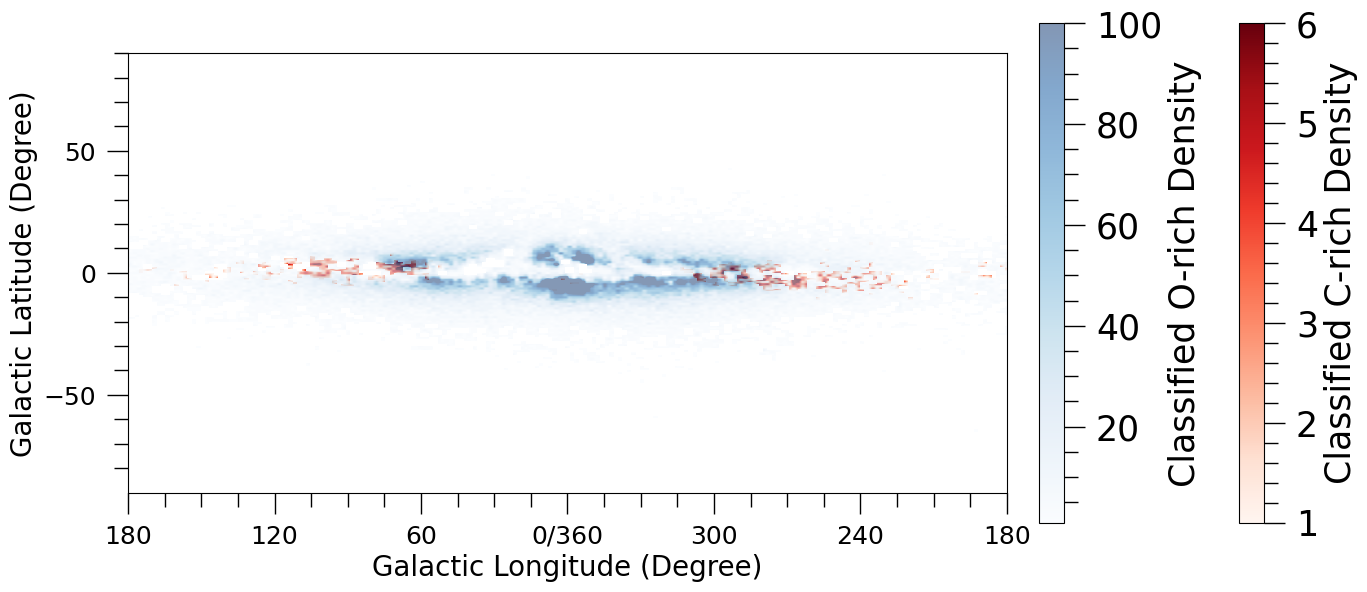}
\caption{Overlaid number-density maps of ML-classified C-rich (red) and O-rich AGB stars (blue) showing their Galactic distribution. Only stars classified within the AGB branches of the Wesenheit diagram (low-mass, intermediate-mass, and massive O-rich, C-rich, and extreme C-rich) are included; the RGB and faint AGB region is excluded. The maps are displayed on a 200$\times$200 grid with a pixel size of 0.9$^\circ$ in latitude and 1.8$^\circ$ in longitude, and the colour bars show the raw counts per pixel. O-rich AGB stars are strongly concentrated toward the inner Galaxy, while C-rich AGB stars are distributed more uniformly around the Galactic mid-plane and extend into the outer Galaxy.}
\label{c_o_densityplot}
\end{figure*}
 
C-rich and O-rich AGB stars, as intermediate stellar populations, serve as tracers of Galactic structure and evolution.
Their distinct spatial distributions reflect differences in age, metallicity, and stellar evolution processes. In the following analysis, we use only stars classified within the AGB regions in the Wesenheit diagram (low-mass, intermediate-mass, massive O-rich, C-rich and extreme C-rich), excluding the RGB and faint AGB region as discussed in Sect.~\ref{rgb_analysis}.

We present the number density distribution of C-rich (red) and O-rich (blue) AGB stars in Galactic coordinates, Galactic latitude ($b$) and Galactic longitude ($l$) in Fig.~\ref{c_o_densityplot}. Although noted above, we emphasise that `O-rich AGB stars' here refers to both low- to intermediate-mass and massive O-rich AGB stars. They show a strong concentration toward the Galactic plane and inner centre, with over 80\% of sources located within $|b|<$15$\degree$ and predominantly spanning $l\sim$0–90$\degree$ and 270–360$\degree$. The highest densities appear within ($|l|\sim$15$\degree$) from the Galactic centre, where the metal-rich environment favours oxygen-rich stellar evolution \citep[e.g.][]{Habing2003}.

In contrast, C-rich AGB stars exhibit a broader distribution along the Galactic plane, with over 90\% of the stars confined within $|b|\sim$10$\degree$. Their distribution is more uniform, spanning nearly the full longitude range, with notable concentrations in anti-Galactic centre directions. This extended distribution reflects the metallicity-dependent nature of carbon star formation. At lower metallicity, fewer oxygen atoms are present, meaning only a small amount of carbon needs to be synthesised for a star to transition from O-rich to C-rich \citep{Marigo1999}. Additionally, lower metallicity outer regions facilitate more efficient carbon enrichment through third dredge-up processes \citep{Iben1983}. Both effects likely contribute to the observed spatial separation between C-rich and O-rich AGB stars, driven by the underlying metallicity gradient of the Milky Way.
Studies using red giants \citep{Hayden2014} and open clusters \citep{Magrini2023} reported metallicity gradients of approximately $-$0.05 to $-$0.09~dex~kpc$^{-1}$ between 5 and 15~kpc, with higher metallicities toward the inner Galaxy.
Hence, this metallicity gradient produces a dominance of O-rich stars in the inner Galaxy and C-rich stars in the outer Galaxy.

The effects of the interstellar dust extinction are evident close to the Galactic plane $|b|\sim$3$\degree$ near the Galactic centre, where dust obscuration in optical and NIR wavelengths reduces the observed stellar density.

Although the number of C-rich stars decreases sharply toward the Galactic centre ($b=$0$\degree$), we identify 347 C-rich sources within $|l|<$30$\degree$ and $|b|<$10$\degree$. This longitude range is commonly used to represent the bulge direction in previous studies of C-rich AGB stars \citep[e.g.][]{Lewis2020}. Our ML-classified C-rich AGB stars therefore contain roughly twice as many C-rich stars in this region as reported in earlier work \citep[e.g.][]{Matsunaga2017, Lewis2020}. 
Of the 347 inner Galaxy C-rich stars, 225 (65\%) are in common with the 1,616 C-rich LPVs identified by \citet{Sanders2023} in the same region, with the remaining 1,391 Sanders stars absent from our sample due to differences in input catalogue and classification method, as discussed in Sect.~\ref{comparison_lit}. The MW's bar-bulge, due to its old metal-rich stellar population \citep{Bensby2013, Bovy2019, Hasselquist2020}, is expected to host relatively few C-rich LPV stars \citep{Battinelli2013, Sanders2023}, consistent with the small number we identify in this region. Those identified may be products of binary mass transfer rather than single-star third dredge-up, as suggested by \citet{Matsunaga2017} and \citet{Sanders2023}.
We note that some C-rich AGB stars toward the inner Galaxy may belong to the Sagittarius (Sgr) dwarf galaxy, which is known to contain $\sim$100 C-rich AGB stars \citep{Battinelli2013_sgr}. To distinguish Sgr dwarf members from Galactic bulge stars requires kinematic analysis, which is beyond the scope of the current analysis but will be explored in future work using \textit{Gaia} proper motions and radial velocities.

The overdensity of C-rich stars at $l=$280$\degree$ ($\pm$3$\degree$) and $b=$-32 $\degree$ ($\pm$2.5$\degree$) corresponds to the LMC, where the lower metallicity environment (Z $\approx$ 0.008) strongly favours carbon star formation, consistent with theoretical predictions and previous observational studies \citep[e.g.][]{Blanco1983, Cioni2000, Groenewegen2002, Matsuura2009}.

\subsubsection{3D Spatial Distribution of C-rich and O-rich AGB Stars}

\begin{figure*}
\centering
\subfigure{%
    \includegraphics[width=.4\linewidth]{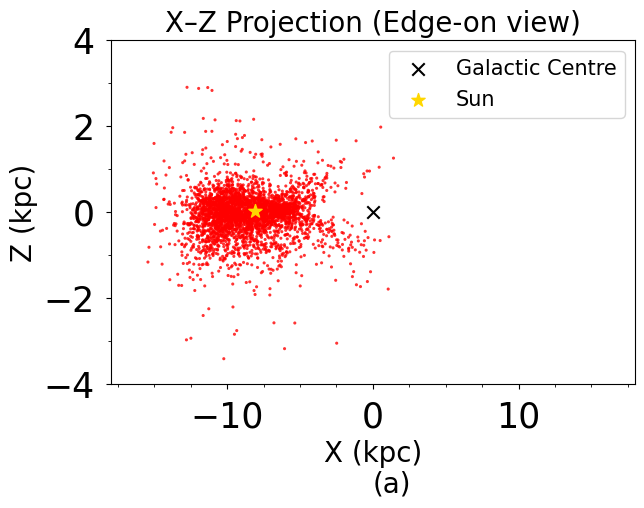}%
}
\hspace{0.01\linewidth}
\subfigure{%
    \includegraphics[width=.4\linewidth]{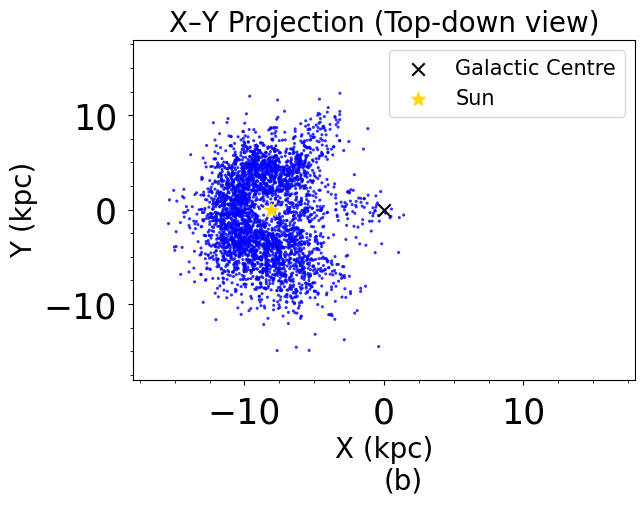}%
}
\subfigure{%
    \includegraphics[width=.4\linewidth]{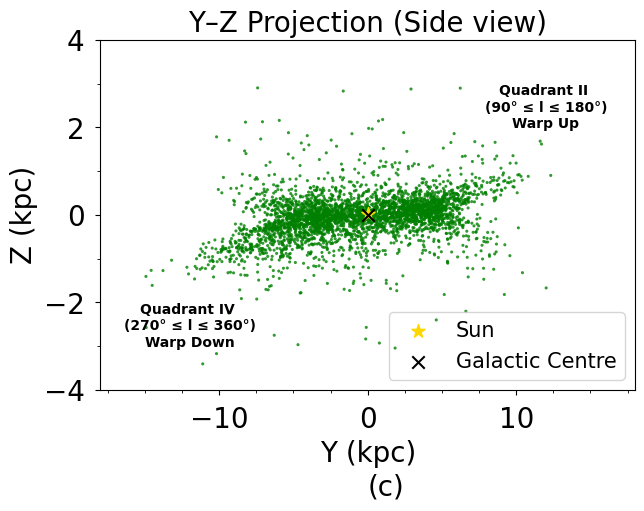}%
}
\caption{%
    Galactocentric 3D projections of ML-classified C-rich AGB stars, showing vertical asymmetries and radial extent of the stars. The Galactic centre (black cross) and the Sun (yellow star) are marked in all panels. Panel (a) Edge-on (X–Z) view showing the vertical structure of the Galactic disc. Panel (b) Top-down (X–Y) view illustrating the Galactocentric radial distribution across the Galactic disc. Panel (c) Side (Y–Z) view showing the distribution with noticeable asymmetry between Galactic quadrants, with upward bending in Quadrant~II and downward bending in~IV. 
}
\label{crich_agb_3d}
\end{figure*}

\begin{figure*}
\centering
\hspace{0.01\linewidth}
\subfigure{%
    \includegraphics[width=.45\linewidth]{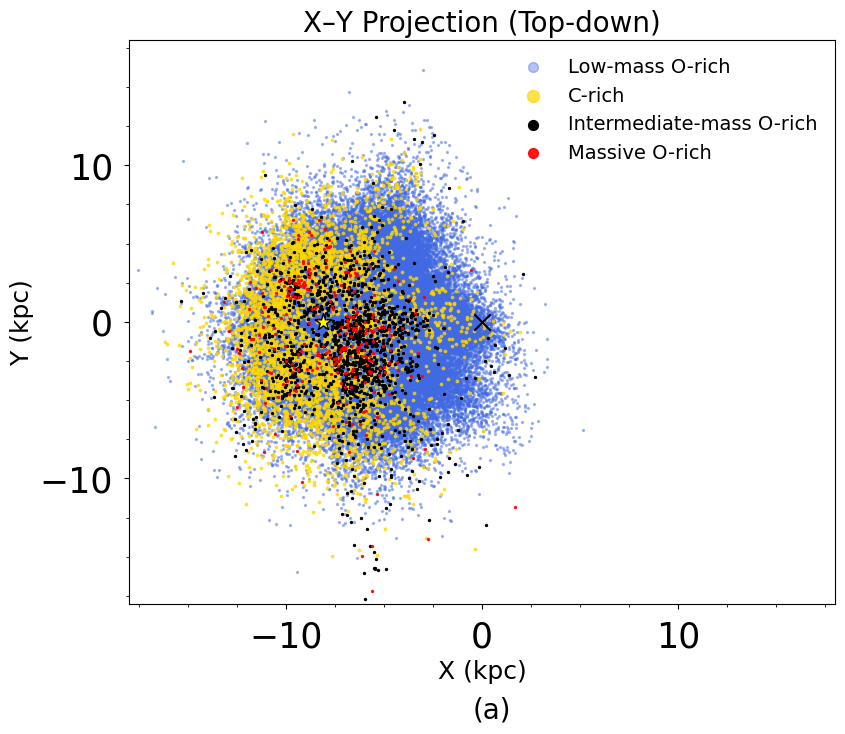}%
}
\hspace{0.2cm}
\subfigure{%
    \includegraphics[width=.45\linewidth]{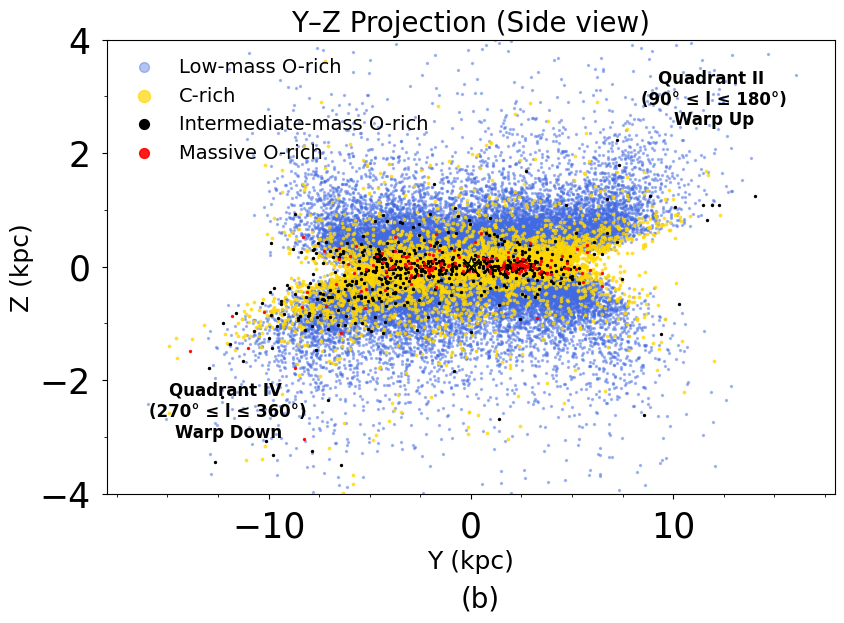}%
}
\caption{%
    Galactocentric X-Y (top-down) and Y-Z (side view) projections of ML-classified C-rich (yellow), low-mass O-rich (blue), intermediate-mass O-rich (black), and massive O-rich (red) AGB stars. These projections show the distinct spatial distributions of C-rich and O-rich AGB stars in the Galactic disc, highlighting their vertical and radial distributions. The Galactic centre (black cross) and the Sun (yellow star) are marked in all panels. Axes span $\pm$18~kpc in X/Y and $\pm$4~kpc in Z. Panel (a) Top-down (X–Y) view illustrating the radial distribution across the Galactic disc, with C-rich AGB stars occupying the outer Galactic regions and O-rich AGB stars occupying the inner Galactic regions. Panel (b) Side (Y–Z) view showing vertical asymmetries of different AGB subgroups, with divergence from the Galactic mid-plane. 
}
\label{all_xy_yz}
\end{figure*}

With geometric distances from \citet{Bailer-Jones2021}, we can build a 3D picture of AGB distribution within the Milky Way. We adopt a Galactocentric coordinate system with the Sun positioned at (X, Y, Z) = (-8.122, 0, 0.02)~kpc \citep{Gravity2018}. Fig.~\ref{crich_agb_3d} presents an example of a 3D projection, displaying C-rich AGB stars. The $X-Y$ (top-down) view shows the Galaxy from above, the $Y-Z$ (side) view shows the Galactic edge-on, and the $X-Z$ (face-on) view shows the Galaxy along a perpendicular plane.
The $X-Y$ plane shows that the C-rich AGB stars are predominantly distributed towards the outer Galaxy, whereas the $X-Z$ projection reveals the effect of interstellar extinction, creating a void toward the Galactic centre. The $Y-Z$ distribution shows a deviation from the Galactic mid-plane for larger $|Y|$, indicative of the Galactic warp discussed in Sect.~\ref{galactic_warp} \citep{2026arXiv260610235K}.

To compare the spatial distribution of different AGB subgroups, we plot the intermediate-mass, low-mass, and massive O-rich and C-rich AGB stars on $X-Y$ and $Y-Z$ projections (Fig.~\ref{all_xy_yz}). The $X-Y$ projection (Fig.~\ref{all_xy_yz}(a)) shows that C-rich stars are located away from the Galactic centre and are mainly concentrated in the outer Galaxy, as found in the 2D distribution (Fig.~\ref{c_o_densityplot}). In contrast, O-rich AGB stars are concentrated toward the inner Galaxy (Fig.~\ref{all_xy_yz}). Low-mass AGB stars show a clear distribution toward the centre and extend outward to $X\sim5$~kpc (Fig.~\ref{all_xy_yz} (a)), whereas massive O-rich AGB stars are found mainly within 5~kpc of the Sun in the $X-Y$ projection. Intermediate-mass O-rich AGB stars are mainly concentrated within $-10<Y<10$~kpc, with some sources extending to $|Y|\sim15$~kpc. 

The $Y-Z$ projection (Fig.~\ref{all_xy_yz}(b)) shows the vertical distribution of the subgroups relative to the Galactic plane. Since O-rich AGB stars are distributed closer to the Galactic plane, they are more strongly affected by extinction. Although extinction corrections have already been applied (Appendix~\ref{ism_extinction_correction}), these corrections are likely underestimated near the plane. Therefore, for the analysis of the $Y-Z$ projection, we removed stars within a radius of 7~kpc from the Galactic centre.
Again, we see a clear distinction between C-rich and O-rich AGB stars. C-rich AGB stars within $|Y|<8$~kpc lie within $|Z|<0.6$~kpc. For $|Y|>8$~kpc, they gradually diverge from the inner Galactic plane, reaching $|Z|\sim2$~kpc at $|Y|\sim14$~kpc. The vertical divergence at larger radii is discussed in more detail in Sect.~\ref{galactic_warp}. 
Among O-rich AGB subgroups, massive O-rich stars are mostly confined within $|Y|\sim8$~kpc and $|Z|\sim0.2$~kpc in the Galactocentric $Y-Z$ frame, indicating that they populate the Galactic mid-plane. Intermediate-mass and low-mass O-rich stars extend to larger vertical heights, up to $|Z|\sim1$~kpc and $|Z|\sim3$~kpc, respectively. Both subgroups also show a gradual divergence from the mid-plane for $|Y|>7$~kpc.

\subsubsection{Normalised vertical profiles of AGB Stars}

\begin{figure}
\centering     
\includegraphics[trim={.3cm, .3cm, .1cm, .1cm}, clip, width=.95\linewidth]{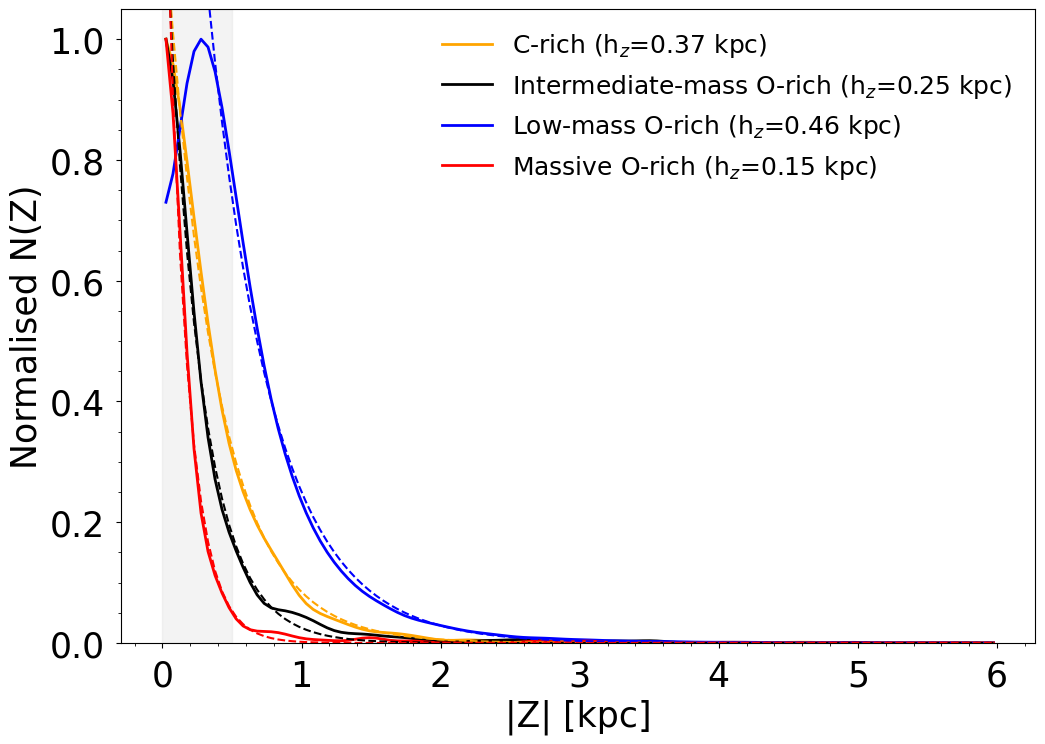}
\caption{Normalised number of stars vs. $|Z|$ for C-rich (yellow), low-mass O-rich (blue), intermediate-mass O-rich (black), and massive O-rich (red) AGB stars. The curves show Gaussian-smoothed distributions normalised to their maxima, with exponential fits (dashed lines) applied to the high-$|Z|$ tails to quantify the vertical scale heights ($h_Z$) of each subgroup.}
\label{normalisedNvsZ}
\end{figure}

To compare the vertical distribution of different subgroups of classified AGB stars, we examine the normalised number of stars as a function of absolute vertical height above the Galactic plane ($|Z|$) in Fig.~\ref{normalisedNvsZ}. The histograms were smoothed with a Gaussian filter and normalised to their maximum to reduce noise and allow a direct comparison between populations. We quantify this vertical distributions by fitting the high-$|Z|$ tail of each population with an exponential function \citep{Claussen1987}:

\begin{equation}
    N(Z) = N_0\ exp\left[-\frac{|Z-Z_0|}{h_Z}\right] 
\end{equation}

where $N_0$ is a normalisation constant, $Z_0$ is the peak of the distribution, and $h_Z$ is the vertical scale height, defined as the distance over which the population declines by a factor of $e$. To reduce biases associated with extinction and crowding near the Galactic plane, we fitted the profile above $Z_{min}=0.27$~kpc only for low-mass O-rich AGB stars (corresponding to their peak) and $Z_{min}=0.05$~kpc for all other subgroups. 

The resulting scale heights reveal clear differences between the AGB sub-populations. Massive O-rich AGB stars are tightly confined to the mid-plane, with a scale height of $h_Z=150$~pc. This is consistent with their young ages and short evolutionary timescales, which limit the extent to which they migrate from their birth molecular clouds. Their distribution decreases sharply, reaching a plateau near $|Z|\sim 0.8$~kpc. Low-mass O-rich AGB stars are more vertically extended ($h_Z=460$~pc), while the C-rich AGB stars have an intermediate scale height of $h_Z=370$~pc, and decline to nearly zero by $|Z|\sim2$~kpc. 
Intermediate-mass O-rich AGB stars have a scale height of  $h_Z=250$~pc, falling between the massive O-rich and C-rich AGB stars, and declining to nearly zero by $|Z|\sim1.5$~kpc.  We note that the low-mass AGB stars show a dip at $|Z|\sim0$ before reaching the peak at $|Z|\sim0.27$~kpc, likely due to interstellar dust extinction toward the inner Galaxy. In addition, the low-mass O-rich AGB subgroup may include some RGB stars near the tip of the RGB (Sect.~\ref{rgb_analysis}). As older RGB stars are expected to be more vertically extended, this contamination could contribute to its larger scale height.

These values are broadly consistent with previous determinations for AGB populations. IRAS-selected AGB candidates have typical scale heights of $\sim$0.3~kpc \citep{Jackson2002}, comparable to our measurement for C-rich stars. Recent {\it Gaia} DR3-based work by \citet{Uttenthaler2025} reported scale heights of $h_Z\approx0.22$~kpc for C-rich Miras and $h_Z\approx0.5$~kpc for M-type Miras, both very close to the values derived here ($0.37$~kpc and $0.46$~kpc, respectively). In contrast, the confinement of massive O-rich AGB stars to $\sim$100--200~pc is consistent with expectations for young and massive AGB stars that remain close to the Galactic mid-plane.

Overall, despite differences in sample selection, classification method, and spatial coverage, our scale height measurements are based on a large sample of AGB stars spanning a wide range of Galactocentric distances and are consistent with previously reported scale heights for Galactic AGB stars.

\subsubsection{AGB stars as tracers of the Galactic warp}\label{galactic_warp}

\begin{figure}
\centering
\hspace{0.01\linewidth}
\subfigure{%
    \includegraphics[trim={.3cm, .3cm, .1cm, .1cm}, clip, width=.9\linewidth]{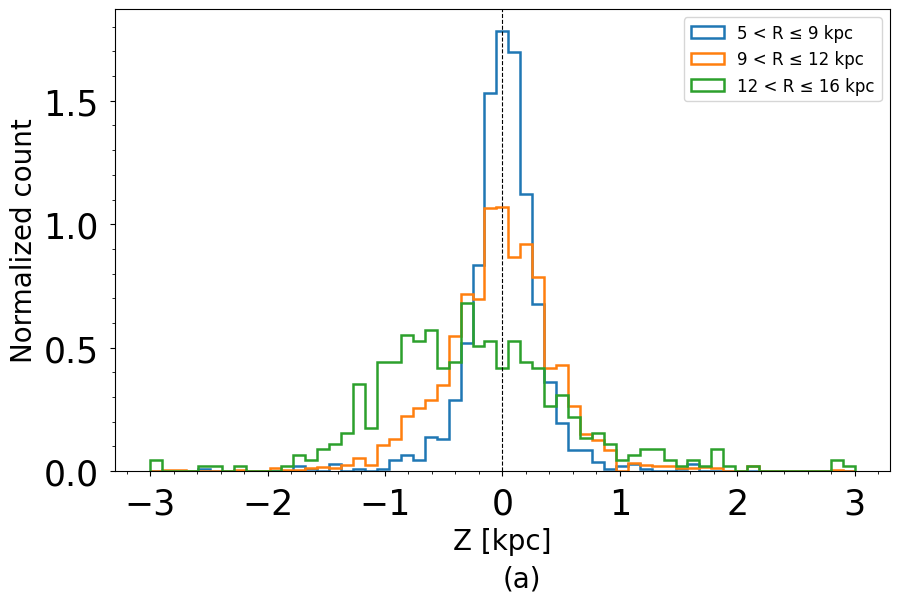}%
}
\subfigure{%
    \includegraphics[trim={.3cm, .3cm, .1cm, .1cm}, clip, width=.9\linewidth]{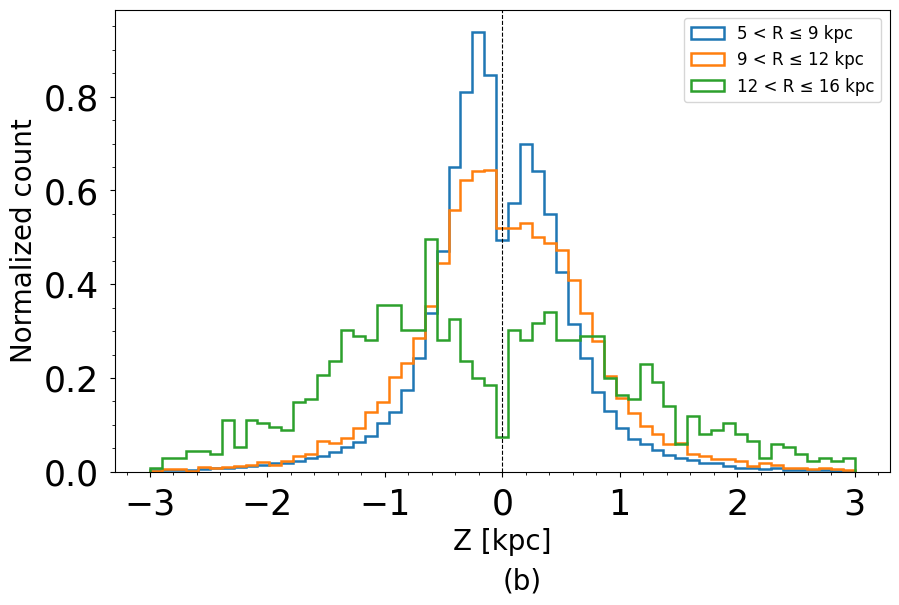}%
}
\caption{Histograms of the normalised vertical distribution ($Z$) of C-rich (a) and intermediate-mass O-rich (b) AGB stars in three Galactocentric radial slices. These distributions show differences in the vertical structure with radius, including deviations from the Galactic plane and the effects of interstellar extinction, which cause a dip in the count near $Z=0$.}
\label{NcountvsZ}
\end{figure}

A warp is an undulation of the Galactic disc in which the mid-plane deviates from $Z=0$ at larger radii due to gravitational torque or recent perturbation in the Galaxy. Warp structure has been observed in many external galaxies \citep{Reshetnikov2025} and is present in more than half of all spiral galaxies \citep{Sanchez1990}, including the Milky Way. The Galactic warp has been traced using \HI~gas \citep{Levine2006, Kalberla2007}, multiple populations, such as old RR Lyrae stars ($>10$~Gyr; \citealt{Cabrera2025}), young Cepheids ($\sim$300~Myr; \citealt{ChenX2019, Skowron2019, Poggio2020, Cabrera2024, Poggio2025, Zhou2025}), red clump (RC) stars \citep{Momany2006, Wang2020}, and dust \citep{Marshall2006}. Earlier kinematic studies of C-rich AGB stars \citep{Battinelli2013} provided tentative evidence for warp-like signatures in their rotational velocity distribution, based on a small sample of $\sim$100 stars.

We examine the vertical distribution of C-rich and intermediate-mass O-rich AGB stars in different Galactocentric radial slices to investigate deviations from the Galactic mid-plane in the outer disc. Fig.~\ref{NcountvsZ} shows the histogram of $Z$ for C-rich and O-rich in three radial slices of $R$: $5<R<9$~kpc, $9<R<12$~kpc, and $12<R<16$~kpc. The slices reveal distinct features for the two populations. In the outermost radial slices, the observed increase in vertical extent could reflect disc flaring (thickening of the Galactic disc) or the Galactic warp. 

For C-rich (Fig.~\ref{NcountvsZ}(a)), the  $5<R<9$~kpc slice shows the highest count near the Galactic plane, with the number of stars decreasing toward larger $|Z|$, while the $9<R<12$~kpc slice also peaks near the Galactic plane but shows a broader distribution toward larger Z. The $12<R<16$~kpc slice shows the distribution shifting toward negative $Z$, with an asymmetric broadening reaching $Z\sim-$2~kpc and $Z\sim$1~kpc, indicative of a possible warp or disc flaring effect.

The vertical distribution of intermediate-mass O-rich AGB stars (Fig.~\ref{NcountvsZ}(b)) shows a different behaviour. For the $5<R<9$~kpc slice, there is a dip likely caused by higher ISM extinction toward the Galactic centre compared to the outer Galaxy. In the $9<R<12$~kpc slice, the dip is still present, possibly because extinction varies differently along different lines of sight at these large distances. The sharp increase in the number of stars at larger vertical extent ($|Z|>\sim0.5$~kpc) for $12<R<16$~kpc could be due to extinction at ``the Split'', a dust lane in the Galactic plane between $l$=30--60$^\circ$ \citep{Zhang2025}, although the presence of the Galactic warp or flaring may also contribute, as seen more clearly for C-rich stars. The outer slice distribution is limited by the available sample size at larger Galactocentric radii.

\begin{figure}
\centering     
\includegraphics[trim={12.8cm 5.3cm 12.8cm 2.cm}, clip, width=1.\linewidth]{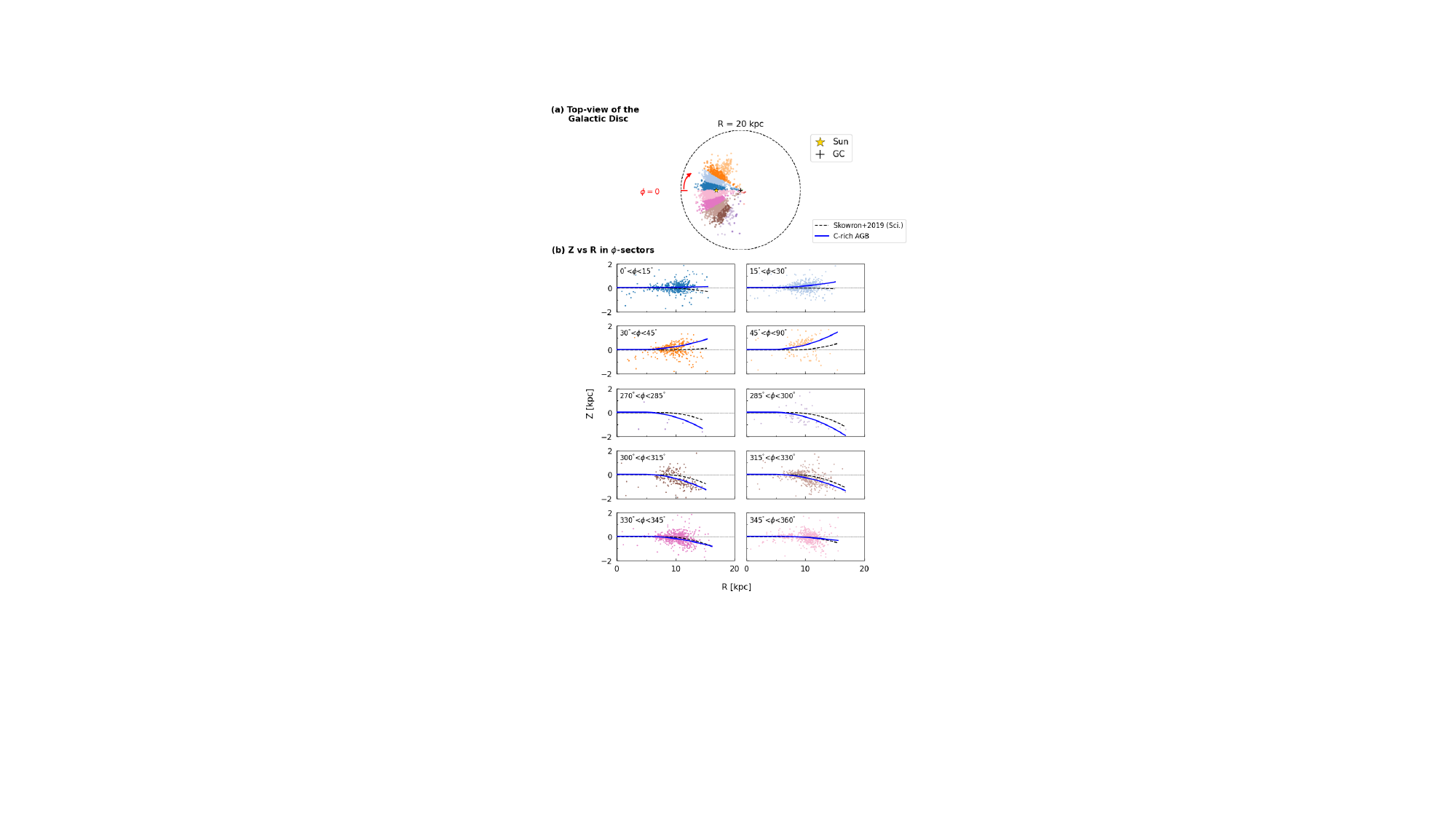}
\caption{Galactic warp analysis using C-rich AGB stars. Panel (a) shows a top-down view of C-rich AGB stars, coloured by azimuthal angle $\phi$ (measured in the direction of Galactic rotation). The Sun (yellow star) and the Galactic centre (black cross) are indicated. Panel (b) shows the vertical distribution $Z$ vs. Galactocentric radius $R$ for the same azimuthal sectors. Colours in panel (b) correspond to the sectors in panel (a), and the $\phi$ sectors are labelled in each subplot. 
The blue solid line shows the best-fitting warp model to the C-rich AGB stars, and the black dashed line shows the Cepheid warp model from \citet{Skowron2019}. The C-rich AGB fit appears to trace larger warp amplitudes than the Cepheid warp model across the key azimuthal bins \citep{2026arXiv260610235K}.}
\label{crich_warp}
\end{figure}

Following the analysis of the Galactic warp by Cepheids presented (Fig.~2 of \citet{Skowron2019}), we constructed an analogous diagram using our AGB sample to investigate whether intermediate-age stellar populations trace a similar warp signature. Initial results for C-rich AGB stars are presented in \citet{2026arXiv260610235K}; here, we briefly summarise the key results for both C-rich and intermediate-mass O-rich AGB stars. We divided the Galaxy into 10 sectors of unequal azimuthal width in Galactocentric polar coordinates, with the azimuth $\phi$ increasing in the direction of the Galactic rotation. We adopt a clockwise azimuthal convention, which is opposite to the anti-clockwise convention used by \citet{Skowron2019}.

\begin{figure}
\centering     
\includegraphics[trim={13.cm 5.7cm 12.8cm 2.cm}, clip, width=1.\linewidth]{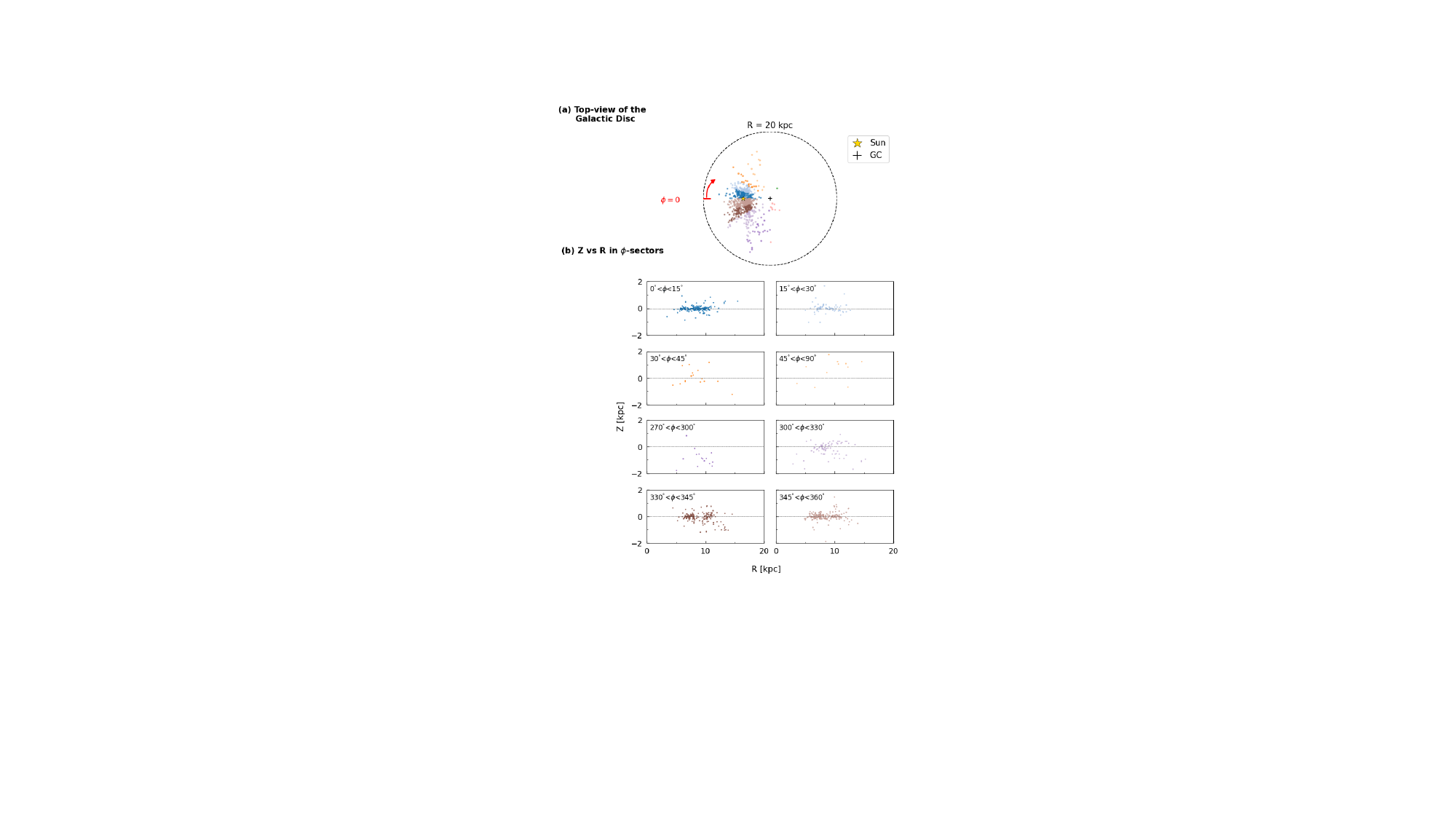}
\caption{Panel (a) shows top-down and panel (b) shows vertical distribution ($Z$ vs. $R$) of intermediate-mass O-rich AGB stars in azimuthal bins, similar to Fig.~\ref{crich_warp} for C-rich stars. Unlike C-rich AGB stars, the sample size of intermediate-mass O-rich AGB stars is insufficient to derive a robust warp fit. Nevertheless, the plots reveal tentative vertical deviations from the Galactic plane, suggesting a possible warp signature in the intermediate-mass O-rich AGB stars (ages of $\sim$100--300~Myr).}
\label{into_warp}
\end{figure}

Panel (a) of Figs.~\ref{crich_warp} and \ref{into_warp} shows the top-down distribution of ML-classified C-rich and intermediate-mass O-rich AGB stars, respectively, colour coded by azimuthal bin. The vertical deviations, which start at outer Galaxy $R\sim$8--10~kpc, are clearly visible. In panel (b) of Fig.~\ref{crich_warp}, the sectors $0^{\circ}<\phi<15^{\circ}$ and $15^{\circ}<\phi<30^{\circ}$ (corresponding to $330^{\circ}<\phi<345^{\circ}$ panel and $345^{\circ}<\phi<360^{\circ}$ in \citealt{Skowron2019}), do not show a clear warp deviation with stars remaining close to the Galactic mid-plane. The sector $30^{\circ}<\phi<45^{\circ}$ (equivalent to $315^{\circ}<\phi<330^{\circ}$ in \citealt{Skowron2019}) shows an upward deviation, with C-rich stars reaching $Z\sim0.7$~kpc at $R\sim$14~kpc, larger than those traced by Cepheids ($Z\sim0.5$~kpc at $R\sim$14~kpc; \citealt{Skowron2019}). A similar upward deviation is seen in the $45^{\circ}<\phi<90^{\circ}$ sector. A downward deviation is apparent in the $270^{\circ}<\phi<360^{\circ}$ sectors, where the deviation from the mid-plane begins around $R\sim$8--10~kpc and reaches $Z\sim-$1~kpc at $R\sim$14~kpc. The C-rich AGB stars appear to trace larger warp amplitudes than Cepheids. We fitted the simplified warp model of \citet{Skowron2019} to the C-rich AGB stars, and the best-fitting model confirms this trend, lying above the Cepheid-based models in the upward bins and below in the downward bins. Full details of the model fitting are provided in \citet{2026arXiv260610235K}.

For intermediate-mass O-rich AGB stars (Fig.~\ref{into_warp}), the sample size is insufficient to draw firm conclusions in most azimuthal bins. A tentative downward deviation is detected in the $270^{\circ}<\phi<345^{\circ}$, consistent with the downward warp seen in C-rich AGB stars and Cepheids. No clear upward warp signal is detected, likely due to the limited number of stars in the upward warp bins.

Overall, the $Z$ vs. $R$ trends demonstrate that both the onset radius and amplitude of the warp vary across azimuth. C-rich AGB stars appear to trace larger warp amplitudes than Cepheids, suggesting a possible dependence of warp amplitude on stellar age and metallicity \citep{2026arXiv260610235K}. These differences may also reflect the distinct spatial distributions of C-rich and intermediate-mass O-rich AGB stars, as well as variations in interstellar extinction associated with their different Galactic locations. Together, these results demonstrate that AGB stars can serve as effective tracers of the Galactic warp at intermediate stellar ages, providing new constraints on the warp amplitude across different stellar populations.

\subsubsection{Comparison with the Galactic spiral arm models}

\begin{figure*}
    \centering
    \includegraphics[width=1.\linewidth]{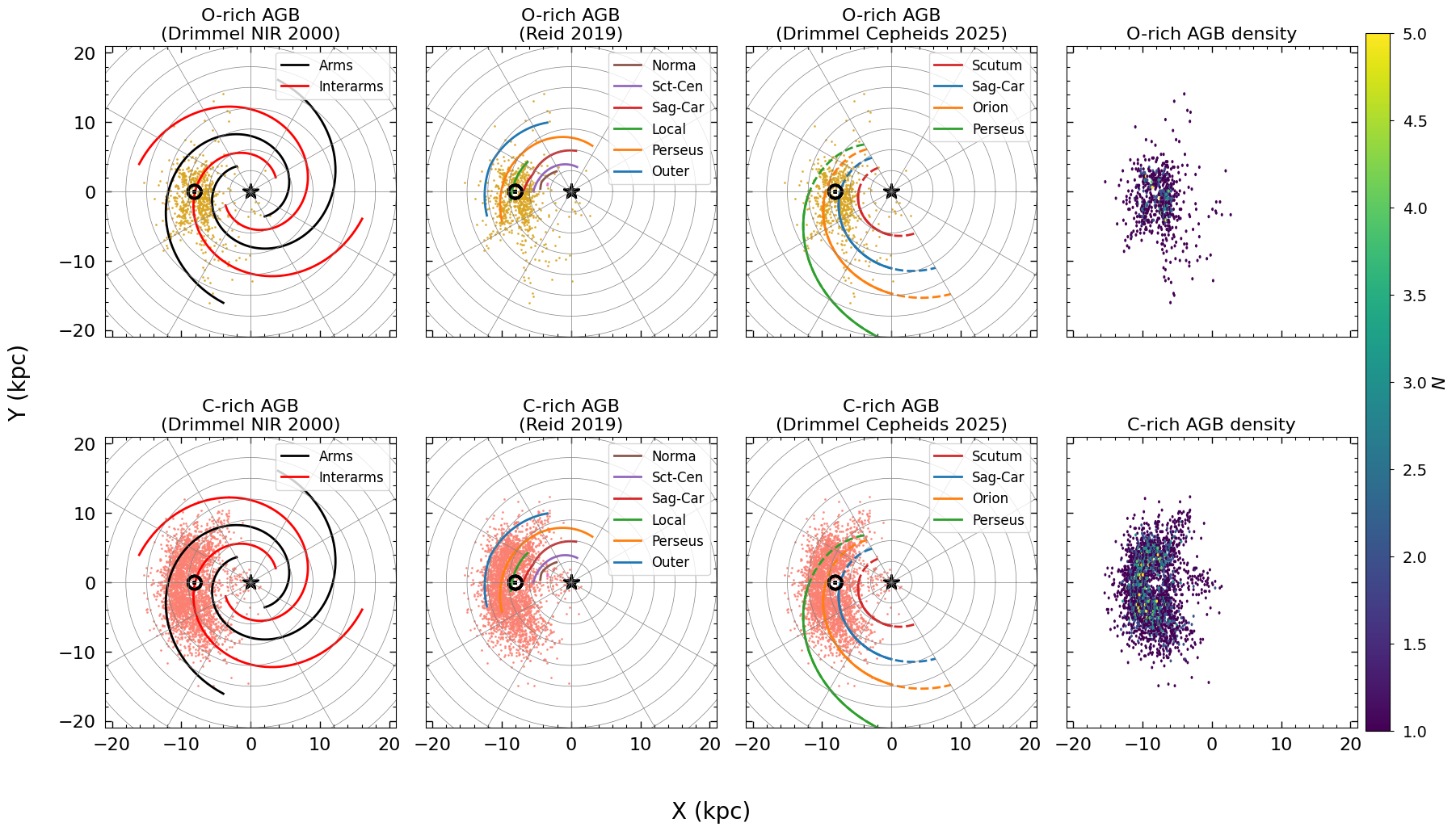}
    \caption{Spiral arm models from \citep{Prusty2025arXiv}. The top row shows intermediate-mass O-rich AGB stars (gold) and the bottom row shows C-rich AGB stars (salmon). The first column displays the model from \citep{Drimmel2000}, based on K-band/240~$\mu$m emission, showing arms (black) and inter-arm regions (red). The second column shows the maser-based spiral arm model from \citep{Reid2019}, including the Norma (brown), Scutum-Centaurus (violet), Sagittarius–Carina (red), Local (green), Perseus (orange), and Outer (blue) arms. The third column presents the Cepheid-based model from \citep{Drimmel2025}, showing the Scutum–Centaurus (red), Sagittarius–Carina (blue), Orion (orange), and Perseus (green) arms. The rightmost column shows the 2D density maps of the intermediate-mass O-rich and C-rich AGB stars, using a common linear colour scale. Coordinates are Galactocentric, with the Sun at 8.122~kpc (marked by $\odot$) \citep{Gravity2018} and the Galactic centre at (0,0) (marked by $\star$), with a polar grid shown in the background. These comparisons illustrate how intermediate-mass O-rich and C-rich AGB stars trace Galactic spiral arm structure: overdensities are particularly visible along the Perseus arm, with additional indications of alignment with the Local arm, whose exact nature remains debated.}
    \label{spiral_arm_models}
\end{figure*}

The Galactic spiral arms have been traced by stellar populations of both young and older ages \citep{Churchwell2009, Hou2021}. To investigate whether AGB stars trace the Milky Way spiral structure, we compare the spatial distributions of C-rich and O-rich AGB stars with several well-established spiral arm models. In addition to the direct spatial overlays, we present 2D number density maps of the intermediate-mass O-rich and C-rich AGB stars (rightmost column of Fig.~\ref{spiral_arm_models}) showing large-scale overdensities. The spiral arm models were generated using the \texttt{SpiralMap} Python library \citep{Prusty2025arXiv}, which compiles parametrised spiral structures derived from multi-wavelength observations. We selected three models for comparison with older stellar populations: \citet{Drimmel2000}, \citet{Reid2019}, and \citet{Drimmel2025}.

The \citet{Drimmel2000} model is based on K-band and 240~$\mu$m emission profiles of the Galactic plane from the \textit{COsmic Background Explorer} (COBE), identifying two major spiral arms (black) and two inter-arm features (red). The \citet{Reid2019} model is derived from trigonometric parallaxes and proper motions of high-mass star-forming regions observed with Very Long Baseline Interferometry (VLBI). These radio masers are unaffected by extinction and provide improved mapping of the nearby Galactic quadrants. Their model defines seven arms: the 3-kpc, Norma, Scutum-Centaurus, Sagittarius-Carina, Local, Perseus, and Outer arms, each shown in different colours (pink, brown, violet, red, green, orange, and blue) in the plots. However, the number of masers remains limited in the third and fourth Galactic quadrants, particularly in the southern hemisphere, resulting in sparser coverage in this region. Most recently, \citet{Drimmel2025} mapped the spiral structure using $\sim$3,600 classical Cepheids (young; $\sim200$~Myr), using mid-infrared WISE data. Their model traces four arms: Scutum (red), Sagittarius-Carina (blue), Orion/Local (orange), and Perseus (green). To visualise these models, the spatial distributions of the intermediate-mass O-rich and C-rich AGB stars are shown together with the three spiral arm models in Fig.~\ref{spiral_arm_models}.

The intermediate-mass O-rich AGB stars appear to tentatively trace parts of the inner spiral arm structure, with some spatial correspondence along the near-side spiral arms in the \citet{Drimmel2000} model (black curves). The arm structures on the far side of the Galaxy remain difficult to interpret due to limited coverage. In the \citet{Reid2019} model, some O-rich stars show spatial overlap with the Local (green), Sagittarius-Carina (red), and Perseus (orange) arms. We caution, however, that this model is derived from high-mass star-forming regions, and its coverage is sparser in the third and fourth Galactic quadrants. Some sources may also lie near the Outer (blue), Scutum-Centaurus (violet), and Norma (brown) arms, consistent with results found for RC stars of similar ages \citep{Uppal2023}. When compared with the \citet{Drimmel2025} Cepheid model, the O-rich AGB stars show tentative correspondence with the Local (Orion) arm (orange) and possible alignment with the Sagittarius-Carina (blue) arm. 

The spatial distribution of C-rich AGB stars differs markedly from that of the intermediate-mass O-rich AGB stars. The C-rich AGB stars show a smoother and more uniform distribution, particularly around $Y\sim-10$~kpc, suggesting they are more evenly dispersed and less strongly confined to the spiral structure. In the \citet{Drimmel2000} model, the C-rich AGB stars show some spatial overlap with one of the spiral arm features (black), although their broad distribution makes clear associations with inter-arms uncertain. In the \citet{Reid2019} model, some C-rich AGB stars overlap regions corresponding to the Local (green) and Perseus (orange), but the correspondence is weak. Comparison with the \citet{Drimmel2025} Cepheid model, similarly, shows only marginal overlap with the Local (Orion) arm (orange) and Sagittarius-Carina (blue) arm. Overall, the C-rich AGB distribution is too smooth to reliably identify spiral arm features, and a deeper infrared-selected sample less affected by interstellar extinction would be required to draw robust conclusions.

Together, these analyses suggest that intermediate-mass O-rich AGB stars show tentative correspondence with Galactic spiral arm features, while the smoother distribution of C-rich AGB stars makes clear associations with individual arms difficult to establish with the current \textit{Gaia} DR3 sample. 

\section{Conclusions}\label{Conclusions}

We have identified and classified Galactic AGB candidates from {\it Gaia} DR3 as C-rich and O-rich using a two-step approach, classifying 304,090 O-rich and 4,227 C-rich AGB candidates. Our main conclusions are:

- Small, high-quality training samples suffice for accurate classification, even in massive datasets like {\it Gaia} DR3. Cosine similarity effectively selects candidate AGB stars, and combining spectroscopic and photometric features for classification with XGBoost achieves very high classification metrics, outperforming single-feature approaches. This demonstrates that ML can efficiently identify rare populations within millions of sources.

- The ML-classified C-rich and O-rich AGB stars occupy distinct regions in the CMD and {\it Gaia}-2MASS diagram, consistent with stellar evolutionary models. The analysis revealed dusty O-rich and C-rich AGB stars tracing intermediate-age populations (1--3~Gyr). Candidates in the RGB and faint AGB region include a substantial fraction of RGB stars. The Galactic structure analysis is therefore restricted to the AGB regions of the \textit{Gaia}-2MASS diagram. A minor adjustment to the Wesenheit separation curve is suggested to improve the distinction near the C-/O-rich separation boundary.

- The 2D and 3D spatial distributions show that O-rich AGB stars dominate the inner Galaxy, which has relatively higher metallicity, while C-rich AGB stars are concentrated in the outer disc and anti-centre regions. 

- Vertical distribution analysis shows that massive AGB stars are confined near the Galactic mid-plane, while low- and intermediate-mass stars, as well as C-rich stars, are more vertically extended. Derived scale heights are consistent with previous studies \citep{Jackson2002, Uttenthaler2025}.

- C-rich AGB stars trace the Galactic warp at intermediate ages \citep{2026arXiv260610235K}, with warp amplitudes that appear larger than those traced by Cepheids \citep{Skowron2019}. Intermediate-mass O-rich AGB stars show tentative warp signatures but may be affected by extinction.

- Intermediate-mass O-rich AGB stars show tentative spatial correspondence with several spiral arm features, including the Local, Sagittarius-Carina, and Perseus arms across multiple models. The C-rich AGB distribution is too smooth to reliably identify spiral arm features with the current \textit{Gaia} DR3 optically-selected sample.

These findings demonstrate that AGB stars serve as powerful tracers of stellar evolution, Galactic structure, and dynamical processes, offering new insights into the Milky Way’s structure and history. Future work will focus on quantifying mass loss from the classified sources to assess their contribution of dust and gas to the interstellar medium. Higher-quality observations toward the Galactic centre and forthcoming {\it Gaia} DR4 data will further refine these insights.

\section*{Acknowledgements}

TK acknowledges PhD studentship from the UKRI Centre for Doctoral Training in Artificial Intelligence, Machine Learning \& Advanced Computing (AIMLAC) - Grant Ref: EP/S023992/1. M.M. and R.W. acknowledge support from the STFC Consolidated grant (ST/W000830/1). RW acknowledges support from the Research Ireland Pathway programme under grant number 21/PATH-S/9360. JH acknowledges the support of a UKRI Ernest Rutherford Fellowship ST/Z510245/1. DK acknowledges MWGaiaDN, a Horizon Europe Marie Sk\l{}odowska-Curie Actions Doctoral Network funded under grant agreement no. 101072454 and also funded by UK Research and Innovation (EP/X031756/1). This work has made use of data from the European Space Agency (ESA) mission {\it Gaia} (\url{https://www.cosmos.esa.int/gaia}), processed by the {\it Gaia} Data Processing and Analysis Consortium (DPAC, \url{https://www.cosmos.esa.int/web/gaia/dpac/consortium}). Funding for the DPAC has been provided by national institutions, in particular, the institutions participating in the {\it Gaia} Multilateral Agreement. This work has made use of the Python package {\sc GaiaXPy}, developed and maintained by members of the {\it Gaia} Data Processing and Analysis Consortium (DPAC) and, in particular, Coordination Unit 5 (CU5), and the Data Processing Centre located at the Institute of Astronomy, Cambridge, UK (DPCI). This publication makes use of data products from the Two Micron All Sky Survey, which is a joint project of the University of Massachusetts and the Infrared Processing and Analysis Center/California Institute of Technology, funded by the National Aeronautics and Space Administration and the National Science Foundation. This publication makes use of data products from the Wide-field Infrared Survey Explorer, which is a joint project of the University of California, Los Angeles, and the Jet Propulsion Laboratory/California Institute of Technology, and {\it NEOWISE}, which is a project of the Jet Propulsion Laboratory/California Institute of Technology. {\it WISE} and {\it NEOWISE} are funded by the National Aeronautics and Space Administration.

\section*{Data Availability}

The full catalogue of 308,317 ML-classified AGB candidates, including coordinates, parallaxes, photometric magnitudes in \textit{Gaia}, 2MASS and WISE bands, Bailer-Jones geometric distances, ML classification probabilities, and Wesenheit diagram region labels, is available in machine-readable format on Zenodo at \url{https://doi.org/10.5281/zenodo.20336870}, and will also be made available via the CDS VizieR service. A sample of the catalogue is presented in Appendix~\ref{appendix:catalogue} (Table~\ref{tab:catalogue}).



\bibliographystyle{mnras}
\bibliography{example} 

\begin{thebibliography}{}
\makeatletter
\relax
\def\mn@urlcharsother{\let\do\@makeother \do\$\do\&\do\#\do\^\do\_\do\%\do\~}
\def\mn@doi{\begingroup\mn@urlcharsother \@ifnextchar [ {\mn@doi@} {\mn@doi@[]}}
\def\mn@doi@[#1]#2{\def\@tempa{#1}\ifx\@tempa\@empty \href {http://dx.doi.org/#2} {doi:#2}\else \href {http://dx.doi.org/#2} {#1}\fi \endgroup}
\def\mn@eprint#1#2{\mn@eprint@#1:#2::\@nil}
\def\mn@eprint@arXiv#1{\href {http://arxiv.org/abs/#1} {{\tt arXiv:#1}}}
\def\mn@eprint@dblp#1{\href {http://dblp.uni-trier.de/rec/bibtex/#1.xml} {dblp:#1}}
\def\mn@eprint@#1:#2:#3:#4\@nil{\def\@tempa {#1}\def\@tempb {#2}\def\@tempc {#3}\ifx \@tempc \@empty \let \@tempc \@tempb \let \@tempb \@tempa \fi \ifx \@tempb \@empty \def\@tempb {arXiv}\fi \@ifundefined {mn@eprint@\@tempb}{\@tempb:\@tempc}{\expandafter \expandafter \csname mn@eprint@\@tempb\endcsname \expandafter{\@tempc}}}

\bibitem[\protect\citeauthoryear{{Abia}, {de Laverny}, {Cristallo}, {Kordopatis}  \& {Straniero}}{{Abia} et~al.}{2020}]{Abia2020}
{Abia} C.,  {de Laverny} P.,  {Cristallo} S.,  {Kordopatis} G.,   {Straniero} O.,  2020, \mn@doi [\aap] {10.1051/0004-6361/201936831}, \href {https://ui.adsabs.harvard.edu/abs/2020A&A...633A.135A} {633, A135}

\bibitem[\protect\citeauthoryear{{Abia}, {de Laverny}, {Romero-G{\'o}mez}  \& {Figueras}}{{Abia} et~al.}{2022}]{Abia2022}
{Abia} C.,  {de Laverny} P.,  {Romero-G{\'o}mez} M.,   {Figueras} F.,  2022, \mn@doi [\aap] {10.1051/0004-6361/202243595}, \href {https://ui.adsabs.harvard.edu/abs/2022A&A...664A..45A} {664, A45}

\bibitem[\protect\citeauthoryear{{Ahmad} \& {Mazzara}}{{Ahmad} \& {Mazzara}}{2024}]{Ahmad2024}
{Ahmad} M.,  {Mazzara} M.,  2024, \mn@doi [IEEE Geoscience and Remote Sensing Letters] {10.1109/LGRS.2024.3365611}, \href {https://ui.adsabs.harvard.edu/abs/2024IGRSL..2165611A} {21, 3365611}

\bibitem[\protect\citeauthoryear{{Akras}, {Leal-Ferreira}, {Guzman-Ramirez}  \& {Ramos-Larios}}{{Akras} et~al.}{2019}]{Akras2019}
{Akras} S.,  {Leal-Ferreira} M.~L.,  {Guzman-Ramirez} L.,   {Ramos-Larios} G.,  2019, \mn@doi [\mnras] {10.1093/mnras/sty3359}, \href {https://ui.adsabs.harvard.edu/abs/2019MNRAS.483.5077A} {483, 5077}

\bibitem[\protect\citeauthoryear{{Alksnis} \& {Alksne}}{{Alksnis} \& {Alksne}}{1962}]{Alksnis1962}
{Alksnis} A.,  {Alksne} Z.,  1962, Peremennye Zvezdy, \href {https://ui.adsabs.harvard.edu/abs/1962PZ.....14..208A} {14, 208}

\bibitem[\protect\citeauthoryear{{Alksnis}, {Balklavs}, {Dzervitis}, {Eglitis}, {Paupers}  \& {Pundure}}{{Alksnis} et~al.}{2001}]{Alksnis2001}
{Alksnis} A.,  {Balklavs} A.,  {Dzervitis} U.,  {Eglitis} I.,  {Paupers} O.,   {Pundure} I.,  2001, \mn@doi [Baltic Astronomy] {10.1515/astro-2001-1-202}, \href {https://ui.adsabs.harvard.edu/abs/2001BaltA..10....1A} {10, 1}

\bibitem[\protect\citeauthoryear{{Andrae}, {Rix}  \& {Chandra}}{{Andrae} et~al.}{2023}]{Andrae2023}
{Andrae} R.,  {Rix} H.-W.,   {Chandra} V.,  2023, \mn@doi [\apjs] {10.3847/1538-4365/acd53e}, \href {https://ui.adsabs.harvard.edu/abs/2023ApJS..267....8A} {267, 8}

\bibitem[\protect\citeauthoryear{{Bailer-Jones}, {Rybizki}, {Fouesneau}, {Demleitner}  \& {Andrae}}{{Bailer-Jones} et~al.}{2021}]{Bailer-Jones2021}
{Bailer-Jones} C.~A.~L.,  {Rybizki} J.,  {Fouesneau} M.,  {Demleitner} M.,   {Andrae} R.,  2021, \mn@doi [\aj] {10.3847/1538-3881/abd806}, \href {https://ui.adsabs.harvard.edu/abs/2021AJ....161..147B} {161, 147}

\bibitem[\protect\citeauthoryear{{Ball} \& {Brunner}}{{Ball} \& {Brunner}}{2010}]{Ball2010}
{Ball} N.~M.,  {Brunner} R.~J.,  2010, \mn@doi [International Journal of Modern Physics D] {10.1142/S0218271810017160}, \href {https://ui.adsabs.harvard.edu/abs/2010IJMPD..19.1049B} {19, 1049}

\bibitem[\protect\citeauthoryear{{Baron}}{{Baron}}{2019}]{Baron2019}
{Baron} D.,  2019, \mn@doi [arXiv e-prints] {10.48550/arXiv.1904.07248}, \href {https://ui.adsabs.harvard.edu/abs/2019arXiv190407248B} {p. arXiv:1904.07248}

\bibitem[\protect\citeauthoryear{{Battinelli} \& {Demers}}{{Battinelli} \& {Demers}}{2013}]{Battinelli2013_sgr}
{Battinelli} P.,  {Demers} S.,  2013, \mn@doi [\aap] {10.1051/0004-6361/201321059}, \href {https://ui.adsabs.harvard.edu/abs/2013A&A...553A..93B} {553, A93}

\bibitem[\protect\citeauthoryear{{Battinelli}, {Demers}, {Rossi}  \& {Gigoyan}}{{Battinelli} et~al.}{2013}]{Battinelli2013}
{Battinelli} P.,  {Demers} S.,  {Rossi} C.,   {Gigoyan} K.~S.,  2013, \mn@doi [Astrophysics] {10.1007/s10511-013-9268-7}, \href {https://ui.adsabs.harvard.edu/abs/2013Ap.....56...68B} {56, 68}

\bibitem[\protect\citeauthoryear{{Bensby} et~al.,}{{Bensby} et~al.}{2013}]{Bensby2013}
{Bensby} T.,  et~al., 2013, \mn@doi [\aap] {10.1051/0004-6361/201220678}, \href {https://ui.adsabs.harvard.edu/abs/2013A&A...549A.147B} {549, A147}

\bibitem[\protect\citeauthoryear{Bergstra \& Bengio}{Bergstra \& Bengio}{2012}]{Bergstra2012}
Bergstra J.,  Bengio Y.,  2012, J. Mach. Learn. Res., 13, 281–305

\bibitem[\protect\citeauthoryear{{Bessell} \& {Brett}}{{Bessell} \& {Brett}}{1988}]{Bessel1988}
{Bessell} M.~S.,  {Brett} J.~M.,  1988, \mn@doi [\pasp] {10.1086/132281}, \href {https://ui.adsabs.harvard.edu/abs/1988PASP..100.1134B} {100, 1134}

\bibitem[\protect\citeauthoryear{{Bethapudi} \& {Desai}}{{Bethapudi} \& {Desai}}{2018}]{Bethapudi2018}
{Bethapudi} S.,  {Desai} S.,  2018, \mn@doi [Astronomy and Computing] {10.1016/j.ascom.2018.02.002}, \href {https://ui.adsabs.harvard.edu/abs/2018A&C....23...15B} {23, 15}

\bibitem[\protect\citeauthoryear{{Blanco} \& {McCarthy}}{{Blanco} \& {McCarthy}}{1983}]{Blanco1983}
{Blanco} V.~M.,  {McCarthy} M.~F.,  1983, \mn@doi [\aj] {10.1086/113433}, \href {https://ui.adsabs.harvard.edu/abs/1983AJ.....88.1442B} {88, 1442}

\bibitem[\protect\citeauthoryear{{Bland-Hawthorn} \& {Gerhard}}{{Bland-Hawthorn} \& {Gerhard}}{2016}]{Bland-Hawthorn2016}
{Bland-Hawthorn} J.,  {Gerhard} O.,  2016, \mn@doi [\araa] {10.1146/annurev-astro-081915-023441}, \href {https://ui.adsabs.harvard.edu/abs/2016ARA&A..54..529B} {54, 529}

\bibitem[\protect\citeauthoryear{{Bloecker}}{{Bloecker}}{1995}]{Blocker1995}
{Bloecker} T.,  1995, \aap, \href {https://ui.adsabs.harvard.edu/abs/1995A&A...297..727B} {297, 727}

\bibitem[\protect\citeauthoryear{{Blum} et~al.,}{{Blum} et~al.}{2006}]{Blum2006}
{Blum} R.~D.,  et~al., 2006, \mn@doi [\aj] {10.1086/508227}, \href {https://ui.adsabs.harvard.edu/abs/2006AJ....132.2034B} {132, 2034}

\bibitem[\protect\citeauthoryear{{Bovy}, {Leung}, {Hunt}, {Mackereth}, {Garc{\'\i}a-Hern{\'a}ndez}  \& {Roman-Lopes}}{{Bovy} et~al.}{2019}]{Bovy2019}
{Bovy} J.,  {Leung} H.~W.,  {Hunt} J. A.~S.,  {Mackereth} J.~T.,  {Garc{\'\i}a-Hern{\'a}ndez} D.~A.,   {Roman-Lopes} A.,  2019, \mn@doi [\mnras] {10.1093/mnras/stz2891}, \href {https://ui.adsabs.harvard.edu/abs/2019MNRAS.490.4740B} {490, 4740}

\bibitem[\protect\citeauthoryear{{Boyer} et~al.,}{{Boyer} et~al.}{2013}]{Boyer2013}
{Boyer} M.~L.,  et~al., 2013, \mn@doi [\apj] {10.1088/0004-637X/774/1/83}, \href {https://ui.adsabs.harvard.edu/abs/2013ApJ...774...83B} {774, 83}

\bibitem[\protect\citeauthoryear{{Boyer} et~al.,}{{Boyer} et~al.}{2019}]{Boyer2019}
{Boyer} M.~L.,  et~al., 2019, \mn@doi [\apj] {10.3847/1538-4357/ab24e2}, \href {https://ui.adsabs.harvard.edu/abs/2019ApJ...879..109B} {879, 109}

\bibitem[\protect\citeauthoryear{{Bressan}, {Marigo}, {Girardi}, {Salasnich}, {Dal Cero}, {Rubele}  \& {Nanni}}{{Bressan} et~al.}{2012}]{Bressan2012}
{Bressan} A.,  {Marigo} P.,  {Girardi} L.,  {Salasnich} B.,  {Dal Cero} C.,  {Rubele} S.,   {Nanni} A.,  2012, \mn@doi [\mnras] {10.1111/j.1365-2966.2012.21948.x}, \href {https://ui.adsabs.harvard.edu/abs/2012MNRAS.427..127B} {427, 127}

\bibitem[\protect\citeauthoryear{{Brett}}{{Brett}}{1989}]{Brett1989}
{Brett} J.~M.,  1989, \mn@doi [\mnras] {10.1093/mnras/241.2.247}, \href {https://ui.adsabs.harvard.edu/abs/1989MNRAS.241..247B} {241, 247}

\bibitem[\protect\citeauthoryear{{Brett}}{{Brett}}{1990}]{Brett1990}
{Brett} J.~M.,  1990, \aap, \href {https://ui.adsabs.harvard.edu/abs/1990A&A...231..440B} {231, 440}

\bibitem[\protect\citeauthoryear{{Cabrera-Gadea}, {Mateu}, {Ramos}, {Romero-G{\'o}mez}, {Antoja}  \& {Aguilar}}{{Cabrera-Gadea} et~al.}{2024}]{Cabrera2024}
{Cabrera-Gadea} M.,  {Mateu} C.,  {Ramos} P.,  {Romero-G{\'o}mez} M.,  {Antoja} T.,   {Aguilar} L.,  2024, \mn@doi [\mnras] {10.1093/mnras/stae308}, \href {https://ui.adsabs.harvard.edu/abs/2024MNRAS.528.4409C} {528, 4409}

\bibitem[\protect\citeauthoryear{{Cabrera-Gadea}, {Mateu}  \& {Ramos}}{{Cabrera-Gadea} et~al.}{2025}]{Cabrera2025}
{Cabrera-Gadea} M.,  {Mateu} C.,   {Ramos} P.,  2025, \mn@doi [\aap] {10.1051/0004-6361/202452736}, \href {https://ui.adsabs.harvard.edu/abs/2025A&A...701A.136C} {701, A136}

\bibitem[\protect\citeauthoryear{{Cabrera-Lavers} \& {Garz{\'o}n}}{{Cabrera-Lavers} \& {Garz{\'o}n}}{2003}]{Cabrera2003}
{Cabrera-Lavers} A.,  {Garz{\'o}n} F.,  2003, \mn@doi [\aap] {10.1051/0004-6361:20030360}, \href {https://ui.adsabs.harvard.edu/abs/2003A&A...403..383C} {403, 383}

\bibitem[\protect\citeauthoryear{{Cardelli}, {Clayton}  \& {Mathis}}{{Cardelli} et~al.}{1989}]{Cardelli1989}
{Cardelli} J.~A.,  {Clayton} G.~C.,   {Mathis} J.~S.,  1989, \mn@doi [\apj] {10.1086/167900}, \href {https://ui.adsabs.harvard.edu/abs/1989ApJ...345..245C} {345, 245}

\bibitem[\protect\citeauthoryear{{Carrasco} et~al.,}{{Carrasco} et~al.}{2021}]{Carrasco2021}
{Carrasco} J.~M.,  et~al., 2021, \mn@doi [\aap] {10.1051/0004-6361/202141249}, \href {https://ui.adsabs.harvard.edu/abs/2021A&A...652A..86C} {652, A86}

\bibitem[\protect\citeauthoryear{{Chen} \& {Yang}}{{Chen} \& {Yang}}{2012}]{Chen2012}
{Chen} P.~S.,  {Yang} X.~H.,  2012, \mn@doi [\aj] {10.1088/0004-6256/143/2/36}, \href {https://ui.adsabs.harvard.edu/abs/2012AJ....143...36C} {143, 36}

\bibitem[\protect\citeauthoryear{{Chen}, {Yang}  \& {Zhang}}{{Chen} et~al.}{2007}]{Chen2007}
{Chen} P.-S.,  {Yang} X.-H.,   {Zhang} P.,  2007, \mn@doi [\aj] {10.1086/518367}, \href {https://ui.adsabs.harvard.edu/abs/2007AJ....134..214C} {134, 214}

\bibitem[\protect\citeauthoryear{{Chen}, {Bressan}, {Girardi}, {Marigo}, {Kong}  \& {Lanza}}{{Chen} et~al.}{2015}]{Chen2015}
{Chen} Y.,  {Bressan} A.,  {Girardi} L.,  {Marigo} P.,  {Kong} X.,   {Lanza} A.,  2015, \mn@doi [\mnras] {10.1093/mnras/stv1281}, \href {https://ui.adsabs.harvard.edu/abs/2015MNRAS.452.1068C} {452, 1068}

\bibitem[\protect\citeauthoryear{{Chen}, {Wang}, {Deng}, {de Grijs}, {Liu}  \& {Tian}}{{Chen} et~al.}{2019}]{ChenX2019}
{Chen} X.,  {Wang} S.,  {Deng} L.,  {de Grijs} R.,  {Liu} C.,   {Tian} H.,  2019, \mn@doi [Nature Astronomy] {10.1038/s41550-018-0686-7}, \href {https://ui.adsabs.harvard.edu/abs/2019NatAs...3..320C} {3, 320}

\bibitem[\protect\citeauthoryear{{Churchwell} et~al.,}{{Churchwell} et~al.}{2009}]{Churchwell2009}
{Churchwell} E.,  et~al., 2009, \mn@doi [\pasp] {10.1086/597811}, \href {https://ui.adsabs.harvard.edu/abs/2009PASP..121..213C} {121, 213}

\bibitem[\protect\citeauthoryear{{Cioni} \& {Habing}}{{Cioni} \& {Habing}}{2003}]{Cioni2003}
{Cioni} M. R.~L.,  {Habing} H.~J.,  2003, \mn@doi [\aap] {10.1051/0004-6361:20030226}, \href {https://ui.adsabs.harvard.edu/abs/2003A&A...402..133C} {402, 133}

\bibitem[\protect\citeauthoryear{{Cioni}, {van der Marel}, {Loup}  \& {Habing}}{{Cioni} et~al.}{2000}]{Cioni2000}
{Cioni} M. R.~L.,  {van der Marel} R.~P.,  {Loup} C.,   {Habing} H.~J.,  2000, \mn@doi [\aap] {10.48550/arXiv.astro-ph/0003223}, \href {https://ui.adsabs.harvard.edu/abs/2000A&A...359..601C} {359, 601}

\bibitem[\protect\citeauthoryear{{Cioni}, {Girardi}, {Marigo}  \& {Habing}}{{Cioni} et~al.}{2006}]{Cioni2006}
{Cioni} M. R.~L.,  {Girardi} L.,  {Marigo} P.,   {Habing} H.~J.,  2006, \mn@doi [\aap] {10.1051/0004-6361:20053933}, \href {https://ui.adsabs.harvard.edu/abs/2006A&A...448...77C} {448, 77}

\bibitem[\protect\citeauthoryear{{Cioni} et~al.,}{{Cioni} et~al.}{2009}]{Cioni2009}
{Cioni} M. R.~L.,  et~al., 2009, \mn@doi [\aap] {10.1051/0004-6361/200809366e}, \href {https://ui.adsabs.harvard.edu/abs/2009A&A...500.1025C} {500, 1025}

\bibitem[\protect\citeauthoryear{{Cioni} et~al.,}{{Cioni} et~al.}{2014}]{Cioni2014}
{Cioni} M. R.~L.,  et~al., 2014, \mn@doi [\aap] {10.1051/0004-6361/201322100}, \href {https://ui.adsabs.harvard.edu/abs/2014A&A...562A..32C} {562, A32}

\bibitem[\protect\citeauthoryear{{Claussen}, {Kleinmann}, {Joyce}  \& {Jura}}{{Claussen} et~al.}{1987}]{Claussen1987}
{Claussen} M.~J.,  {Kleinmann} S.~G.,  {Joyce} R.~R.,   {Jura} M.,  1987, \mn@doi [\apjs] {10.1086/191229}, \href {https://ui.adsabs.harvard.edu/abs/1987ApJS...65..385C} {65, 385}

\bibitem[\protect\citeauthoryear{{Covey} et~al.,}{{Covey} et~al.}{2007}]{Covey2007}
{Covey} K.~R.,  et~al., 2007, \mn@doi [\aj] {10.1086/522052}, \href {https://ui.adsabs.harvard.edu/abs/2007AJ....134.2398C} {134, 2398}

\bibitem[\protect\citeauthoryear{{Cunha} \& {Humphrey}}{{Cunha} \& {Humphrey}}{2022}]{Cunha2022}
{Cunha} P.~A.~C.,  {Humphrey} A.,  2022, \mn@doi [\aap] {10.1051/0004-6361/202243135}, \href {https://ui.adsabs.harvard.edu/abs/2022A&A...666A..87C} {666, A87}

\bibitem[\protect\citeauthoryear{{De Angeli}, {Montegriffo}, {Palaversa}  \& {Ruz-Mieres}}{{De Angeli} et~al.}{2022}]{gaiaxpy}
{De Angeli} F.,  {Montegriffo} P.,  {Palaversa} L.,   {Ruz-Mieres} D.,  2022, \url{https://gaia-dpci.github.io/GaiaXPy-website/}, \mn@doi{DOI v2.0.1: 10.5281/zenodo.7566303}

\bibitem[\protect\citeauthoryear{{De Angeli} et~al.,}{{De Angeli} et~al.}{2023}]{Deangeli2023}
{De Angeli} F.,  et~al., 2023, \mn@doi [\aap] {10.1051/0004-6361/202243680}, \href {https://ui.adsabs.harvard.edu/abs/2023A&A...674A...2D} {674, A2}

\bibitem[\protect\citeauthoryear{{Deason} \& {Belokurov}}{{Deason} \& {Belokurov}}{2024}]{Deason2024}
{Deason} A.~J.,  {Belokurov} V.,  2024, \mn@doi [\nar] {10.1016/j.newar.2024.101706}, \href {https://ui.adsabs.harvard.edu/abs/2024NewAR..9901706D} {99, 101706}

\bibitem[\protect\citeauthoryear{{Doherty}, {Gil-Pons}, {Siess}  \& {Lattanzio}}{{Doherty} et~al.}{2017}]{Doherty2017}
{Doherty} C.~L.,  {Gil-Pons} P.,  {Siess} L.,   {Lattanzio} J.~C.,  2017, \mn@doi [\pasa] {10.1017/pasa.2017.52}, \href {https://ui.adsabs.harvard.edu/abs/2017PASA...34...56D} {34, e056}

\bibitem[\protect\citeauthoryear{{Drimmel}}{{Drimmel}}{2000}]{Drimmel2000}
{Drimmel} R.,  2000, \mn@doi [\aap] {10.48550/arXiv.astro-ph/0005241}, \href {https://ui.adsabs.harvard.edu/abs/2000A&A...358L..13D} {358, L13}

\bibitem[\protect\citeauthoryear{{Drimmel}, {Khanna}, {Poggio}  \& {Skowron}}{{Drimmel} et~al.}{2025}]{Drimmel2025}
{Drimmel} R.,  {Khanna} S.,  {Poggio} E.,   {Skowron} D.~M.,  2025, \mn@doi [\aap] {10.1051/0004-6361/202451100}, \href {https://ui.adsabs.harvard.edu/abs/2025A&A...698A.230D} {698, A230}

\bibitem[\protect\citeauthoryear{Dwek \& Scalo}{Dwek \& Scalo}{1980}]{Dwek1980}
Dwek E.,  Scalo J.~M.,  1980, \mn@doi [Astrophysical Journal] {10.1086/158100}, 239, 193

\bibitem[\protect\citeauthoryear{{Dyck}, {Benson}, {van Belle}  \& {Ridgway}}{{Dyck} et~al.}{1996}]{Dyck1996}
{Dyck} H.~M.,  {Benson} J.~A.,  {van Belle} G.~T.,   {Ridgway} S.~T.,  1996, \mn@doi [\aj] {10.1086/117910}, \href {https://ui.adsabs.harvard.edu/abs/1996AJ....111.1705D} {111, 1705}

\bibitem[\protect\citeauthoryear{{Ferreira} \& {Costa}}{{Ferreira} \& {Costa}}{2022}]{Ferreira2022}
{Ferreira} T.~A.,  {Costa} C.~A.,  2022, \mn@doi [Classical and Quantum Gravity] {10.1088/1361-6382/ac813d}, \href {https://ui.adsabs.harvard.edu/abs/2022CQGra..39p5013F} {39, 165013}

\bibitem[\protect\citeauthoryear{{Freeman} \& {Bland-Hawthorn}}{{Freeman} \& {Bland-Hawthorn}}{2002}]{Freeman2002}
{Freeman} K.,  {Bland-Hawthorn} J.,  2002, \mn@doi [\araa] {10.1146/annurev.astro.40.060401.093840}, \href {https://ui.adsabs.harvard.edu/abs/2002ARA&A..40..487F} {40, 487}

\bibitem[\protect\citeauthoryear{{GRAVITY Collaboration} et~al.,}{{GRAVITY Collaboration} et~al.}{2018}]{Gravity2018}
{GRAVITY Collaboration} et~al., 2018, \mn@doi [\aap] {10.1051/0004-6361/201833718}, \href {https://ui.adsabs.harvard.edu/abs/2018A&A...615L..15G} {615, L15}

\bibitem[\protect\citeauthoryear{{Gaia Collaboration} et~al.,}{{Gaia Collaboration} et~al.}{2016}]{Prusti2016}
{Gaia Collaboration} et~al., 2016, \mn@doi [\aap] {10.1051/0004-6361/201629272}, \href {https://ui.adsabs.harvard.edu/abs/2016A&A...595A...1G} {595, A1}

\bibitem[\protect\citeauthoryear{{Gaia Collaboration} et~al.,}{{Gaia Collaboration} et~al.}{2018}]{Gaia2018}
{Gaia Collaboration} et~al., 2018, \mn@doi [\aap] {10.1051/0004-6361/201832843}, \href {https://ui.adsabs.harvard.edu/abs/2018A&A...616A..10G} {616, A10}

\bibitem[\protect\citeauthoryear{{Gaia Collaboration} et~al.,}{{Gaia Collaboration} et~al.}{2023a}]{Valenari2023}
{Gaia Collaboration} et~al., 2023a, \mn@doi [\aap] {10.1051/0004-6361/202243940}, \href {https://ui.adsabs.harvard.edu/abs/2023A&A...674A...1G} {674, A1}

\bibitem[\protect\citeauthoryear{{Gaia Collaboration} et~al.,}{{Gaia Collaboration} et~al.}{2023b}]{GaiaDR32023}
{Gaia Collaboration} et~al., 2023b, \mn@doi [\aap] {10.1051/0004-6361/202243940}, \href {https://ui.adsabs.harvard.edu/abs/2023A&A...674A...1G} {674, A1}

\bibitem[\protect\citeauthoryear{{Girardi}, {Groenewegen}, {Hatziminaoglou}  \& {da Costa}}{{Girardi} et~al.}{2005}]{Girardi2005}
{Girardi} L.,  {Groenewegen} M.~A.~T.,  {Hatziminaoglou} E.,   {da Costa} L.,  2005, \mn@doi [\aap] {10.1051/0004-6361:20042352}, \href {https://ui.adsabs.harvard.edu/abs/2005A&A...436..895G} {436, 895}

\bibitem[\protect\citeauthoryear{{Gonneau} et~al.,}{{Gonneau} et~al.}{2016}]{Gonneau2016}
{Gonneau} A.,  et~al., 2016, \mn@doi [\aap] {10.1051/0004-6361/201526292}, \href {https://ui.adsabs.harvard.edu/abs/2016A&A...589A..36G} {589, A36}

\bibitem[\protect\citeauthoryear{{Groenewegen}}{{Groenewegen}}{2002}]{Groenewegen2002}
{Groenewegen} M.~A.~T.,  2002, \mn@doi [arXiv e-prints] {10.48550/arXiv.astro-ph/0208449}, \href {https://ui.adsabs.harvard.edu/abs/2002astro.ph..8449G} {pp astro--ph/0208449}

\bibitem[\protect\citeauthoryear{{Groenewegen}}{{Groenewegen}}{2006}]{Groenewegen2006}
{Groenewegen} M.~A.~T.,  2006, \mn@doi [\aap] {10.1051/0004-6361:20054163}, \href {https://ui.adsabs.harvard.edu/abs/2006A&A...448..181G} {448, 181}

\bibitem[\protect\citeauthoryear{{Groenewegen}}{{Groenewegen}}{2022}]{Groenewegen2022}
{Groenewegen} M.~A.~T.,  2022, \mn@doi [\aap] {10.1051/0004-6361/202142648}, \href {https://ui.adsabs.harvard.edu/abs/2022A&A...659A.145G} {659, A145}

\bibitem[\protect\citeauthoryear{{Groenewegen} \& {Sloan}}{{Groenewegen} \& {Sloan}}{2018}]{Groenewegen2018}
{Groenewegen} M.~A.~T.,  {Sloan} G.~C.,  2018, \mn@doi [\aap] {10.1051/0004-6361/201731089}, \href {https://ui.adsabs.harvard.edu/abs/2018A&A...609A.114G} {609, A114}

\bibitem[\protect\citeauthoryear{{Gruendl}, {Chu}, {Seale}, {Matsuura}, {Speck}, {Sloan}  \& {Looney}}{{Gruendl} et~al.}{2008}]{Gruendl2008}
{Gruendl} R.~A.,  {Chu} Y.~H.,  {Seale} J.~P.,  {Matsuura} M.,  {Speck} A.~K.,  {Sloan} G.~C.,   {Looney} L.~W.,  2008, \mn@doi [\apjl] {10.1086/593979}, \href {https://ui.adsabs.harvard.edu/abs/2008ApJ...688L...9G} {688, L9}

\bibitem[\protect\citeauthoryear{{H. Gomaa} \& {A. Fahmy}}{{H. Gomaa} \& {A. Fahmy}}{2013}]{Gomaa2013}
{H. Gomaa} W.,  {A. Fahmy} A.,  2013, \mn@doi [International Journal of Computer Applications] {10.5120/11638-7118}, \href {https://ui.adsabs.harvard.edu/abs/2013IJCA...68m..13H} {68, 13}

\bibitem[\protect\citeauthoryear{{Habing}}{{Habing}}{1996}]{Habing1996}
{Habing} H.~J.,  1996, \mn@doi [\aapr] {10.1007/PL00013287}, \href {https://ui.adsabs.harvard.edu/abs/1996A&ARv...7...97H} {7, 97}

\bibitem[\protect\citeauthoryear{{Habing} \& {Olofsson}}{{Habing} \& {Olofsson}}{2003}]{Habing2003}
{Habing} H.~J.,  {Olofsson} H.,  eds, 2003, {Asymptotic giant branch stars}

\bibitem[\protect\citeauthoryear{{Habing} \& {Olofsson}}{{Habing} \& {Olofsson}}{2004}]{Habing2004}
{Habing} H.~J.,  {Olofsson} H.,  2004, {Asymptotic Giant Branch Stars}, \mn@doi{10.1007/978-1-4757-3876-6.
}

\bibitem[\protect\citeauthoryear{{Hasselquist} et~al.,}{{Hasselquist} et~al.}{2020}]{Hasselquist2020}
{Hasselquist} S.,  et~al., 2020, \mn@doi [\apj] {10.3847/1538-4357/abaeee}, \href {https://ui.adsabs.harvard.edu/abs/2020ApJ...901..109H} {901, 109}

\bibitem[\protect\citeauthoryear{{Hayden} et~al.,}{{Hayden} et~al.}{2014}]{Hayden2014}
{Hayden} M.~R.,  et~al., 2014, \mn@doi [\aj] {10.1088/0004-6256/147/5/116}, \href {https://ui.adsabs.harvard.edu/abs/2014AJ....147..116H} {147, 116}

\bibitem[\protect\citeauthoryear{He \& Garcia}{He \& Garcia}{2009}]{Haibo2009}
He H.,  Garcia E.~A.,  2009, \mn@doi [IEEE Transactions on Knowledge and Data Engineering] {10.1109/TKDE.2008.239}, 21, 1263

\bibitem[\protect\citeauthoryear{{He}, {Cao}, {Deng}, {Wang}, {Mei}  \& {Tan}}{{He} et~al.}{2024}]{He2024}
{He} Y.,  {Cao} Z.,  {Deng} H.,  {Wang} F.,  {Mei} Y.,   {Tan} L.,  2024, \mn@doi [\apjs] {10.3847/1538-4365/ad6261}, \href {https://ui.adsabs.harvard.edu/abs/2024ApJS..274....6H} {274, 6}

\bibitem[\protect\citeauthoryear{{Helmi}}{{Helmi}}{2020}]{Helmi2020}
{Helmi} A.,  2020, \mn@doi [\araa] {10.1146/annurev-astro-032620-021917}, \href {https://ui.adsabs.harvard.edu/abs/2020ARA&A..58..205H} {58, 205}

\bibitem[\protect\citeauthoryear{{Hernandez}, {Srinivasan}  \& {Marshall}}{{Hernandez} et~al.}{2021}]{Hernandez2021}
{Hernandez} E.~J.,  {Srinivasan} S.,   {Marshall} J.,  2021, in American Astronomical Society Meeting Abstracts. p. 541.06

\bibitem[\protect\citeauthoryear{{Hou}}{{Hou}}{2021}]{Hou2021}
{Hou} L.~G.,  2021, \mn@doi [Frontiers in Astronomy and Space Sciences] {10.3389/fspas.2021.671670}, \href {https://ui.adsabs.harvard.edu/abs/2021FrASS...8..103H} {8, 103}

\bibitem[\protect\citeauthoryear{{Iben} \& {Renzini}}{{Iben} \& {Renzini}}{1983}]{Iben1983}
{Iben} I. J.,  {Renzini} A.,  1983, \mn@doi [\araa] {10.1146/annurev.aa.21.090183.001415}, \href {https://ui.adsabs.harvard.edu/abs/1983ARA&A..21..271I} {21, 271}

\bibitem[\protect\citeauthoryear{{Ishihara}, {Kaneda}, {Onaka}, {Ita}, {Matsuura}  \& {Matsunaga}}{{Ishihara} et~al.}{2011}]{Ishihara2011}
{Ishihara} D.,  {Kaneda} H.,  {Onaka} T.,  {Ita} Y.,  {Matsuura} M.,   {Matsunaga} N.,  2011, \mn@doi [\aap] {10.1051/0004-6361/201117626}, \href {https://ui.adsabs.harvard.edu/abs/2011A&A...534A..79I} {534, A79}

\bibitem[\protect\citeauthoryear{{Jackson}, {Ivezi{\'c}}  \& {Knapp}}{{Jackson} et~al.}{2002}]{Jackson2002}
{Jackson} T.,  {Ivezi{\'c}} {\v{Z}}.,   {Knapp} G.~R.,  2002, \mn@doi [\mnras] {10.1046/j.1365-8711.2002.05980.x}, \href {https://ui.adsabs.harvard.edu/abs/2002MNRAS.337..749J} {337, 749}

\bibitem[\protect\citeauthoryear{{Ji}, {Cui}, {Liu}, {Luo}, {Zhao}  \& {Zhang}}{{Ji} et~al.}{2016}]{Ji2016}
{Ji} W.,  {Cui} W.,  {Liu} C.,  {Luo} A.,  {Zhao} G.,   {Zhang} B.,  2016, \mn@doi [\apjs] {10.3847/0067-0049/226/1/1}, \href {https://ui.adsabs.harvard.edu/abs/2016ApJS..226....1J} {226, 1}

\bibitem[\protect\citeauthoryear{{Kalberla}, {Dedes}, {Kerp}  \& {Haud}}{{Kalberla} et~al.}{2007}]{Kalberla2007}
{Kalberla} P.~M.~W.,  {Dedes} L.,  {Kerp} J.,   {Haud} U.,  2007, \mn@doi [\aap] {10.1051/0004-6361:20066362}, \href {https://ui.adsabs.harvard.edu/abs/2007A&A...469..511K} {469, 511}

\bibitem[\protect\citeauthoryear{{Karakas} \& {Lattanzio}}{{Karakas} \& {Lattanzio}}{2007}]{Karakas2007}
{Karakas} A.,  {Lattanzio} J.~C.,  2007, \mn@doi [\pasa] {10.1071/AS07021}, \href {https://ui.adsabs.harvard.edu/abs/2007PASA...24..103K} {24, 103}

\bibitem[\protect\citeauthoryear{{Karakas} \& {Lattanzio}}{{Karakas} \& {Lattanzio}}{2014}]{Karakas2014}
{Karakas} A.~I.,  {Lattanzio} J.~C.,  2014, \mn@doi [\pasa] {10.1017/pasa.2014.21}, \href {https://ui.adsabs.harvard.edu/abs/2014PASA...31...30K} {31, e030}

\bibitem[\protect\citeauthoryear{{Keenan}}{{Keenan}}{1954}]{Keenan1954}
{Keenan} P.~C.,  1954, \mn@doi [\apj] {10.1086/145937}, \href {https://ui.adsabs.harvard.edu/abs/1954ApJ...120..484K} {120, 484}

\bibitem[\protect\citeauthoryear{{Keenan} \& {Boeshaar}}{{Keenan} \& {Boeshaar}}{1980}]{Keenan1980}
{Keenan} P.~C.,  {Boeshaar} P.~C.,  1980, \mn@doi [\apjs] {10.1086/190673}, \href {https://ui.adsabs.harvard.edu/abs/1980ApJS...43..379K} {43, 379}

\bibitem[\protect\citeauthoryear{{Kim}, {Min}, {Kim}, {Cho}, {Oh}, {Ha}  \& {Yi}}{{Kim} et~al.}{2019}]{Kim2019}
{Kim} B.,  {Min} C.,  {Kim} H.,  {Cho} S.,  {Oh} J.,  {Ha} S.-H.,   {Yi} J.-h.,  2019, \mn@doi [Sensors] {10.3390/s19143047}, \href {https://ui.adsabs.harvard.edu/abs/2019Senso..19.3047K} {19, 3047}

\bibitem[\protect\citeauthoryear{{Knapp}, {Pourbaix}  \& {Jorissen}}{{Knapp} et~al.}{2001}]{Knapp2001}
{Knapp} G.,  {Pourbaix} D.,   {Jorissen} A.,  2001, \mn@doi [\aap] {10.1051/0004-6361:20010348}, \href {https://ui.adsabs.harvard.edu/abs/2001A&A...371..222K} {371, 222}

\bibitem[\protect\citeauthoryear{{Kobayashi}, {Karakas}  \& {Lugaro}}{{Kobayashi} et~al.}{2020}]{Kobayashi2020}
{Kobayashi} C.,  {Karakas} A.~I.,   {Lugaro} M.,  2020, \mn@doi [\apj] {10.3847/1538-4357/abae65}, \href {https://ui.adsabs.harvard.edu/abs/2020ApJ...900..179K} {900, 179}

\bibitem[\protect\citeauthoryear{Krawczyk}{Krawczyk}{2016}]{Krawczyk2016}
Krawczyk B.,  2016, \mn@doi [Progress in Artificial Intelligence] {10.1007/s13748-016-0094-0}, 5, 221

\bibitem[\protect\citeauthoryear{{Kushwahaa}, {Matsuura}, {Hunt}, {Kawata}, {Wesson}  \& {Davis}}{{Kushwahaa} et~al.}{2026}]{2026arXiv260610235K}
{Kushwahaa} T.,  {Matsuura} M.,  {Hunt} J. A.~S.,  {Kawata} D.,  {Wesson} R.,   {Davis} T.~A.,  2026, \mn@doi [arXiv e-prints] {10.48550/arXiv.2606.10235}, \href {https://ui.adsabs.harvard.edu/abs/2026arXiv260610235K} {p. arXiv:2606.10235}

\bibitem[\protect\citeauthoryear{{Ku{\v{c}}inskas}, {Hauschildt}, {Ludwig}, {Brott}, {Vansevi{\v{c}}ius}, {Lindegren}, {Tanab{\'e}}  \& {Allard}}{{Ku{\v{c}}inskas} et~al.}{2005}]{Kucinskas2005}
{Ku{\v{c}}inskas} A.,  {Hauschildt} P.~H.,  {Ludwig} H.-G.,  {Brott} I.,  {Vansevi{\v{c}}ius} V.,  {Lindegren} L.,  {Tanab{\'e}} T.,   {Allard} F.,  2005, \mn@doi [\aap] {10.1051/0004-6361:20053028}, \href {https://ui.adsabs.harvard.edu/abs/2005A&A...442..281K} {442, 281}

\bibitem[\protect\citeauthoryear{{Lan{\c{c}}on} \& {Wood}}{{Lan{\c{c}}on} \& {Wood}}{2000}]{Lancon2000}
{Lan{\c{c}}on} A.,  {Wood} P.~R.,  2000, \mn@doi [\aaps] {10.1051/aas:2000269}, \href {https://ui.adsabs.harvard.edu/abs/2000A&AS..146..217L} {146, 217}

\bibitem[\protect\citeauthoryear{{Lebzelter}, {Mowlavi}, {Marigo}, {Pastorelli}, {Trabucchi}, {Wood}  \& {Lecoeur-Ta{\"\i}bi}}{{Lebzelter} et~al.}{2018}]{Lebzelter2018}
{Lebzelter} T.,  {Mowlavi} N.,  {Marigo} P.,  {Pastorelli} G.,  {Trabucchi} M.,  {Wood} P.~R.,   {Lecoeur-Ta{\"\i}bi} I.,  2018, \mn@doi [\aap] {10.1051/0004-6361/201833615}, \href {https://ui.adsabs.harvard.edu/abs/2018A&A...616L..13L} {616, L13}

\bibitem[\protect\citeauthoryear{{Lebzelter}, {Trabucchi}, {Mowlavi}, {Wood}, {Marigo}, {Pastorelli}  \& {Lecoeur-Ta{\"\i}bi}}{{Lebzelter} et~al.}{2019}]{Lebzelter2019}
{Lebzelter} T.,  {Trabucchi} M.,  {Mowlavi} N.,  {Wood} P.~R.,  {Marigo} P.,  {Pastorelli} G.,   {Lecoeur-Ta{\"\i}bi} I.,  2019, \mn@doi [\aap] {10.1051/0004-6361/201936395}, \href {https://ui.adsabs.harvard.edu/abs/2019A&A...631A..24L} {631, A24}

\bibitem[\protect\citeauthoryear{{Levine}, {Blitz}  \& {Heiles}}{{Levine} et~al.}{2006}]{Levine2006}
{Levine} E.~S.,  {Blitz} L.,   {Heiles} C.,  2006, \mn@doi [\apj] {10.1086/503091}, \href {https://ui.adsabs.harvard.edu/abs/2006ApJ...643..881L} {643, 881}

\bibitem[\protect\citeauthoryear{{Lewis}, {Pihlstr{\"o}m}, {Sjouwerman}, {Stroh}, {Morris}  \& {BAaDE Collaboration}}{{Lewis} et~al.}{2020}]{Lewis2020}
{Lewis} M.~O.,  {Pihlstr{\"o}m} Y.~M.,  {Sjouwerman} L.~O.,  {Stroh} M.~C.,  {Morris} M.~R.,   {BAaDE Collaboration} 2020, \mn@doi [\apj] {10.3847/1538-4357/ab7920}, \href {https://ui.adsabs.harvard.edu/abs/2020ApJ...892...52L} {892, 52}

\bibitem[\protect\citeauthoryear{{Li} et~al.,}{{Li} et~al.}{2018}]{Li2018}
{Li} Y.-B.,  et~al., 2018, \mn@doi [\apjs] {10.3847/1538-4365/aaa415}, \href {https://ui.adsabs.harvard.edu/abs/2018ApJS..234...31L} {234, 31}

\bibitem[\protect\citeauthoryear{{Li}, {Liu}, {Zhang}, {Tian}, {Qiu}  \& {Tian}}{{Li} et~al.}{2021}]{Li2021}
{Li} J.,  {Liu} C.,  {Zhang} B.,  {Tian} H.,  {Qiu} D.,   {Tian} H.,  2021, \mn@doi [\apjs] {10.3847/1538-4365/abe1c1}, \href {https://ui.adsabs.harvard.edu/abs/2021ApJS..253...45L} {253, 45}

\bibitem[\protect\citeauthoryear{{Liu} \& {Xu}}{{Liu} \& {Xu}}{2024}]{Liu2024}
{Liu} A.,  {Xu} J.~C.,  2024, \mn@doi [Research Notes of the American Astronomical Society] {10.3847/2515-5172/ad34b2}, \href {https://ui.adsabs.harvard.edu/abs/2024RNAAS...8...83L} {8, 83}

\bibitem[\protect\citeauthoryear{{Loidl}, {Lan{\c{c}}on}  \& {J{\o}rgensen}}{{Loidl} et~al.}{2001}]{Loidl2001}
{Loidl} R.,  {Lan{\c{c}}on} A.,   {J{\o}rgensen} U.~G.,  2001, \mn@doi [\aap] {10.1051/0004-6361:20010400}, \href {https://ui.adsabs.harvard.edu/abs/2001A&A...371.1065L} {371, 1065}

\bibitem[\protect\citeauthoryear{{Lucey} et~al.,}{{Lucey} et~al.}{2023}]{Lucey2023}
{Lucey} M.,  et~al., 2023, \mn@doi [\mnras] {10.1093/mnras/stad1675}, \href {https://ui.adsabs.harvard.edu/abs/2023MNRAS.523.4049L} {523, 4049}

\bibitem[\protect\citeauthoryear{{Magrini} et~al.,}{{Magrini} et~al.}{2023}]{Magrini2023}
{Magrini} L.,  et~al., 2023, \mn@doi [\aap] {10.1051/0004-6361/202244957}, \href {https://ui.adsabs.harvard.edu/abs/2023A&A...669A.119M} {669, A119}

\bibitem[\protect\citeauthoryear{{Marigo}, {Girardi}  \& {Bressan}}{{Marigo} et~al.}{1999}]{Marigo1999}
{Marigo} P.,  {Girardi} L.,   {Bressan} A.,  1999, \mn@doi [\aap] {10.48550/arXiv.astro-ph/9901235}, \href {https://ui.adsabs.harvard.edu/abs/1999A&A...344..123M} {344, 123}

\bibitem[\protect\citeauthoryear{{Marigo}, {Bressan}, {Girardi}, {Aringer}, {Gullieuszik}  \& {Groenewegen}}{{Marigo} et~al.}{2011}]{Marigo2011}
{Marigo} P.,  {Bressan} A.,  {Girardi} L.,  {Aringer} B.,  {Gullieuszik} M.,   {Groenewegen} M.~A.~T.,  2011, in {Kerschbaum} F.,  {Lebzelter} T.,   {Wing} R.~F.,  eds,  Astronomical Society of the Pacific Conference Series Vol. 445, Why Galaxies Care about AGB Stars II: Shining Examples and Common Inhabitants. p.~431

\bibitem[\protect\citeauthoryear{{Marigo}, {Bressan}, {Nanni}, {Girardi}  \& {Pumo}}{{Marigo} et~al.}{2013}]{Marigo2013}
{Marigo} P.,  {Bressan} A.,  {Nanni} A.,  {Girardi} L.,   {Pumo} M.~L.,  2013, \mn@doi [\mnras] {10.1093/mnras/stt1034}, \href {https://ui.adsabs.harvard.edu/abs/2013MNRAS.434..488M} {434, 488}

\bibitem[\protect\citeauthoryear{{Marigo} et~al.,}{{Marigo} et~al.}{2017}]{Marigo2017}
{Marigo} P.,  et~al., 2017, \mn@doi [\apj] {10.3847/1538-4357/835/1/77}, \href {https://ui.adsabs.harvard.edu/abs/2017ApJ...835...77M} {835, 77}

\bibitem[\protect\citeauthoryear{{Marshall}, {Robin}, {Reyl{\'e}}, {Schultheis}  \& {Picaud}}{{Marshall} et~al.}{2006}]{Marshall2006}
{Marshall} D.~J.,  {Robin} A.~C.,  {Reyl{\'e}} C.,  {Schultheis} M.,   {Picaud} S.,  2006, \mn@doi [\aap] {10.1051/0004-6361:20053842}, \href {https://ui.adsabs.harvard.edu/abs/2006A&A...453..635M} {453, 635}

\bibitem[\protect\citeauthoryear{{Marton} et~al.,}{{Marton} et~al.}{2019}]{Marton2019}
{Marton} G.,  et~al., 2019, \mn@doi [\mnras] {10.1093/mnras/stz1301}, \href {https://ui.adsabs.harvard.edu/abs/2019MNRAS.487.2522M} {487, 2522}

\bibitem[\protect\citeauthoryear{{Matsunaga}, {Menzies}, {Feast}, {Whitelock}, {Onozato}, {Barway}  \& {Aydi}}{{Matsunaga} et~al.}{2017}]{Matsunaga2017}
{Matsunaga} N.,  {Menzies} J.~W.,  {Feast} M.~W.,  {Whitelock} P.~A.,  {Onozato} H.,  {Barway} S.,   {Aydi} E.,  2017, \mn@doi [\mnras] {10.1093/mnras/stx1213}, \href {https://ui.adsabs.harvard.edu/abs/2017MNRAS.469.4949M} {469, 4949}

\bibitem[\protect\citeauthoryear{{Matsuura} et~al.,}{{Matsuura} et~al.}{2009}]{Matsuura2009}
{Matsuura} M.,  et~al., 2009, \mn@doi [\mnras] {10.1111/j.1365-2966.2009.14743.x}, \href {https://ui.adsabs.harvard.edu/abs/2009MNRAS.396..918M} {396, 918}

\bibitem[\protect\citeauthoryear{{Momany}, {Zaggia}, {Gilmore}, {Piotto}, {Carraro}, {Bedin}  \& {de Angeli}}{{Momany} et~al.}{2006}]{Momany2006}
{Momany} Y.,  {Zaggia} S.,  {Gilmore} G.,  {Piotto} G.,  {Carraro} G.,  {Bedin} L.~R.,   {de Angeli} F.,  2006, \mn@doi [\aap] {10.1051/0004-6361:20054081}, \href {https://ui.adsabs.harvard.edu/abs/2006A&A...451..515M} {451, 515}

\bibitem[\protect\citeauthoryear{{Nikolaev} \& {Weinberg}}{{Nikolaev} \& {Weinberg}}{2000}]{Nikolaev2000}
{Nikolaev} S.,  {Weinberg} M.~D.,  2000, \mn@doi [\apj] {10.1086/317048}, \href {https://ui.adsabs.harvard.edu/abs/2000ApJ...542..804N} {542, 804}

\bibitem[\protect\citeauthoryear{{Pastorelli} et~al.,}{{Pastorelli} et~al.}{2019}]{Pastorelli2019}
{Pastorelli} G.,  et~al., 2019, \mn@doi [\mnras] {10.1093/mnras/stz725}, \href {https://ui.adsabs.harvard.edu/abs/2019MNRAS.485.5666P} {485, 5666}

\bibitem[\protect\citeauthoryear{{Pastorelli} et~al.,}{{Pastorelli} et~al.}{2020}]{Pastorelli2020}
{Pastorelli} G.,  et~al., 2020, \mn@doi [\mnras] {10.1093/mnras/staa2565}, \href {https://ui.adsabs.harvard.edu/abs/2020MNRAS.498.3283P} {498, 3283}

\bibitem[\protect\citeauthoryear{Pedregosa et~al.,}{Pedregosa et~al.}{2011}]{scikit-learn2011}
Pedregosa F.,  et~al., 2011, Journal of Machine Learning Research, 12, 2825

\bibitem[\protect\citeauthoryear{{Perryman} et~al.,}{{Perryman} et~al.}{1997}]{Perryman1997}
{Perryman} M.~A.~C.,  et~al., 1997, \aap, \href {https://ui.adsabs.harvard.edu/abs/1997A&A...323L..49P} {323, L49}

\bibitem[\protect\citeauthoryear{{Phillips}}{{Phillips}}{1948}]{Phillip1948}
{Phillips} J.~G.,  1948, \mn@doi [\apj] {10.1086/145023}, \href {https://ui.adsabs.harvard.edu/abs/1948ApJ...107..389P} {107, 389}

\bibitem[\protect\citeauthoryear{{Phillips}}{{Phillips}}{1969}]{Phillip1969}
{Phillips} J.~G.,  1969, \mn@doi [\apj] {10.1086/150079}, \href {https://ui.adsabs.harvard.edu/abs/1969ApJ...157..449P} {157, 449}

\bibitem[\protect\citeauthoryear{{Pietrzy{\'n}ski} et~al.,}{{Pietrzy{\'n}ski} et~al.}{2019}]{Pietrzynski2019}
{Pietrzy{\'n}ski} G.,  et~al., 2019, \mn@doi [\nat] {10.1038/s41586-019-0999-4}, \href {https://ui.adsabs.harvard.edu/abs/2019Natur.567..200P} {567, 200}

\bibitem[\protect\citeauthoryear{{Poelarends}, {Herwig}, {Langer}  \& {Heger}}{{Poelarends} et~al.}{2008}]{Poelarends2008}
{Poelarends} A.~J.~T.,  {Herwig} F.,  {Langer} N.,   {Heger} A.,  2008, \mn@doi [\apj] {10.1086/520872}, \href {https://ui.adsabs.harvard.edu/abs/2008ApJ...675..614P} {675, 614}

\bibitem[\protect\citeauthoryear{{Poggio}, {Drimmel}, {Andrae}, {Bailer-Jones}, {Fouesneau}, {Lattanzi}, {Smart}  \& {Spagna}}{{Poggio} et~al.}{2020}]{Poggio2020}
{Poggio} E.,  {Drimmel} R.,  {Andrae} R.,  {Bailer-Jones} C.~A.~L.,  {Fouesneau} M.,  {Lattanzi} M.~G.,  {Smart} R.~L.,   {Spagna} A.,  2020, \mn@doi [Nature Astronomy] {10.1038/s41550-020-1017-3}, \href {https://ui.adsabs.harvard.edu/abs/2020NatAs...4..590P} {4, 590}

\bibitem[\protect\citeauthoryear{{Poggio} et~al.,}{{Poggio} et~al.}{2025}]{Poggio2025}
{Poggio} E.,  et~al., 2025, \mn@doi [\aap] {10.1051/0004-6361/202451668}, \href {https://ui.adsabs.harvard.edu/abs/2025A&A...699A.199P} {699, A199}

\bibitem[\protect\citeauthoryear{{Prusty} \& {Khanna}}{{Prusty} \& {Khanna}}{2025}]{Prusty2025arXiv}
{Prusty} A.~K.,  {Khanna} S.,  2025, \mn@doi [arXiv e-prints] {10.48550/arXiv.2506.11383}, \href {https://ui.adsabs.harvard.edu/abs/2025arXiv250611383P} {p. arXiv:2506.11383}

\bibitem[\protect\citeauthoryear{{Reid} et~al.,}{{Reid} et~al.}{2019}]{Reid2019}
{Reid} M.~J.,  et~al., 2019, \mn@doi [\apj] {10.3847/1538-4357/ab4a11}, \href {https://ui.adsabs.harvard.edu/abs/2019ApJ...885..131R} {885, 131}

\bibitem[\protect\citeauthoryear{{Reshetnikov} et~al.,}{{Reshetnikov} et~al.}{2025}]{Reshetnikov2025}
{Reshetnikov} V.~P.,  et~al., 2025, \mn@doi [\aap] {10.1051/0004-6361/202554941}, \href {https://ui.adsabs.harvard.edu/abs/2025A&A...697L...1R} {697, L1}

\bibitem[\protect\citeauthoryear{{Romano}, {Karakas}, {Tosi}  \& {Matteucci}}{{Romano} et~al.}{2010}]{Romano2010}
{Romano} D.,  {Karakas} A.~I.,  {Tosi} M.,   {Matteucci} F.,  2010, \mn@doi [\aap] {10.1051/0004-6361/201014483}, \href {https://ui.adsabs.harvard.edu/abs/2010A&A...522A..32R} {522, A32}

\bibitem[\protect\citeauthoryear{{Samus'}, {Kazarovets}, {Durlevich}, {Kireeva}  \& {Pastukhova}}{{Samus'} et~al.}{2017}]{Samus2017}
{Samus'} N.~N.,  {Kazarovets} E.~V.,  {Durlevich} O.~V.,  {Kireeva} N.~N.,   {Pastukhova} E.~N.,  2017, \mn@doi [Astronomy Reports] {10.1134/S1063772917010085}, \href {https://ui.adsabs.harvard.edu/abs/2017ARep...61...80S} {61, 80}

\bibitem[\protect\citeauthoryear{{S{\'a}nchez-Saavedra}, {Battaner}  \& {Florido}}{{S{\'a}nchez-Saavedra} et~al.}{1990}]{Sanchez1990}
{S{\'a}nchez-Saavedra} M.~L.,  {Battaner} E.,   {Florido} E.,  1990, \mnras, \href {https://ui.adsabs.harvard.edu/abs/1990MNRAS.246..458S} {246, 458}

\bibitem[\protect\citeauthoryear{{Sanders} \& {Matsunaga}}{{Sanders} \& {Matsunaga}}{2023}]{Sanders2023}
{Sanders} J.~L.,  {Matsunaga} N.,  2023, \mn@doi [\mnras] {10.1093/mnras/stad574}, \href {https://ui.adsabs.harvard.edu/abs/2023MNRAS.521.2745S} {521, 2745}

\bibitem[\protect\citeauthoryear{{Skowron} et~al.,}{{Skowron} et~al.}{2019}]{Skowron2019}
{Skowron} D.~M.,  et~al., 2019, \mn@doi [Science] {10.1126/science.aau3181}, \href {https://ui.adsabs.harvard.edu/abs/2019Sci...365..478S} {365, 478}

\bibitem[\protect\citeauthoryear{{Skrutskie} et~al.,}{{Skrutskie} et~al.}{2006}]{Skrutskie2006}
{Skrutskie} M.~F.,  et~al., 2006, \mn@doi [\aj] {10.1086/498708}, \href {https://ui.adsabs.harvard.edu/abs/2006AJ....131.1163S} {131, 1163}

\bibitem[\protect\citeauthoryear{{Sloan}, {Little-Marenin}  \& {Price}}{{Sloan} et~al.}{1998}]{Sloan1998}
{Sloan} G.~C.,  {Little-Marenin} I.~R.,   {Price} S.~D.,  1998, \mn@doi [\aj] {10.1086/300205}, \href {https://ui.adsabs.harvard.edu/abs/1998AJ....115..809S} {115, 809}

\bibitem[\protect\citeauthoryear{{Sloan} et~al.,}{{Sloan} et~al.}{2016}]{Sloan2016}
{Sloan} G.~C.,  et~al., 2016, \mn@doi [\apj] {10.3847/0004-637X/826/1/44}, \href {https://ui.adsabs.harvard.edu/abs/2016ApJ...826...44S} {826, 44}

\bibitem[\protect\citeauthoryear{{Son}, {Cho}, {Kim}, {Baek}, {Yoon}  \& {Suh}}{{Son} et~al.}{2025}]{Son2025}
{Son} S.-M.,  {Cho} S.-H.,  {Kim} J.,  {Baek} H.,  {Yoon} D.-H.,   {Suh} K.-W.,  2025, \mn@doi [\aj] {10.3847/1538-3881/ae0586}, \href {https://ui.adsabs.harvard.edu/abs/2025AJ....170..266S} {170, 266}

\bibitem[\protect\citeauthoryear{{Soszy{\'n}ski} et~al.,}{{Soszy{\'n}ski} et~al.}{2005}]{Soszynski2005}
{Soszy{\'n}ski} I.,  et~al., 2005, \mn@doi [\actaa] {10.48550/arXiv.astro-ph/0512578}, \href {https://ui.adsabs.harvard.edu/abs/2005AcA....55..331S} {55, 331}

\bibitem[\protect\citeauthoryear{{Stephenson}}{{Stephenson}}{1973}]{Stephenson1973}
{Stephenson} C.~B.,  1973, Publications of the Warner \& Swasey Observatory, \href {https://ui.adsabs.harvard.edu/abs/1973PW&SO...1.....S} {}

\bibitem[\protect\citeauthoryear{{Stephenson}}{{Stephenson}}{1989}]{Stephenson1989}
{Stephenson} C.~B.,  1989, Publications of the Warner \& Swasey Observatory, \href {https://ui.adsabs.harvard.edu/abs/1989PW&SO...3...53S} {3, 53}

\bibitem[\protect\citeauthoryear{{Suh}}{{Suh}}{2021}]{Suh2021}
{Suh} K.-W.,  2021, \mn@doi [\apjs] {10.3847/1538-4365/ac1274}, \href {https://ui.adsabs.harvard.edu/abs/2021ApJS..256...43S} {256, 43}

\bibitem[\protect\citeauthoryear{Swan}{Swan}{1857}]{Swan1857}
Swan W.,  1857, \mn@doi [Transactions of the Royal Society of Edinburgh] {10.1017/S0080456800032233}, 21, 411–429

\bibitem[\protect\citeauthoryear{{Tabur}, {Bedding}, {Kiss}, {Giles}, {Derekas}  \& {Moon}}{{Tabur} et~al.}{2010}]{Tabur2010}
{Tabur} V.,  {Bedding} T.~R.,  {Kiss} L.~L.,  {Giles} T.,  {Derekas} A.,   {Moon} T.~T.,  2010, \mn@doi [\mnras] {10.1111/j.1365-2966.2010.17341.x}, \href {https://ui.adsabs.harvard.edu/abs/2010MNRAS.409..777T} {409, 777}

\bibitem[\protect\citeauthoryear{{Tang}, {Bressan}, {Rosenfield}, {Slemer}, {Marigo}, {Girardi}  \& {Bianchi}}{{Tang} et~al.}{2014}]{Tang2014}
{Tang} J.,  {Bressan} A.,  {Rosenfield} P.,  {Slemer} A.,  {Marigo} P.,  {Girardi} L.,   {Bianchi} L.,  2014, \mn@doi [\mnras] {10.1093/mnras/stu2029}, \href {https://ui.adsabs.harvard.edu/abs/2014MNRAS.445.4287T} {445, 4287}

\bibitem[\protect\citeauthoryear{{Turnshek}, {Turnshek}  \& {Craine}}{{Turnshek} et~al.}{1985}]{Turnshek1985}
{Turnshek} D.~E.,  {Turnshek} D.~A.,   {Craine} E.~R.,  1985, {An atlas of digital spectra of cool stars}

\bibitem[\protect\citeauthoryear{{Uppal}, {Ganesh}  \& {Schultheis}}{{Uppal} et~al.}{2023}]{Uppal2023}
{Uppal} N.,  {Ganesh} S.,   {Schultheis} M.,  2023, \mn@doi [\aap] {10.1051/0004-6361/202244548}, \href {https://ui.adsabs.harvard.edu/abs/2023A&A...673A..99U} {673, A99}

\bibitem[\protect\citeauthoryear{{Uttenthaler} \& {Merchan-Benitez}}{{Uttenthaler} \& {Merchan-Benitez}}{2025}]{Uttenthaler2025}
{Uttenthaler} S.,  {Merchan-Benitez} P.,  2025, \mn@doi [\aap] {10.1051/0004-6361/202453622}, \href {https://ui.adsabs.harvard.edu/abs/2025A&A...700A.187U} {700, A187}

\bibitem[\protect\citeauthoryear{{Valenti}, {Piskunov}  \& {Johns-Krull}}{{Valenti} et~al.}{1998}]{Valenti1998}
{Valenti} J.~A.,  {Piskunov} N.,   {Johns-Krull} C.~M.,  1998, \mn@doi [\apj] {10.1086/305587}, \href {https://ui.adsabs.harvard.edu/abs/1998ApJ...498..851V} {498, 851}

\bibitem[\protect\citeauthoryear{{Valenti}, {Ferraro}  \& {Origlia}}{{Valenti} et~al.}{2004}]{Valenti2004}
{Valenti} E.,  {Ferraro} F.~R.,   {Origlia} L.,  2004, \mn@doi [\mnras] {10.1111/j.1365-2966.2004.08249.x}, \href {https://ui.adsabs.harvard.edu/abs/2004MNRAS.354..815V} {354, 815}

\bibitem[\protect\citeauthoryear{{Valiante}, {Schneider}, {Bianchi}  \& {Andersen}}{{Valiante} et~al.}{2009}]{Valiante2009}
{Valiante} R.,  {Schneider} R.,  {Bianchi} S.,   {Andersen} A.~C.,  2009, \mn@doi [\mnras] {10.1111/j.1365-2966.2009.15076.x}, \href {https://ui.adsabs.harvard.edu/abs/2009MNRAS.397.1661V} {397, 1661}

\bibitem[\protect\citeauthoryear{{Van Eck}, {Jorissen}, {Udry}, {Mayor}  \& {Pernier}}{{Van Eck} et~al.}{1998}]{VanEck1998}
{Van Eck} S.,  {Jorissen} A.,  {Udry} S.,  {Mayor} M.,   {Pernier} B.,  1998, \mn@doi [\aap] {10.48550/arXiv.astro-ph/9708006}, \href {https://ui.adsabs.harvard.edu/abs/1998A&A...329..971V} {329, 971}

\bibitem[\protect\citeauthoryear{{Vassiliadis} \& {Wood}}{{Vassiliadis} \& {Wood}}{1993}]{Vassiliadis1993}
{Vassiliadis} E.,  {Wood} P.~R.,  1993, \mn@doi [\apj] {10.1086/173033}, \href {https://ui.adsabs.harvard.edu/abs/1993ApJ...413..641V} {413, 641}

\bibitem[\protect\citeauthoryear{{Vergely}, {Lallement}  \& {Cox}}{{Vergely} et~al.}{2022}]{Vergely2022}
{Vergely} J.~L.,  {Lallement} R.,   {Cox} N.~L.~J.,  2022, \mn@doi [\aap] {10.1051/0004-6361/202243319}, \href {https://ui.adsabs.harvard.edu/abs/2022A&A...664A.174V} {664, A174}

\bibitem[\protect\citeauthoryear{Wang, Huang, Zhang, López-Corredoira, Cui, Chen, Guo  \& Chang}{Wang et~al.}{2020}]{Wang2020}
Wang H.-F.,  Huang Y.,  Zhang H.-W.,  López-Corredoira M.,  Cui W.-Y.,  Chen B.-Q.,  Guo R.,   Chang J.,  2020, \mn@doi [The Astrophysical Journal] {10.3847/1538-4357/abb3c8}, 902, 70

\bibitem[\protect\citeauthoryear{Weiss \& Provost}{Weiss \& Provost}{2003}]{Weiss2003}
Weiss G.~M.,  Provost F.~J.,  2003, ArXiv, abs/1106.4557

\bibitem[\protect\citeauthoryear{{Wood} et~al.,}{{Wood} et~al.}{1999}]{1999IAUS..191..151W}
{Wood} P.~R.,  et~al., 1999, in {Le Bertre} T.,  {Lebre} A.,   {Waelkens} C.,  eds,  IAU Symposium Vol. 191, Asymptotic Giant Branch Stars. p.~151

\bibitem[\protect\citeauthoryear{{Woods} et~al.,}{{Woods} et~al.}{2011}]{Woods2011}
{Woods} P.~M.,  et~al., 2011, \mn@doi [\mnras] {10.1111/j.1365-2966.2010.17794.x}, \href {https://ui.adsabs.harvard.edu/abs/2011MNRAS.411.1597W} {411, 1597}

\bibitem[\protect\citeauthoryear{{Wright}, {Barlow}, {Greimel}, {Drew}, {Matsuura}, {Unruh}  \& {Zijlstra}}{{Wright} et~al.}{2009}]{Wright2009}
{Wright} N.~J.,  {Barlow} M.~J.,  {Greimel} R.,  {Drew} J.~E.,  {Matsuura} M.,  {Unruh} Y.~C.,   {Zijlstra} A.~A.,  2009, \mn@doi [\mnras] {10.1111/j.1365-2966.2009.15536.x}, \href {https://ui.adsabs.harvard.edu/abs/2009MNRAS.400.1413W} {400, 1413}

\bibitem[\protect\citeauthoryear{{Wright} et~al.,}{{Wright} et~al.}{2010}]{Wright2010}
{Wright} E.~L.,  et~al., 2010, \mn@doi [\aj] {10.1088/0004-6256/140/6/1868}, \href {https://ui.adsabs.harvard.edu/abs/2010AJ....140.1868W} {140, 1868}

\bibitem[\protect\citeauthoryear{{Yao}, {Ji}, {Koposov}  \& {Limberg}}{{Yao} et~al.}{2024}]{Yao2024}
{Yao} Y.,  {Ji} A.~P.,  {Koposov} S.~E.,   {Limberg} G.,  2024, \mn@doi [\mnras] {10.1093/mnras/stad3775}, \href {https://ui.adsabs.harvard.edu/abs/2024MNRAS.52710937Y} {527, 10937}

\bibitem[\protect\citeauthoryear{{Yogish}, {Manjunath}, {Yogish}  \& {Hegadi}}{{Yogish} et~al.}{2020}]{Yogish2020}
{Yogish} D.,  {Manjunath} T.~N.,  {Yogish} H.~K.,   {Hegadi} R.~S.,  2020, \mn@doi [Journal of Computational and Theoretical Nanoscience] {10.1166/jctn.2020.9330}, \href {https://ui.adsabs.harvard.edu/abs/2020JCTN...17.4531Y} {17, 4531}

\bibitem[\protect\citeauthoryear{{Zamora}, {Abia}, {Plez}, {Dom{\'\i}nguez}  \& {Cristallo}}{{Zamora} et~al.}{2009}]{Zamora2009}
{Zamora} O.,  {Abia} C.,  {Plez} B.,  {Dom{\'\i}nguez} I.,   {Cristallo} S.,  2009, \mn@doi [\aap] {10.1051/0004-6361/200912843}, \href {https://ui.adsabs.harvard.edu/abs/2009A&A...508..909Z} {508, 909}

\bibitem[\protect\citeauthoryear{{Zeraatgari}, {Hafezianzadeh}, {Zhang}, {Mei}, {Ayubinia}, {Mosallanezhad}  \& {Zhang}}{{Zeraatgari} et~al.}{2024}]{Zeraatgari2024}
{Zeraatgari} F.~Z.,  {Hafezianzadeh} F.,  {Zhang} Y.,  {Mei} L.,  {Ayubinia} A.,  {Mosallanezhad} A.,   {Zhang} J.,  2024, \mn@doi [\mnras] {10.1093/mnras/stad3436}, \href {https://ui.adsabs.harvard.edu/abs/2024MNRAS.527.4677Z} {527, 4677}

\bibitem[\protect\citeauthoryear{{Zhang} \& {Green}}{{Zhang} \& {Green}}{2025}]{Zhang2025}
{Zhang} X.,  {Green} G.~M.,  2025, \mn@doi [Science] {10.1126/science.ado9787}, \href {https://ui.adsabs.harvard.edu/abs/2025Sci...387.1209Z} {387, 1209}

\bibitem[\protect\citeauthoryear{{Zhou}, {Chen}, {Deng}, {Wang}, {Wang}  \& {Zhang}}{{Zhou} et~al.}{2025}]{Zhou2025}
{Zhou} X.,  {Chen} X.,  {Deng} L.,  {Wang} S.,  {Wang} J.,   {Zhang} J.,  2025, \mn@doi [\apj] {10.3847/1538-4357/adee1b}, \href {https://ui.adsabs.harvard.edu/abs/2025ApJ...989..213Z} {989, 213}

\makeatother
\end{thebibliography}




\appendix

\section{Removal of M-dwarfs}\label{removal_of_mdwarfs}

\begin{figure}
\centering     
\includegraphics[trim={0.2cm 0.2cm 0.2cm 0cm}, clip, width=.8\linewidth]{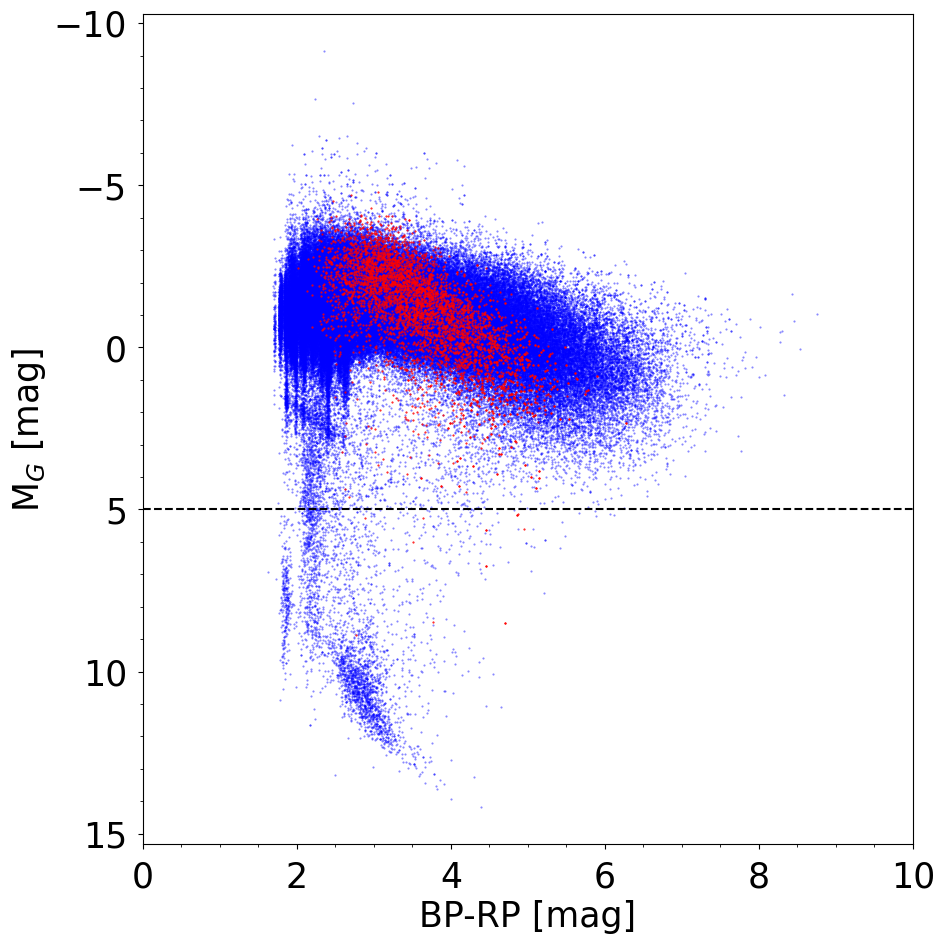}
\caption{Absolute $G$ vs. $BP-RP$ plot of classified C-rich (red) and O-rich (blue) sources. The dashed lines at M$_G$ = 5 separate M-dwarfs at M$_G >$5 from M-giants at M$_G <$5.}
\label{Mdwarfs}
\end{figure}

Fig.~\ref{Mdwarfs} shows a plot of absolute magnitude $G$ versus colour index $BP-RP$ for the classified O-rich sources (blue) and C-rich (red) sources. Two distinct clusters are visible for the O-rich sources. Upon comparison with the study by \citet{Li2021}, we found that the cluster with an absolute \textit{G} magnitude less than 5 represents M-type giants, while the cluster above that value is likely M-dwarfs. The two faint blue clusters, with an absolute $G$ magnitude greater than 5, are contaminated sources identified after binary classification.

Both M giants and M dwarfs exhibit TiO spectral absorption bands. However, we are only interested in the giant stars for this study, so we removed the dwarf sources. When comparing the number of M dwarfs with M giants, we found that the contamination of dwarfs with giants is approximately 1\%.

We also observed a uniformly distributed set of sources between the M-giants and M-dwarfs clusters. These noisy sources likely belong to either M-giants or M-dwarfs, as they display faint TiO features. However, classifying these noisy sources using XP spectra is challenging. Increasing the probability threshold from 90\% to 95\% and then to 99\% helped eliminate sources between M dwarfs and M giants. Nevertheless, the strong similarity between the spectral features of M-giants and M-dwarfs led to the formation of the M-dwarf cluster, even with a high probability cut-off for binary classification.

Increasing the similarity threshold to the 99.999th percentile, which means probing only the top 0.001\% of the most similar sources from the \textit{G2MW} data, significantly reduced the contamination between giants and dwarfs compared to the lower threshold of 99.99th percentile. With a 99.999th percentile similarity threshold, the contamination of M-giants from M-dwarfs decreased to about 3,080 sources ($\sim$1\%). For further analysis, we excluded the M-dwarf sources, reducing our classified O-rich sources to 304,111.

\section{ISM Extinction Correction}\label{ism_extinction_correction}

Extinction correction is essential for accurately assessing the intrinsic properties of AGB stars in the MW, as interstellar dust affects observed magnitudes and colours. To account for this, we obtained the total integrated visual extinctions of our classified C-rich and O-rich AGB stars using the \textit{G-TOMO}\footnote{https://explore-platform.eu/sda/g-tomo} application with the lowest resolution and largest extent map (\textit{v2}) of 50~pc resolution, based on extinction cubes from \citep{Vergely2022}. The map extends 5~kpc from the Sun to its outer edge, with a sampling size of 20~pc. It covers a total volume of 10$\times$10$\times$0.8~kpc, with its height limited to $\pm$400~pc above and below the Galactic plane. 

Using the extinction values obtained from \textit{G-TOMO}, we estimated the effects on various photometric bands. We estimated extinction values for the {\it Gaia} \textit{G}, \textit{BP}, and \textit{RP} magnitudes, as well as the \textit{2MASS} \textit{J}, \textit{H}, and \textit{K$_s$} magnitudes, using equation~1 from \cite{Cardelli1989} with a visual extinction ($A_v$) of 1. We found that the {\it Gaia} DR3 bands are affected by approximately 1.07~mag (\textit{BP}), 0.62~mag (\textit{RP}), and 0.84~mag (\textit{G}), respectively, while the 2MASS bands are affected by roughly 0.29~mag (\textit{J}), 0.18~mag (\textit{H}), and 0.12~mag (\textit{K$_s$}). These extinction effects result in a systematic shift in the CMD. 

It should be noted that the \textit{G-TOMO} extinction cube provides reliable extinction estimates up to approximately 2.5~kpc, with the map extending to 5~kpc from the Sun. As several of our sources have computed distances exceeding 5~kpc, their extinction values are likely underestimated. For distant sources, especially those located toward the inner Galaxy or along the Galactic plane, $A_v$ can reach values as high as 5~mag, meaning that the true extinction may be higher than the cube values. Extinction corrections are applied to all magnitudes presented in the analysis presented in Sect.~\ref{analysis_classified_agb_and_discussions}.

\section{Modified {\it Gaia}-2MASS Diagram}

\begin{figure}
    \centering
    \includegraphics[width=1.\linewidth]{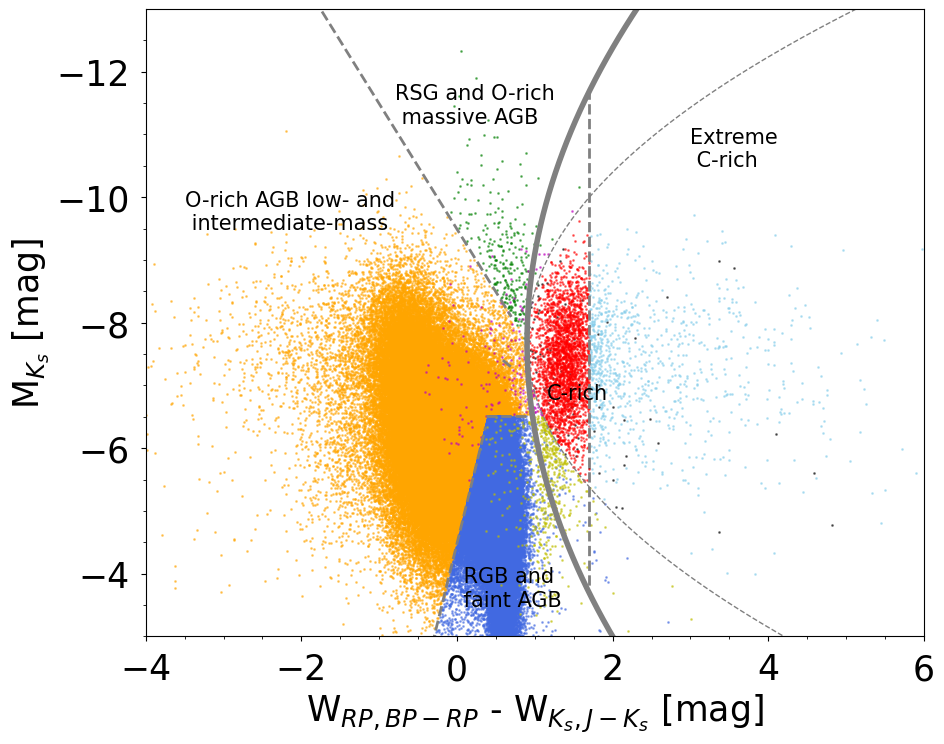}
    \caption{
    \MK vs. \WH diagram showing the classification of C-rich and O-rich AGB stars in the Milky Way. Different colours represent stellar populations occupying distinct regions: orange (low- and intermediate-mass O-rich AGB), green (RSG and massive O-rich AGB), blue (RGB and faint AGB O-rich), red (C-rich AGB), light blue (extreme C-rich AGB), magenta (C-rich stars previously classified as O-rich), and yellow (C-rich stars in the RGB/faint AGB region). The grey solid parabolic line shows our modified classification boundary, while the thin dashed line indicates the original boundary from \citet{Abia2020}.}
    \label{modified_wesenheit}
\end{figure}

We have removed the distinction between the low- and intermediate-mass AGB stars and modified the quadratic equation for the classification curve. The original boundary was defined as $0.9+0.15\times(K_s-10.8)^2$, which we have modified to $0.9+0.05\times(K_s-10.8)^2$ with an increased parabola width. In Figure~\ref{modified_wesenheit}, orange points represent low- and intermediate-mass O-rich AGB stars, green indicates RSG and massive O-rich stars, and blue shows RGB and faint AGB O-rich stars. Red points display C-rich stars, light blue shows extreme C-rich stars, while magenta points indicate C-rich stars that were positioned on the left side of the original curve (shown with a thin dashed line). Similarly, yellow dots represent C-rich stars on the original left side of the curve but occupying the RGB and faint AGB region.

It is evident that many stars situated along the boundary are C-rich according to {\it Gaia} DR3 spectra and SIMBAD identification, and are correctly classified as C-rich by our machine learning algorithm. However, on the diagram, these stars appear slightly outside the c-rich region, indicating a need to modify the boundaries to incorporate them properly. By increasing the width of the parabola (modified curve shown with a solid line), we have incorporated most of the C-rich stars along the classification boundary. As discussed in Sect.~\ref{wesenheit_intro}, some magenta points in the O-rich region likely represent misclassifications due to poor spectral features in the {\it Gaia} DR3 spectra, whereas the black points on the right side indicate O-rich AGB stars that appear redder due to circumstellar dust from mass loss.

\section{AGB Candidate Catalogue}\label{appendix:catalogue}

Table~\ref{tab:catalogue} presents a sample of our ML-classified AGB candidate catalogue, containing 308,317 stars in total (4,227 C-rich and 304,090 O-rich). The catalogue includes \textit{Gaia} DR3 source identifiers, equatorial and Galactic coordinates, parallax, proper motions, ISM extinction from \citet{Vergely2022}, \textit{Gaia}, 2MASS and WISE photometry, Bailer-Jones geometric distances, ML classification probabilities, spectral type (C-rich or O-rich), and Wesenheit diagram region classification. The full catalogue is available in machine-readable format at \url{https://doi.org/10.5281/zenodo.20336870}.

\begin{landscape}
\begin{table}
\scriptsize
\setlength{\tabcolsep}{2pt}
\centering
\caption{Sample of the ML-classified AGB candidate catalogue. The full catalogue of 308,317 stars (4,227 C-rich and 304,090 O-rich) is available in machine-readable format online. Columns are: \textit{Gaia} DR3 source identifier, right ascension (RA), declination (Dec), Galactic longitude ($l$) and latitude ($b$), parallax ($\varpi$) and uncertainty ($\sigma_\varpi$), proper motions in right ascension ($\mu_\alpha$, $\sigma_{\mu_\alpha}$) and declination ($\mu_\delta$, $\sigma_{\mu_\delta}$), ISM extinction $A_0$ from \citet{Vergely2022}, \textit{Gaia} $G$, $G_{BP}$, $G_{RP}$, 2MASS $J$, $H$, $K_s$, and WISE $W1$, $W2$, $W3$ magnitudes, Bailer-Jones geometric distance ($r_{\rm geo}$) with 16th ($b_{\rm geo}$) and 84th ($B_{\rm geo}$) percentile confidence intervals, ML classification probabilities $p(C)$ and $p(O)$, spectral type, and Wesenheit diagram region.}
\label{tab:catalogue}
\begin{tabular}{rrrrrrrrrrrrrrrrrrrrrrrrrrrl}
\hline
Source ID & RA & Dec & $l$ & $b$ &
$\varpi$ & $\sigma_\varpi$ &
$\mu_\alpha$ & $\sigma_{\mu_\alpha}$ &
$\mu_\delta$ & $\sigma_{\mu_\delta}$ &
$A_0$ &
$G$ & $G_{BP}$ & $G_{RP}$ &
$J$ & $H$ & $K_s$ &
$W1$ & $W2$ & $W3$ &
$r_{\rm geo}$ & $b_{\rm geo}$ & $B_{\rm geo}$ &
$p(C)$ & $p(O)$ & Type & Region \\
& (deg) & (deg) & (deg) & (deg) &
(mas) & (mas) &
\multicolumn{2}{c}{(mas~yr$^{-1}$)} &
\multicolumn{2}{c}{(mas~yr$^{-1}$)} &
(mag) &
(mag) & (mag) & (mag) &
(mag) & (mag) & (mag) &
(mag) & (mag) & (mag) &
(pc) & (pc) & (pc) & & & & \\
\hline
5355309406153850112 & 154.1006 & $-$54.9957 & 281.8789 &    1.4099 & 0.196 & 0.028 & $-$4.798 & 0.029 &    3.361 & 0.028 & 1.987 & 12.35 & 14.86 & 10.99 &  8.28 &  6.81 &  6.04 & 5.40 & 5.19 & 4.78 & 5023.9 & 4501.6 & 5709.2 & 0.995 & 0.005 & C & C-rich AGB \\
4156252531085899136 & 277.0697 &  $-$8.5724 &  22.6679 &    1.1924 & 0.414 & 0.057 & $-$1.602 & 0.059 & $-$5.046 & 0.049 & 3.515 & 12.56 & 15.26 & 11.15 &  7.85 &  6.34 &  5.70 & 5.10 & 5.07 & 4.88 & 2385.6 & 2111.5 & 2773.9 & 0.993 & 0.007 & C & C-rich AGB \\
2032045308089973120 & 295.6469 &   29.8888  &  65.0623 &    3.2355 & 0.108 & 0.027 & $-$1.872 & 0.026 & $-$3.708 & 0.030 & 1.329 & 14.57 & 17.14 & 13.17 & 10.13 &  8.68 &  7.88 & 7.30 & 7.38 & 6.77 & 6021.9 & 5297.4 & 7079.2 & 0.994 & 0.006 & C & C-rich AGB \\
3448777031233388672 &  83.5737 &   32.2360  & 175.9098 & $-$0.3100 & 0.184 & 0.053 &    0.526 & 0.060 & $-$0.112 & 0.039 & 2.630 & 13.86 & 16.93 & 12.39 & 10.21 &  8.12 &  6.51 & 4.47 & 3.45 & 2.91 & 4119.4 & 3470.5 & 5067.9 & 0.992 & 0.008 & C & Extreme C-rich \\
 427730505911925760 &  11.4647 &   62.4981  & 122.2879 & $-$0.3666 & 0.143 & 0.043 & $-$1.118 & 0.037 & $-$0.367 & 0.042 & 3.348 & 12.57 & 15.75 & 10.96 &  7.80 &  5.93 &  4.82 & 3.95 & 3.38 & 2.73 & 5881.4 & 4898.3 & 7538.6 & 0.964 & 0.036 & C & Extreme C-rich \\
4658061808357211520 &  81.8765 & $-$69.6318 & 280.2488 & $-$32.5745 & 0.294 & 0.049 &    2.292 & 0.056 &    0.629 & 0.067 & 0.192 & 15.14 & 16.87 & 13.90 & 11.87 & 10.72 & 10.08 & 9.66 & 9.75 & 8.94 & 2912.9 & 2554.6 & 3404.2 & 0.978 & 0.022 & C & RGB and faint AGB (C-rich) \\
 514420228892582784 &  37.0548 &   63.9047  & 133.3323 &    3.0047 & 0.254 & 0.028 & $-$0.654 & 0.018 &    0.819 & 0.028 & 2.139 & 13.57 & 16.61 & 12.14 &  9.17 &  7.86 &  7.34 & 6.97 & 7.20 & 6.84 & 3286.7 & 2949.1 & 3550.9 & 0.977 & 0.023 & C & C-rich in O-rich CMD region \\
5958547905156532992 & 262.5137 & $-$43.9727 & 345.6012 & $-$5.3883  & 0.393 & 0.072 & $-$3.993 & 0.088 & $-$3.398 & 0.072 & 1.112 & 10.79 & 15.29 &  9.08 &  5.68 &  4.47 &  3.97 & 2.42 & 2.61 & 1.85 & 2412.1 & 2016.9 & 2870.6 & 0.013 & 0.987 & O & O-rich AGB low-mass \\
2028999763912876160 & 300.7240 &   28.2664  &  65.9456 & $-$1.4051  & 0.246 & 0.065 & $-$0.815 & 0.053 & $-$2.773 & 0.060 & 3.397 & 13.00 & 17.07 & 11.41 &  7.80 &  6.59 &  5.98 & 5.72 & 5.69 & 4.95 & 3516.4 & 2881.8 & 4513.9 & 0.022 & 0.978 & O & O-rich AGB low-mass \\
2071934795309446400 & 298.7603 &   37.7976  &  73.2181 &    4.9789  & 0.132 & 0.025 & $-$3.293 & 0.026 & $-$5.674 & 0.026 & 0.843 & 11.51 & 13.31 & 10.24 &  8.18 &  7.27 &  6.93 & 6.74 & 7.00 & 6.76 & 6428.0 & 5411.0 & 7449.9 & 0.006 & 0.994 & O & O-rich AGB low-mass \\
2208409103425002368 & 342.7061 &   64.2514  & 110.2900 &    4.3813  & 0.439 & 0.092 & $-$0.200 & 0.108 &    2.637 & 0.103 & 2.391 & 10.54 & 15.97 &  8.78 &  3.83 &  2.60 &  1.89 &$-$0.66 &$-$0.07 & 0.02 & 2215.6 & 1846.4 & 2667.9 & 0.016 & 0.984 & O & O-rich AGB intermediate-mass \\
4056456015422922240 & 267.9086 & $-$29.6640 &   0.0524 & $-$1.4988  & 0.245 & 0.030 &    0.037 & 0.032 & $-$1.761 & 0.021 & 2.815 &  9.78 & 11.37 &  8.58 &  6.49 &  5.50 &  5.16 & 4.66 & 4.87 & 5.02 & 3546.5 & 3209.7 & 3909.3 & 0.011 & 0.989 & O & O-rich AGB intermediate-mass \\
1749624373355974400 & 309.6406 &    9.2476  &  54.3600 & $-$18.8157 & 0.421 & 0.045 &    0.459 & 0.048 &    0.638 & 0.038 & 0.237 & 10.32 & 11.72 &  9.15 &  7.46 &  6.56 &  6.29 & 6.17 & 6.21 & 6.05 & 2152.1 & 1951.3 & 2341.0 & 0.006 & 0.994 & O & RGB and faint AGB (O-rich) \\
3051278357743543168 & 106.3502 &  $-$9.0610 & 222.5805 & $-$1.0485  & 0.277 & 0.023 & $-$1.240 & 0.022 &    1.657 & 0.023 & 1.374 & 12.10 & 13.59 & 10.93 &  8.96 &  7.97 &  7.60 & 7.43 & 7.59 & 7.43 & 3563.0 & 3281.5 & 3820.4 & 0.022 & 0.978 & O & RGB and faint AGB (O-rich) \\
 203352549552907520 &  71.4429 &   41.3299  & 162.8395 & $-$2.7072  & 0.280 & 0.028 & $-$0.307 & 0.035 & $-$1.297 & 0.026 & 1.396 &  8.26 &  9.56 &  7.15 &  5.34 &  4.47 &  4.12 & 4.05 & 3.94 & 3.98 & 3151.3 & 2911.8 & 3415.9 & 0.027 & 0.973 & O & RSG and O-rich massive AGB \\
5862204611892464896 & 199.3586 & $-$63.3202 & 305.8452 & $-$0.5956  & 0.226 & 0.021 & $-$6.180 & 0.017 & $-$1.263 & 0.021 & 2.152 & 10.19 & 11.60 &  9.05 &  7.00 &  6.05 &  5.63 & 5.43 & 5.46 & 5.50 & 3848.4 & 3584.1 & 4148.9 & 0.024 & 0.976 & O & Extreme O-rich (dusty) \\
\multicolumn{28}{c}{(This full table is available in machine-readable form online.)} \\
\hline
\end{tabular}
\end{table}
\end{landscape}

\bsp	
\label{lastpage}
\end{document}